\newcommand{\lab}{{\sc{LoS-ShortRange}}}
\newcommand{\lc}{{\sc{NLoS-ShortRange}}}
\newcommand{\lo}{{\sc{NLoS-LongRange}}}

\newcommand{\WSKG}{{{WSKG}}}

{\color{red}}%
{}

\newcommand{\myitemizebegin}{\begin{list}{$\bullet$}
{
 \setlength{\leftmargin}{0.4cm}
 \setlength{\parsep}{0.0cm}
 \setlength{\itemsep}{0.05cm}
 \setlength{\topsep}{0.0cm}
}}
\newcommand{\myitemizeend}{\end{list}}

\documentclass[transmag]{IEEEtran}
\usepackage{amsthm}
 \usepackage{amsmath}
 \usepackage{amssymb}
\usepackage{mathrsfs}
 \usepackage{mathtools}

\usepackage{cite}

\ifCLASSINFOpdf
   \usepackage[pdftex]{graphicx}
  \DeclareGraphicsExtensions{.pdf,.jpeg,.png}
\else
   \usepackage{graphicx}
   \DeclareGraphicsExtensions{.eps}
\fi

\usepackage{amsmath}
\usepackage{algorithmic}
\usepackage{algorithm}

\usepackage{array}

\ifCLASSOPTIONcompsoc
  \usepackage[caption=false,font=normalsize,labelfont=sf,textfont=sf]{subfig}
\else
\usepackage[caption=false,font=footnotesize]{subfig}
\fi
\usepackage{fixltx2e}

\usepackage{stfloats}
\usepackage{xcolor}
\usepackage{hyperref}
\hypersetup{
    colorlinks=true,
    linkcolor=blue,
    filecolor=magenta,      
    urlcolor=blue,
    pdftitle={Overleaf Example},
    pdfpagemode=FullScreen,
    }

\begin{document}

\title{Wavelet-Based CSI Reconstruction for Improved Wireless Security Through Channel Reciprocity 
\thanks{The accepted version of this article is available at \href{https://doi.org/10.1016/j.cose.2025.104423}{https://doi.org/10.1016/j.cose.2025.104423}.\\
\textcopyright 2025. This manuscript version is made available under the CC-BY-NC-ND 4.0 license \href{https://creativecommons.org/licenses/by-nc-nd/4.0/}{https://creativecommons.org/licenses/by-nc-nd/4.0/}.}
}

\author{Nora Basha${^\dag}$ and Bechir Hamdaoui${^{\dag \ddag}}$\\
 $^\dag$ \small Oregon State University, Corvallis, OR, USA \\
  ${^\ddag}$ \small Hamad Bin Khalifa University, Doha, Qatar
}

\maketitle

\begin{abstract}
The reciprocity of channel state information (CSI) collected by two devices communicating over a wireless channel has been leveraged to provide security solutions to resource-limited  IoT devices. Despite the extensive research that has been done on this topic, much of the focus has been on theoretical and simulation analysis. However, these security solutions face
key implementation challenges, mostly pertaining to limitations of IoT hardware and variations of channel conditions, limiting their practical adoption. To address this research gap, we revisit the channel reciprocity assumption from an experimental standpoint using resource-constrained devices. Our experimental study reveals a significant degradation in channel reciprocity for low-cost devices due to the varying channel conditions. Through
experimental investigations, we first identify key practical causes for the degraded channel reciprocity. We then propose a new wavelet-based CSI reconstruction technique using wavelet coherence and time-lagged cross-correlation to construct CSI data that are consistent between the two participating devices, resulting in significant improvement in channel reciprocity. Additionally, we propose a secret-key generation scheme that exploits the wavelet-based CSI reconstruction, yielding significant increase in the key generation rates. Finally, we propose a technique that exploits CSI temporal variations to enhance device authentication
resiliency through effective detection of replay attacks.
\end{abstract}

\begin{IEEEkeywords}
Wireless key generation, channel reciprocity, wireless \& physical-layer authentication, IoT network security.
\end{IEEEkeywords}

\section{Introduction}
The broadcast nature of wireless communication poses security challenges, demanding robust encryption, secure key management, and authentication. The universal feasibility of Public Key Infrastructure (PKI) and symmetric key cryptography may be limited. Specifically, the increasing number of Internet of Things (IoT) devices face challenges in supporting PKI and symmetric-key cryptography due to factors such as limited power and computing resources and absence of PKI especially for consumer IoT devices, where simplicity and ease of deployment without extensive user intervention are important (Plug and Play) \cite{alladiConsumerIoTSecurity2020}. In response to these challenges, some IoT and wireless devices opt for the utilization of hard-coded cryptographic material, such as encryption keys, authentication tokens, or other cryptographic parameters within the software or firmware of the device. This alternative practice assumes that the cryptographic operations take place in closed, reliable computing environments. Unfortunately, this practice is susceptible to security threats as actual computers and microchips inadvertently leak information about the operations they process \cite{camuratiScreamingChannelsWhen2018} and the pre-stored secrets embedded in the devices can be easily compromised by side-channel attacks \cite{kocherDifferentialPowerAnalysis1999}.

Physical-layer security (PLS) emerges as a potential alternative for ensuring the security of IoT devices by leveraging the characteristics of the wireless channel, such as its reciprocity and temporal variation, to manage the cryptographic material and achieve confidentiality, integrity, and authentication without PKI. 
Among the different approaches in PLS, we highlight techniques that depend on the channel reciprocity assumption, particularly channel reciprocity-based secret-key generation \cite{zhangKeyGenerationWireless2016c} and physical layer authentication \cite{xie_survey_2021}. Channel reciprocity for PLS has been extensively explored in theoretical research. Nevertheless, it is crucial to emphasize that although channel reciprocity presents certain advantages in security applications, its practical adoption necessitates careful consideration of some key challenges that arise from imperfections in IoT devices, variations in environmental conditions, and vulnerabilities that could affect the reciprocity and hinder the actual implementation of the PLS solutions. 

Within this context, this paper reassesses the channel reciprocity assumption experimentally using resource-limited IoT devices, identifies major implementation limitations that diminish channel reciprocity, and proposes metrics to measure and evaluate its quality. Building upon our experimental evaluation of channel reciprocity and proposed metrics, we introduce a Wavelet Transform-based method for enhancing channel reciprocity and showcase its effectiveness by applying it to secret-key generation. 
Additionally, we leverage the temporal variations and time shifts in the collected CSI signals to propose a CSI handshake technique that increases the robustness of device authentication through the detection and prevention of replay (authentication) attacks.


\subsection{Related Works}
PLS taps into the inherent randomness of the wireless channel, electronic circuitry, and radio frequency (RF) hardware components to achieve security functionalities like secret key generation and distribution \cite{li_fast_2022,zoli_physical-layer-security_2020} and physical layer authentication (PLA) \cite{al-meer_physical_2023, basha_channel-resilient_2023}. 
For secret-key generation, significant research effort has focused on generating keys from the randomness of the wireless channel in Time Division Duplex (TDD) systems from the received signal strength (RSS) \cite{zhaoPhysicalLayerKeyGeneration2022,guoLightweightKeyGeneration2021}, channel state information (CSI) \cite{hentilaKeyGenerationSecure2019, delpreteStudyPhysicalLayer2022,chenPhysicalLayerKey2023}, angle of arrival (AoA) and angle of departure (AoD) in MIMO systems \cite{jiaoPhysicalLayerKey2018}, and indices of OFDM subcarriers with the highest gains \cite{furqanNewPhysicalLayer2021}. 
To improve key generation rates, one approach focuses on the quantization step and works on improving the key generation rate by either designing quantization schemes that drop noisy samples causing high-bit mismatch \cite{zhaoPhysicalLayerKeyGeneration2022, junejoLoRaLiSKLightweightShared2022a} or designing adaptive, self-correcting quantization schemes with on-line varying quantization thresholds according to the channel conditions \cite{han_flora_2023, walther_improving_2018}. 
For instance, in \cite{walther_improving_2018} the bits distributions at the output of the quantization step are used to infer biases towards some quantization levels, and hence the quantization thresholds are modified to reduce errors. Another approach is to increase the channel randomness by randomizing the probing signals \cite{huExploitingArtificialRandomness2021}, inducing channel randomness using multiple antennas random scheduling \cite{li_fast_2022}, and optimizing the phases of intelligent reflecting surfaces (IRS) \cite{staatIntelligentReflectingSurfaceAssisted2021}. 

The preprocessing of channel measurements before quantization using curve fitting \cite{zhan_efficient_2018}, the discrete wavelet transform (DWT) \cite{alp_topal_using_2018, kumar_physical_2021} and the fast Fourier transform (FFT) \cite{zhang_efficient_2016} has been proposed to improve key generation performance by using less noisy versions of the channel measurements. However, these techniques assume that only noise and small fluctuations impact channel measurements and degrade key generation performance ignoring the impact of other factors that distort the channel measurements such as fading, delays, and packet loss. Moreover, multiple of those CSI preprocessing techniques drop fixed samples or specific coefficients in the transformed domains to reduce the impact of noise, which may be inefficient given the actual varying channel conditions \cite{zhan_efficient_2018,alp_topal_using_2018, kumar_physical_2021, zhang_efficient_2016}. 

Deep Learning (DL) approaches have also been proposed to increase the key generation rate by training different DL network architectures to learn correlated features between the uplink and downlink and hence generate secret keys between two devices with low bit mismatch. Denoising auto-encoder \cite{hanPhysicalLayerSecret2020c,chenPhysicalLayerAuthentication2024} and Multitask auto-encoder \cite{zhouPhysicalLayerSecret2022} have been used to encode the channel measurements at both devices into highly correlated representations to improve the key generation and boost its security robustness by preventing nearby attackers from generating the same keys.
Bidirectional convolution neural network has been proposed to increase the key generation rate by minimizing the mean squared error between two devices' channel measurements \cite{chenPhysicalLayerSecretKey2023}.
DL approaches have been also proposed to enable secret key generation in Frequency Division Duplex (FDD) systems where the channel is not reciprocal, since the uplink and downlink use two different frequency bands. Early works on DL-based key generation for FDD systems in \cite{zhangDeepLearningBasedPhysicalLayerSecret2022c} proved the existence of a feature mapping function between different frequency bands that could be approximated by a simple feed forward neural network with a single hidden layer to make two users generate correlated channel features in FDD systems. Complex DL models like Generative Adversarial Networks (GANs) \cite{houSecretKeyGeneration2021a} and transfer learning \cite{zhangEnablingDeepLearningBased2024a} have been proposed to enhance the bands feature mapping approximation and enable robust key generation with non-degradable performance in FDD systems in multi-environments.
For PLA, one avenue involves leveraging the inherent manufacturing variations encountered during chip fabrication \cite{liReconfigurableMachineLearning2024} or the RF components of devices to generate physical fingerprints suitable for authentication \cite{ibrahim_mag-auth_2023, basha_channel-resilient_2023}. Another avenue entails utilizing channel-extracted features and employs the disparities in CSI between the authorized channel and the eavesdropping channel for authentication \cite{du_physical_2014, luPhysicalLayerAuthenticationBased2023b, xiePhysicalLayerAuthenticationUsing2021}.
Despite the promising results observed in PLS approaches with multiple wireless protocols such as WiFi \cite{guillaume_bringing_2015,zhang_experimental_2016} and LPWAN \cite{han_flora_2023,han_lora-based_2020,peng_secure_2020}, a substantial portion of the research in secret key generation and utilizing channel-extracted features for 

PLA has predominantly focused on theoretical and simulation analysis \cite{kumar_physical_2021}, or key generation implementation using simulated datasets such as the DeepMIMO dataset \cite{alkhateebDeepMIMOGenericDeep2019}.

While theoretical and simulation works contribute to advancing the comprehension of PLS, it is imperative to substantiate these findings through extensive real-world experiments and practical implementations. Considerations related to practical deployment, hardware limitations, and environmental variables are pivotal in assessing the efficacy of these techniques and the validity of their assumptions in practical settings. 

\subsection{Contributions}
The contributions lie in tackling the gap in PLS testbed implementation and in proposing new techniques for key generation and authentication that overcome these implementation limitations. Specifically, our main contributions are:
\myitemizebegin
    
    \item We expand the channel reciprocity assumption assessment in \cite{10901753, 10901137}, utilizing low-cost microcontroller boards with limited resources. We identify implementation limitations that impact channel reciprocity negatively, and investigate metrics that best measure and quantify reciprocity. Our findings show that Pearson's correlation, the time-lagged cross-correlation, and the Wavelet Coherence (WC) can effectively quantify the impact of the identified limitations on the channel reciprocity quality. We also provide the collected CSI datasets utilized in this study for the research community.
    
    \item We propose a Wavelet Transform (WT)-based CSI reconstruction framework that leverages WC time and frequency information for channel reciprocity enhancement. We also demonstrate the need for CSI data synchronization to provide further reciprocity improvement and show the superiority of the proposed technique to other existing approaches, such as CSI constructions via Golay filtering, FFT, and two of the state-of-the art AI-enabled key generation approaches.        
    
    \item 
%
    We propose a secret-key generation scheme that exploits the proposed wavelet-based CSI data reconstruction and synchronization to improve the bit error bit rate (BER) and the key generation rate (KGR) performances.

    
\myitemizeend
%

The paper is structured as follows: Sec.~\ref{sec:testbed} presents the experimental setup for CSI collection and channel reciprocity evaluation. Sec.~\ref{sec:metrics_analysis} analyzes CSI data behavior, investigates the causes that impact the quality of channel reciprocity, and identifies the metrics that quantify the channel reciprocity. Sec.~\ref{sec:technique} presents a Wavelet Transform-based CSI reconstruction framework that enhances the channel reciprocity. Sec.~\ref{sec:skg} presents our proposed Wavelet-based secret-key generation scheme, and Sec.~\ref{sec:replay} presents our proposed CSI handshake authentication and evaluates its effectiveness and ability in detecting and preventing replay attacks. Finally, Sec.~\ref{sec:conc} concludes the paper.

\section{Experimental Setup and Scenarios}
\label{sec:testbed}
\begin{figure}
\centerline{
\subfloat[Lopy device\label{subfig:device}]{%
\includegraphics[keepaspectratio, height = 3.5 cm]{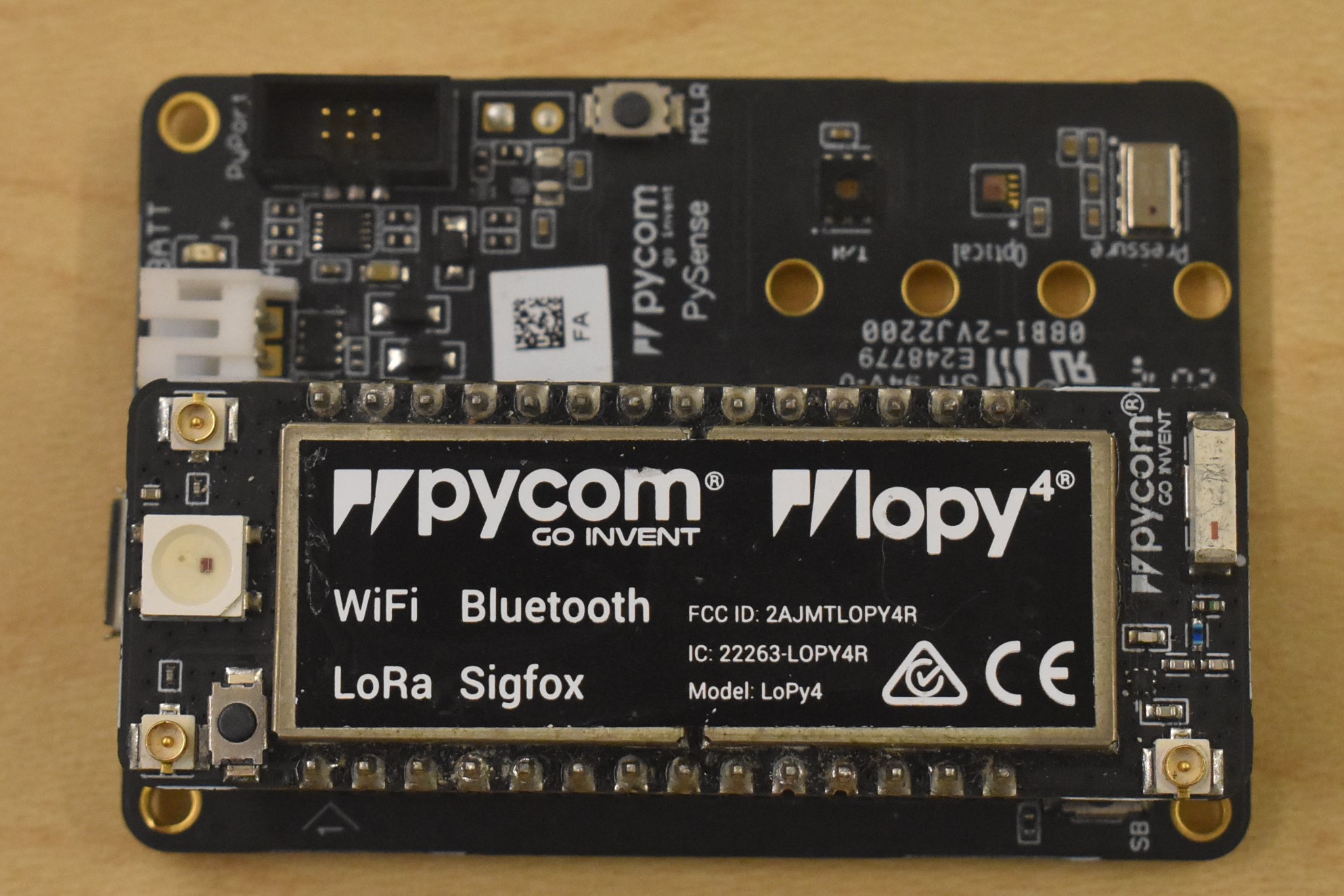} }
\subfloat[CSI collection\label{subfig:CSI_setting}]{%
\includegraphics[keepaspectratio, height=3.5 cm]{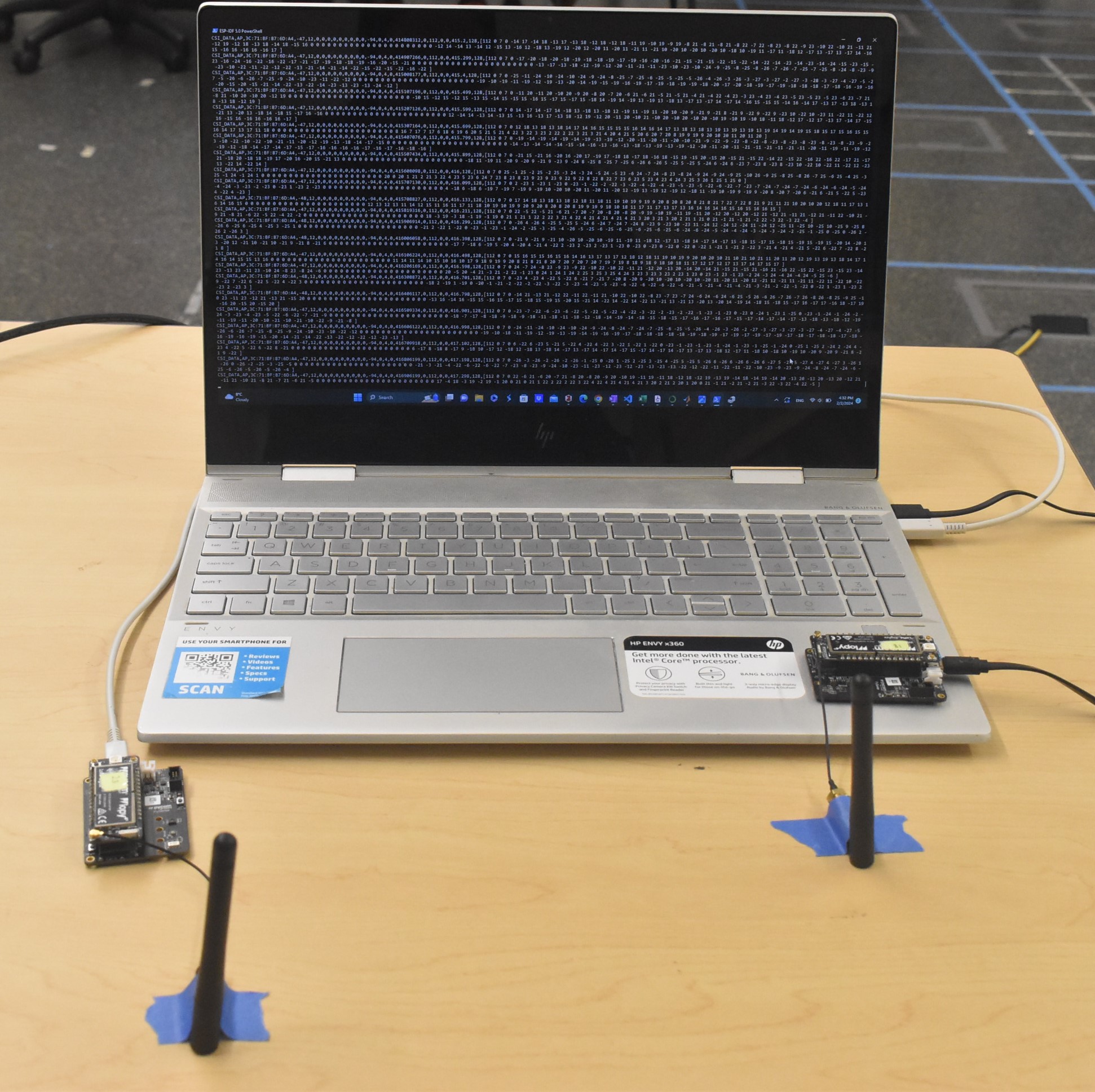}}
}

\caption{Experiment setup.}
\label{fig:exp1}
\end{figure} 

For our experimental evaluations, we use an experimental testbed of two Pycom devices (LoPy model 4 devices connected to PySense extension boards depicted in Fig~\ref{subfig:device}), one serving as an Access Point (AP) and one as a station (STA). AP and STA communicate using IEEE 802.11n WiFi protocol at $2.427$ GHz.
Both AP and STA exchange packets at a rate of 10 packets per second and collect CSI (Channel State Information) using the ESP23 CSI toolkit \cite{Hern2006:Lightweight}.

The reason we fixed the packet rate to $10$ packets per second is merely due to hardware limitations. We observed that packet rates exceeding $10$ packets per second for Pycom boards during CSI collection resulted in high packet loss (missing CSI for multiple packets) that limited key generation and severely impaired channel reciprocity.
Moreover, even though the rate of $10$ packets per seconds does not reflect the varying traffic patterns for IoT networks, CSI data collection for security purposes in IoT networks must be done at rates that ensure the reliability of the security solution even if the rates are different from those used for data communication. Our measurements indicate that this can be achieved by a rate of $10$ packets per second.

In our experiments, both devices are connected to a Windows machine for data collection and processing via USB ports at a baud rate of $115200$ as depicted in Fig.~\ref{subfig:CSI_setting}.
The devices estimate CSI using both the non-HT Legacy Long Training field (L-LTF) and the High Throughput Long Training Field (HT-LTF) of the WiFi physical layer frame, i.e., AP and STA receive $10$ channel measurements per second, with each measurement consisting of $64$ L-LTF CSI complex values and 128 HT-LTF CSI complex values corresponding to the In-phase and the Quadrature components (IQ) of the $64$ OFDM subcarriers. 
The CSI datasets were collected on the same building floor, considering 3 location scenarios:
\myitemizebegin
    \item \lab: A line-of-sight (LoS) scenario, where both AP and STA are located in a room of size $9$ meters $\times$ $9.6$ meters and separated by a distance of $6.5$ meters.
    \item \lc: A non-line of sight (NLoS) scenario, where AP is located in a room and STA is located in an adjacent corridor with scarce human movement.
    \item \lo: A NLoS scenario, where AP and STA are in two different rooms, and the two rooms are about $13$ meters apart with frequent human movement.
\myitemizeend
%
Four experiments were performed for each scenario, each lasting $1$ hour. The reported results represent the averages from these experiments. The IQ values of the CSI are used to determine the CSI magnitude. Throughout the paper, the focus is on studying channel reciprocity specifically for subcarrier index 6 of the CSI. While experiments were conducted across multiple subcarriers, and all yielded similar and consistent results, we present findings for a single subcarrier (index 6) to avoid redundancy, with the understanding that the metrics and techniques discussed can be extended to other subcarriers and transmission modes \cite{zhang_efficient_2016}. 
The collected CSI datasets used in this paper are accessible for use by the research community and available to download at NetSTAR lab at \href{https://research.engr.oregonstate.edu/hamdaoui/datasets}{https://research.engr.oregonstate.edu/hamdaoui/datasets}.
%


\section{Channel Reciprocity Assessment}
\label{sec:metrics_analysis}
\begin{figure}
    \centering
    \includegraphics[width=1\linewidth]{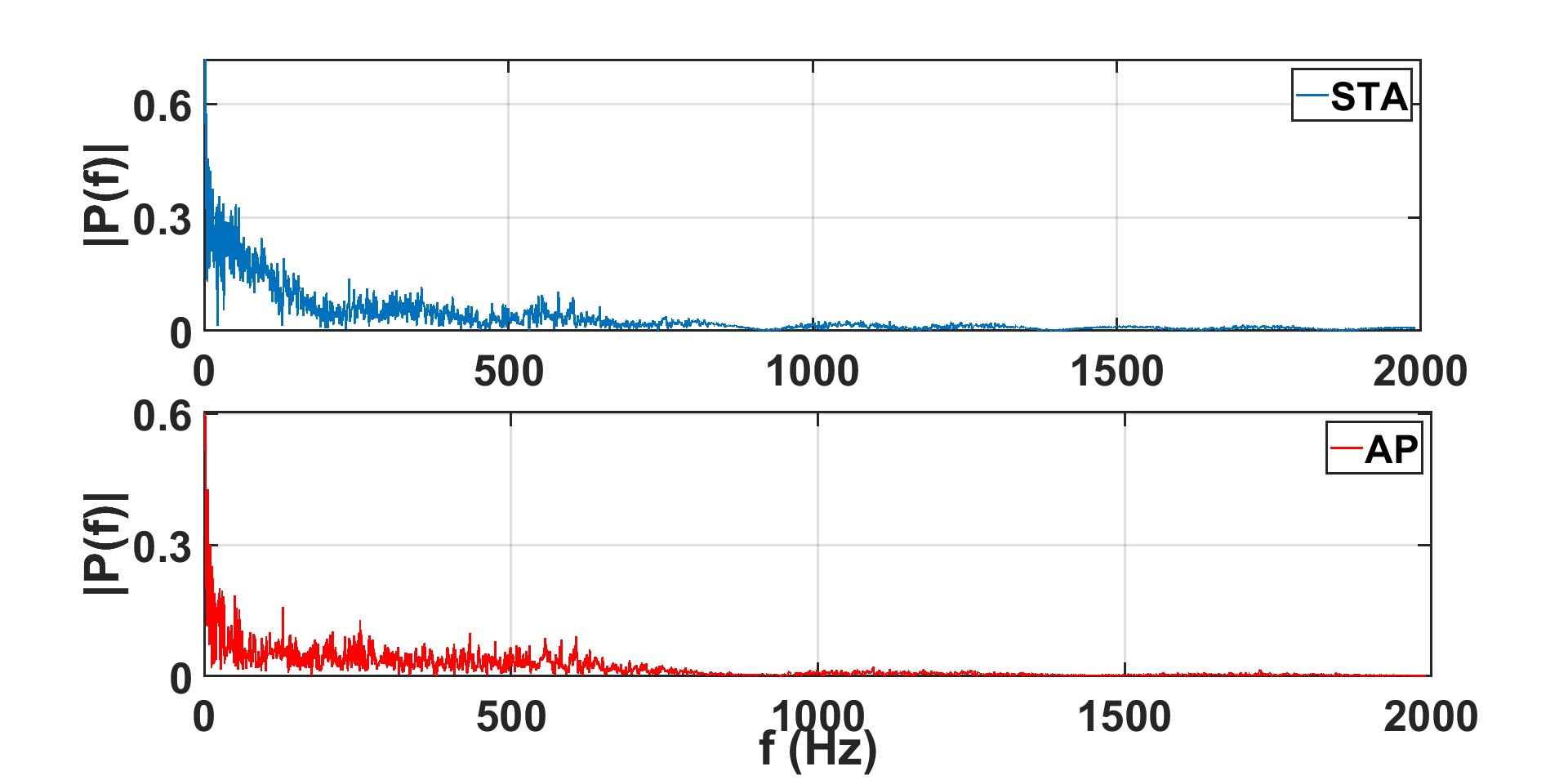}
    \caption{    
    Power spectrum of CSI at AP and STA: \lab~scenario.}
    \label{fig:fft}
\end{figure}

Channel reciprocity refers to the symmetry property of the wireless channel between a transmitter-receiver pair, like the STA-AP pair in our case. Thus, a perfect reciprocity corresponds to when the channel characteristics from STA to AP and from AP to STA are the same. However, 
in practice, channel reciprocity may not be perfect due to various factors like the presence of obstacles, movement, fading, noise, and device impairments, which may affect, either entirely or partially, the symmetry property of the channel. 
Therefore, to leverage this symmetry property for designing practical PLS solutions, there is a need for metrics that can effectively quantify the wireless channel reciprocity through collected CSI data.
In this section, we focus on studying and figuring out which metrics could best serve this purpose.
For this, we analyze and compare various metrics while using raw CSI as well as preprocessed CSI data using the following previously proposed denoising techniques in PLS:
\myitemizebegin
    \item \textbf{Golay Filtering \cite{schaferWhatSavitzkyGolayFilter2011}}: This technique proposed to improve secret key generation in \cite{junejoLoRaLiSKLightweightShared2022a} smooths noisy CSI signals based on local least-squares polynomial approximation, eliminates noise spikes at AP and STA and at the same time preserves the CSI signals high-frequency variations. 
    \item \textbf{FFT-based Reconstruction with Low Frequency Components \cite{zhang_efficient_2016}}: The single-sided power spectrum (Fig.~\ref{fig:fft}) reveals that low-frequency components ($0-800$ Hz) dominate CSI at both AP and STA. Removing high-frequency components facilitates the reconstruction of less noisy, more reciprocal CSI signals. As a preprocessing step, both AP and STA reconstruct the CSI signal (via inverse FFT), considering only low-frequency components contributing to the CSI power above a predefined threshold.
\myitemizeend
\subsection{Reciprocity Assessment}
In this section, four reciprocity assessment metrics (Pearson's correlation, Jeffrey's divergence, Wasserstein distance, and wavelet coherence) are experimentally evaluated for their effectiveness in capturing the impact of channel impairments on channel reciprocity.
Given that our goal is to leverage channel reciprocity for enabling security, we choose to compare the studied reciprocity assessment metrics using key generation performance metrics. 
We aim to find which reciprocity metric yields results consistent with the performance of key generation that is based on the same channel. 
Accurate reciprocity metrics would measure high reciprocity (which could be high correlation, low divergence, or small distance) when high key generation rate and low bit error rate are observed under raw and preprocessed CSI and in different location scenarios.

Recall that in channel reciprocity-based key generation, each of the two communicating devices, STA and AP, first estimates its CSI based on its observed channel and then converts it into a stream of bits or keys using a quantization scheme \cite{zhangKeyGenerationWireless2016c}. 

The secret key generation flow for the raw CSI, Golay filtering, and FFT-reconstruction does not involve computing Pearson's correlation, Jeffrey's divergence, or Wasserstein distance.

%
For our channel reciprocity assessment, we use the following two secret-key generation metrics:
\myitemizebegin
\item \textbf{Bit Error Rate (BER)}: denotes the number of mismatched bits between the two keys divided by the key length.

\item \textbf{Key Generation Rate (KGR)}: denotes the number of bits generated per packet/measurement. The higher the rate of keys whose mismatched bit rates (or BERs) do not exceed a certain bit-error threshold, the higher the KGR.
For the sake of evaluation, in this section, KGR and BER are calculated at a bit-error threshold of $20$ bits and a key length of $200$ bits.
\myitemizeend

\subsubsection{Pearson's Correlation~\cite{asuero_correlation_2006}}
Fig. \ref{subfig:pearson}, which depicts Pearson's correlation of the CSI for three locations scenarios under raw and preprocessed CSI, and Fig.~\ref{fig:SKG_metrics}, which depicts the corresponding BER and KGR performance, illustrate that the correlation between AP and STA for raw CSI can drop as low as $0.15$ indicating that, in practical scenarios, channel reciprocity is severely diminished which limits KGR, and increases BER.   
The figures demonstrate that Pearson's correlation gauges channel reciprocity accurately. It captures the linear dependency between AP's CSI and STA's CSI and correlates the effects of CSI preprocessing and AP's and STA's locations concisely with BER and KGR, with high correlation values corresponding to low BERs and high KGRs. 
%
Figs. \ref{subfig:pearson} and \ref{fig:SKG_metrics} also show that any CSI preprocessing technique that increases the correlation improves the BER and KGR metrics. For example, using the Golay-filtered CSI data increases the correlation to $0.65$ under \lab~and increases KGR from $5\times 10^{-4}$ to $2 \times 10^{-3}$, as shown in Fig. \ref{subfig:KGR_metrics}. The same trend is observed for the FFT-reconstructed CSI.
\subsubsection{Jeffrey's Divergence~\cite{jeffreysInvariantFormPrior1997}} 
\begin{figure} 
    \centerline{
  \subfloat[Pearson\label{subfig:pearson}]{%
       \includegraphics[height=0.1\textwidth,width=0.17\textwidth]{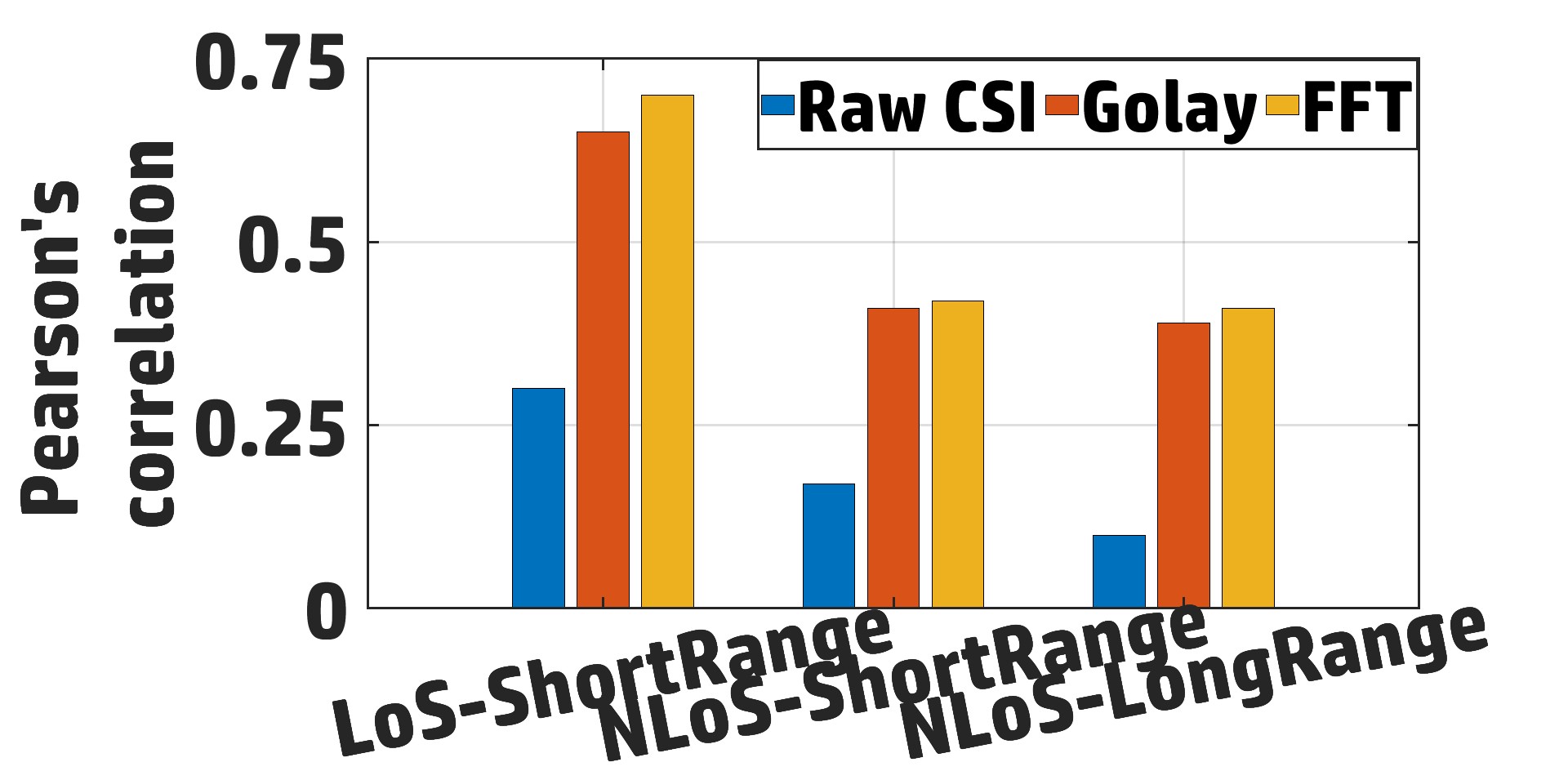}}
    \hspace{-0.1in}
  \subfloat[Jeffery\label{subfig:jeff}]{%
        \includegraphics[height=0.1\textwidth,width=0.17\textwidth]{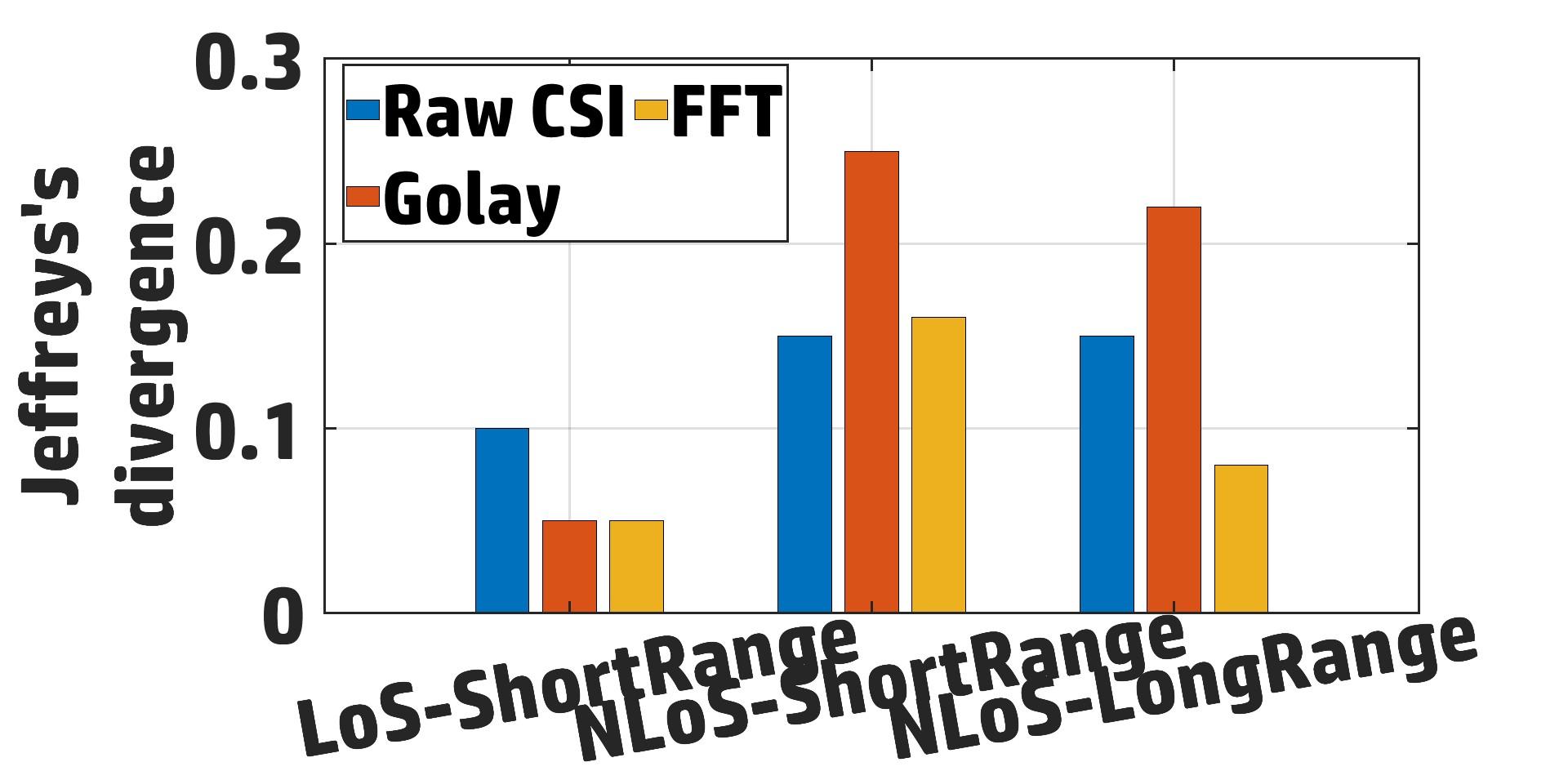}}
   \hspace{-0.1in}
  \subfloat[Wasserstein\label{subfig:wass}]{%
        \includegraphics[height=0.1\textwidth,width=0.17\textwidth]{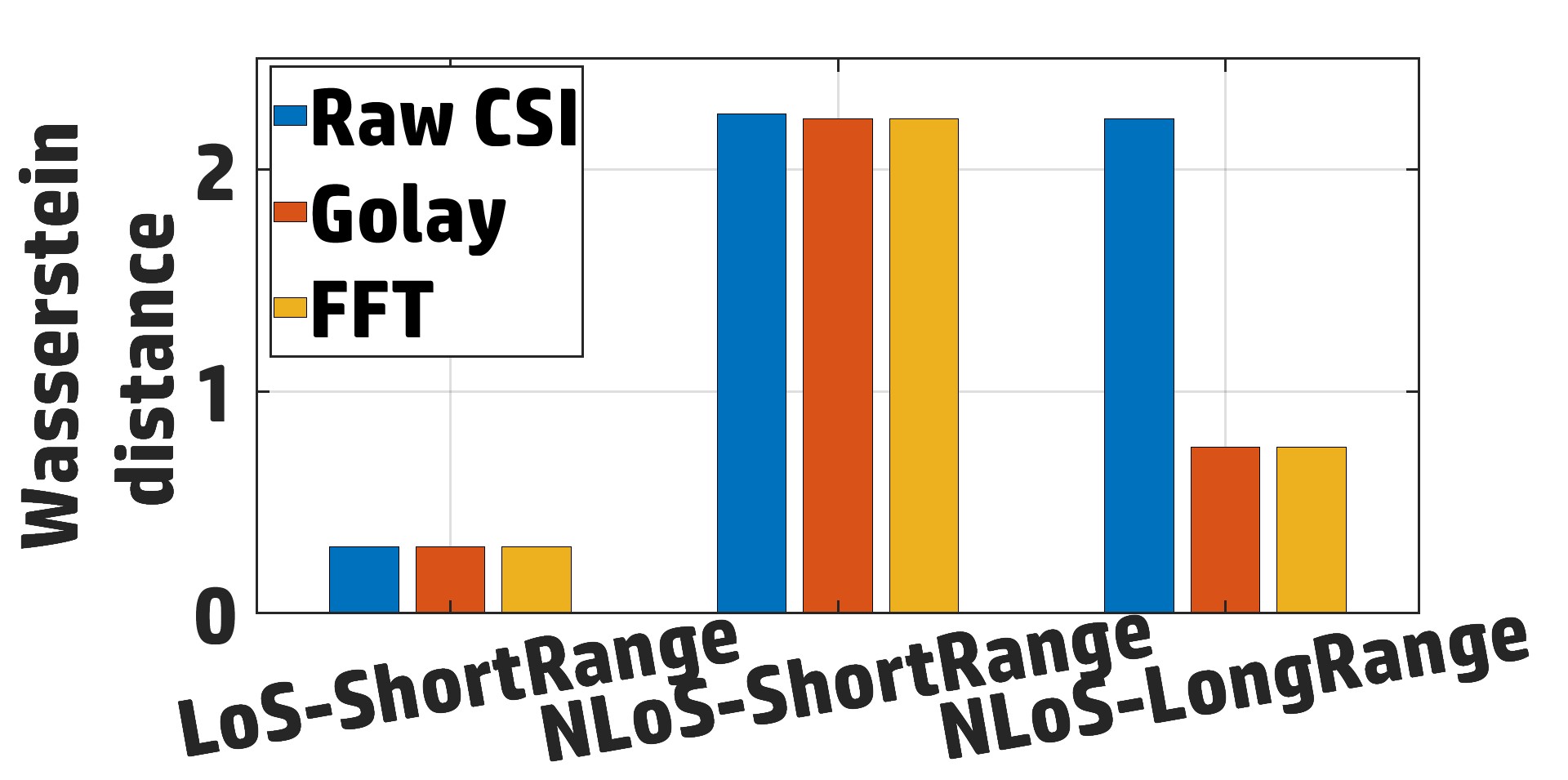}}}
        
    
   \caption{  Channel reciprocity assessment metrics. }
  \label{fig:rec_metrics} 
\end{figure}

\begin{figure} 
        \centerline{
  \subfloat[BER\label{subfig:BER_metrics}]{%
        \includegraphics[keepaspectratio, width=0.25\textwidth]{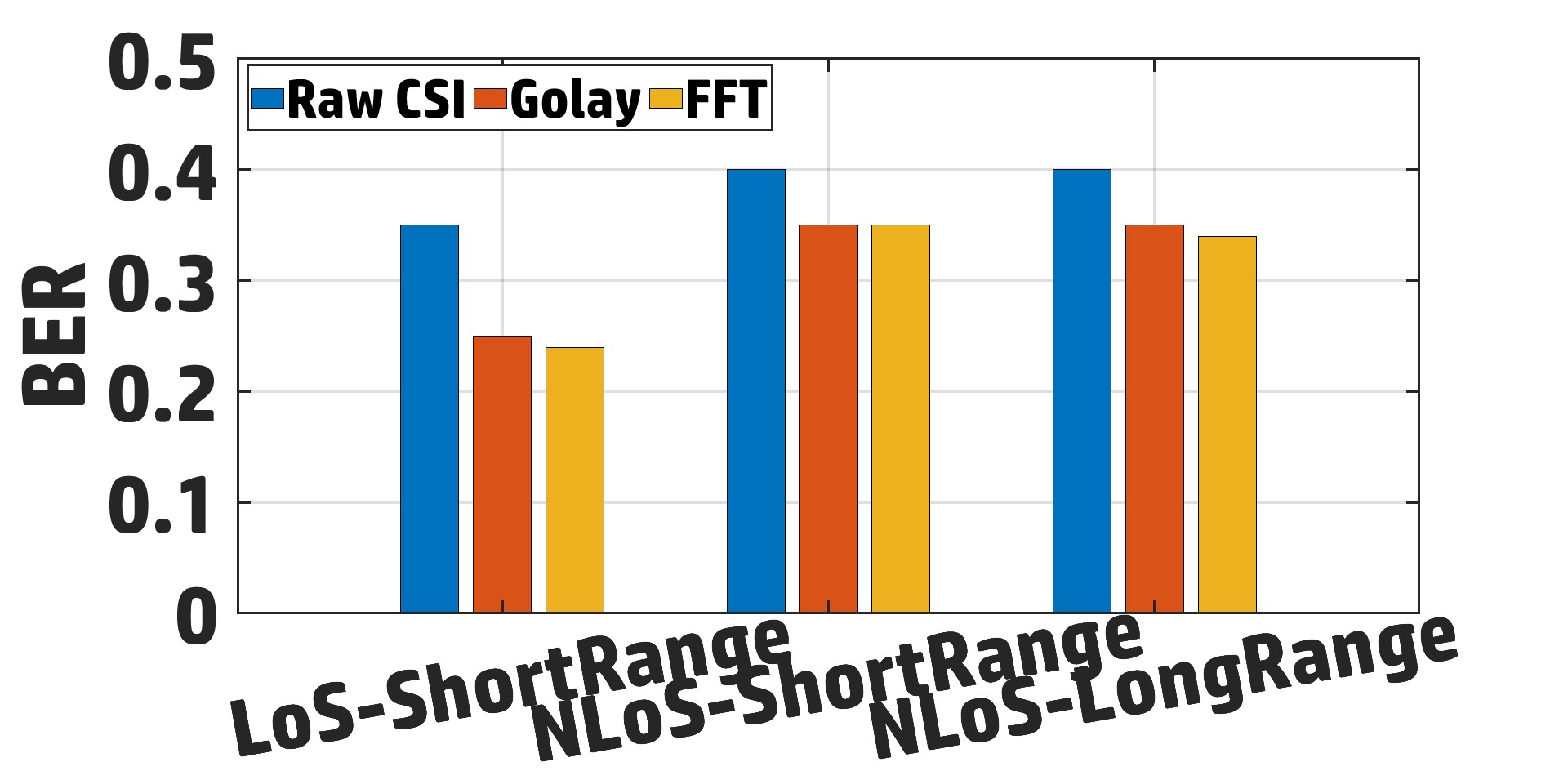}}
\hspace{-0.1in}
    
  \subfloat[KGR\label{subfig:KGR_metrics}]{%
        \includegraphics[keepaspectratio, width=0.25\textwidth]{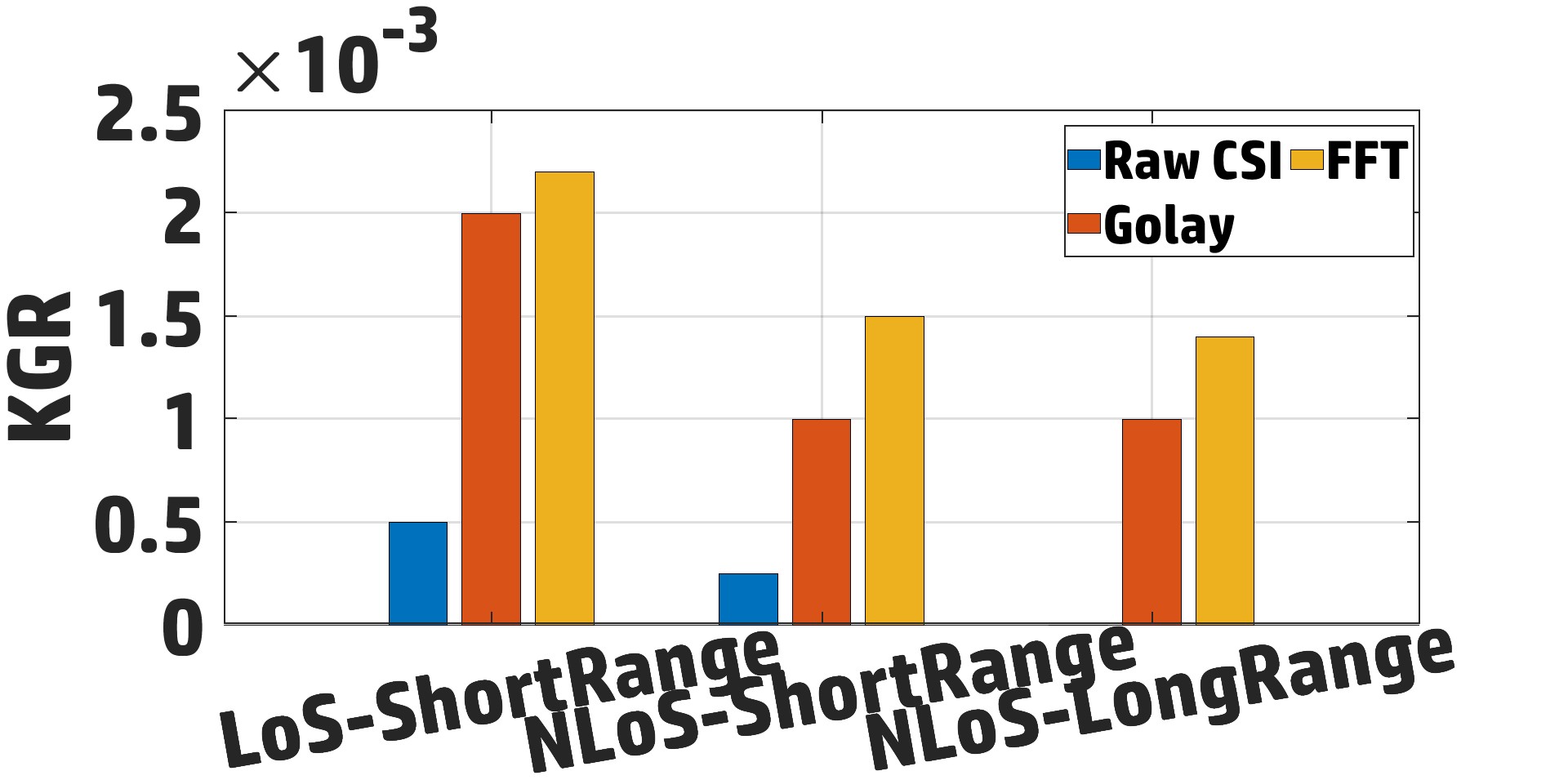}}}
        
   \caption{  Key generation performance metrics.  }
  \label{fig:SKG_metrics} 
\end{figure}

%
Fig.~\ref{subfig:jeff} shows Jeffrey's divergence measured under raw and preprocessed CSI data, and Fig.~\ref{fig:SKG_metrics} shows the corresponding BER and KGR.
Jeffrey's divergence ($D_J$) is the arithmetic symmetrization of the Kullback-leibler (KL) divergence \cite{kullback_information_1951, murphy2012machine}:
    \begin{equation*}
    D_{J} = \frac{D_{KL}(P||Q)+ D_{KL}(Q||P)}{2}
    \end{equation*}
where $D_{KL}(P||Q) = \sum_X P(X) \log \frac{P(X)}{Q(X)}$ is the KL divergence of the distribution of the channel measurements $X$ at AP $P(X)$ and the channel measurements distribution at STA $Q(X)$.
The figures illustrate that Jeffrey's divergence, which measures the disparity between the CSI distributions, can somewhat capture the variations in CSI caused by changes in the distance between the AP and STA, but falls short in capturing the impact of the CSI preprocessing techniques.
For instance, in Fig. \ref{subfig:jeff}, high divergence values under raw CSI correspond to low KGR (Fig. \ref{subfig:KGR_metrics}) and high BER (Fig. \ref{subfig:BER_metrics}) for each of the three locations scenarios. 
However, Fig. \ref{subfig:jeff} indicates that Golay filtering increases the divergence under \lc, suggesting a worsening of channel reciprocity. Nonetheless, KGR and BER under \lc, shown in Fig. \ref{fig:SKG_metrics}, reveal that Golay filtering decreases BER, and increases KGR.
This inconsistency between the BER and KGR metrics and the divergence results is also observed for \lo~and FFT-reconstructed CSI, indicating that the divergence is not a reliable measure for assessing channel reciprocity. Moreover, improving or minimizing the divergence between AP's CSI and STA's CSI does not necessarily enhance channel reciprocity for PLS solutions.

%

\subsubsection{Wasserstein Distance~\cite{panaretos_statistical_2019}} 
In Fig.~\ref{subfig:wass}, we plot the Wasserstein distance between the CSI data collected at AP and STA for the three locations scenarios using both raw and preprocessed CSI data. The figure demonstrates that Wasserstein distance captures CSI variations caused by the locations of AP and STA and fails to gauge the impact of CSI preprocessing. Figs. \ref{subfig:wass} and \ref{fig:SKG_metrics} show that the increase of Wasserstein distance with the distance between AP and STA is also reciprocated in the key generation metrics only for the raw CSI. For raw CSI, higher Wasserstein distance corresponds to higher physical distance between AP and STA, higher KGR and lower BER. However, it fails entirely to capture the impact of Golay filtering and FFT reconstruction despite their improvement in KGR and BER depicted in Fig. \ref{fig:SKG_metrics}.
%
Our experimental findings indicate that Wasserstein distance measures the distance between the probability distributions of the collected CSI at AP and STA and exhibits a significant dependency on the mean values of these distributions, particularly when the CSI data at AP and STA conform to normal distributions with comparable variances but distinct mean values. Consequently, even though Pearson's correlation is high and suggests reciprocity between AP's CSI and STA's CSI, Wasserstein distance registers a high value due to the difference between the mean values, leading to a misleading evaluation of channel reciprocity.
\begin{figure*} 
    \centerline{
  \subfloat[\lab\label{subfig:wc_room}]{%
       \includegraphics[width=0.3\textwidth]{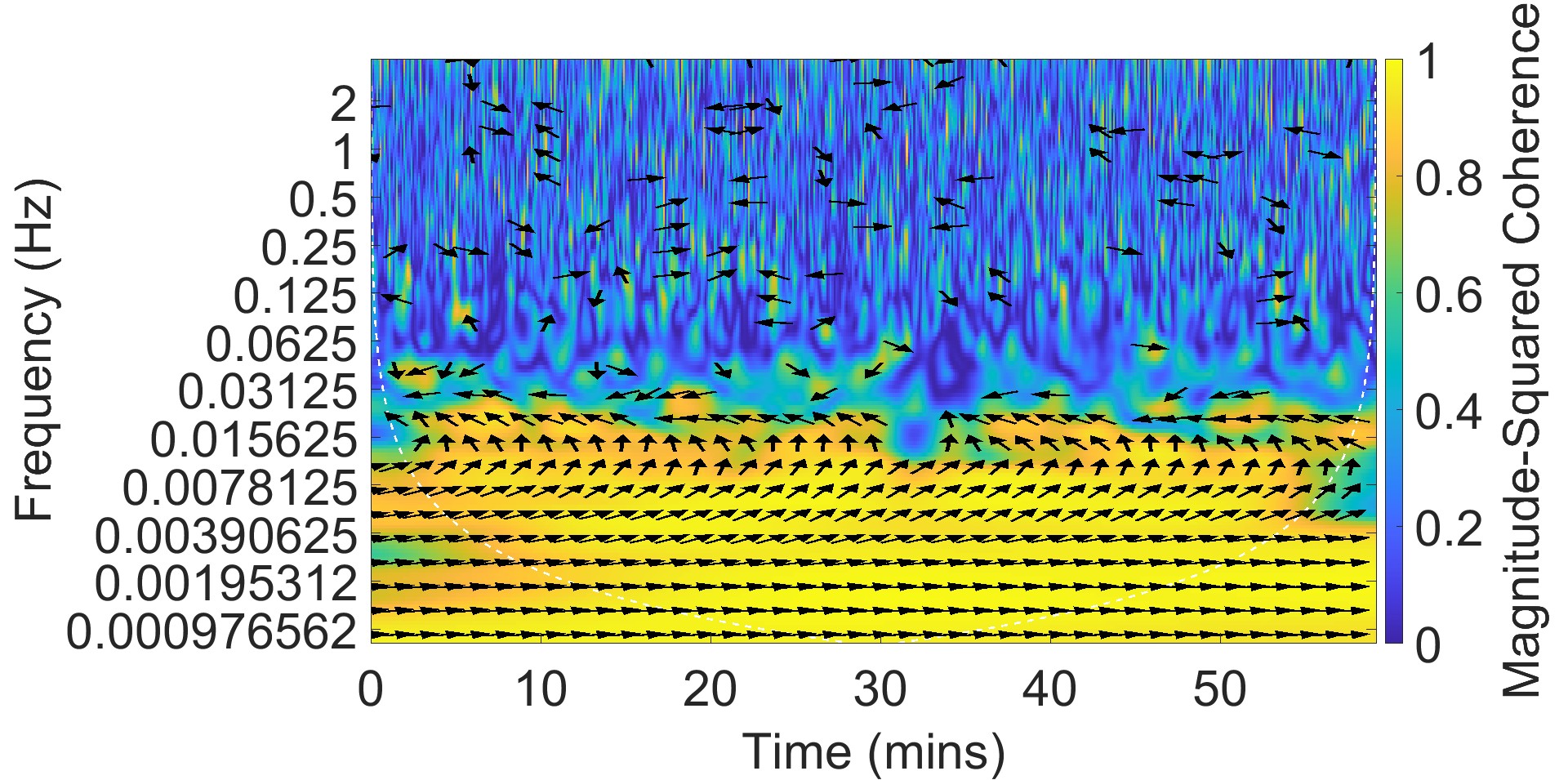}}
  \subfloat[\lc\label{subfig:wc_close}]{%
        \includegraphics[width=0.3\textwidth]{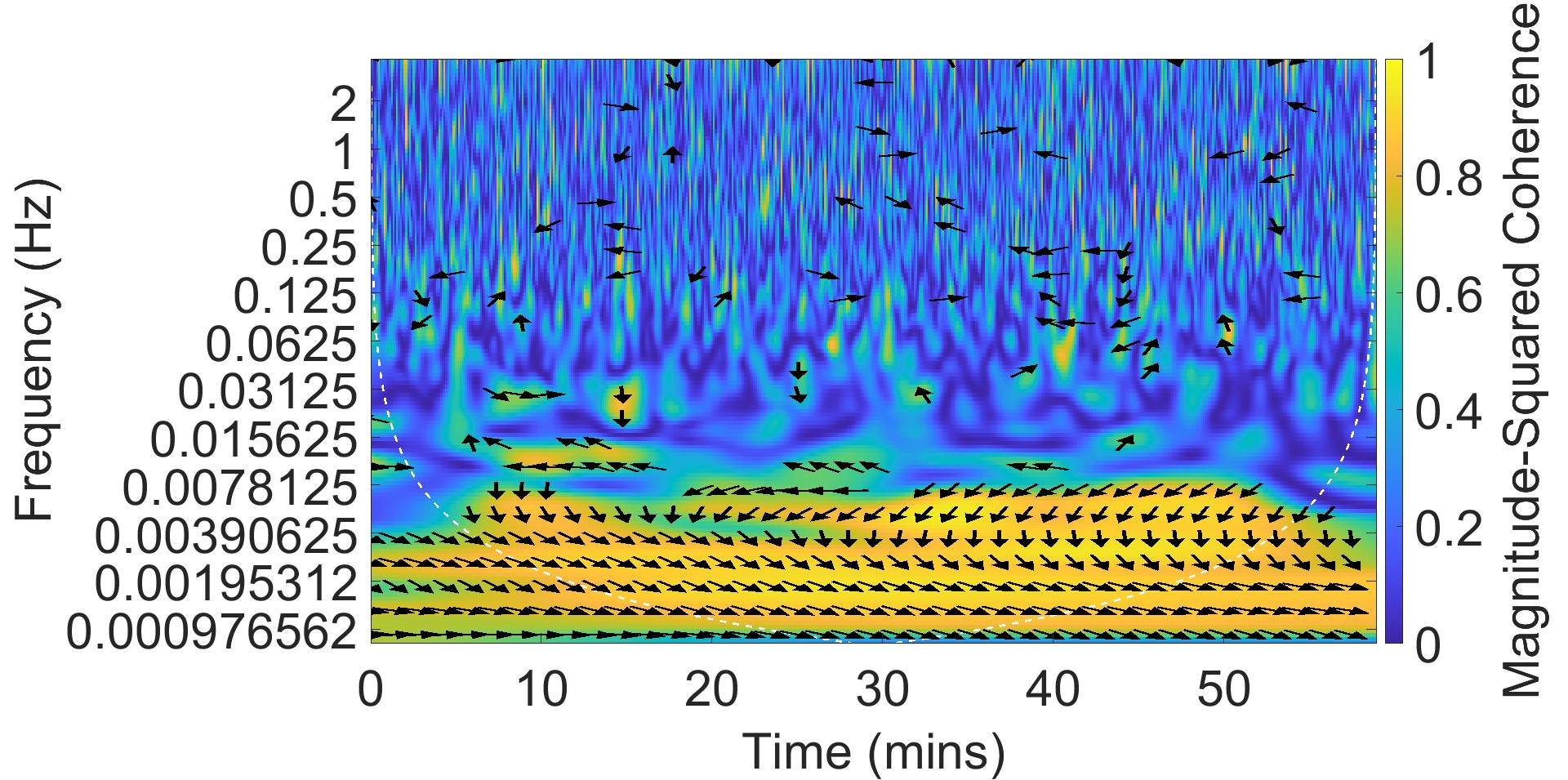}}
  \subfloat[\lo\label{subfig:wc_far}]{%
        \includegraphics[width=0.3\textwidth]{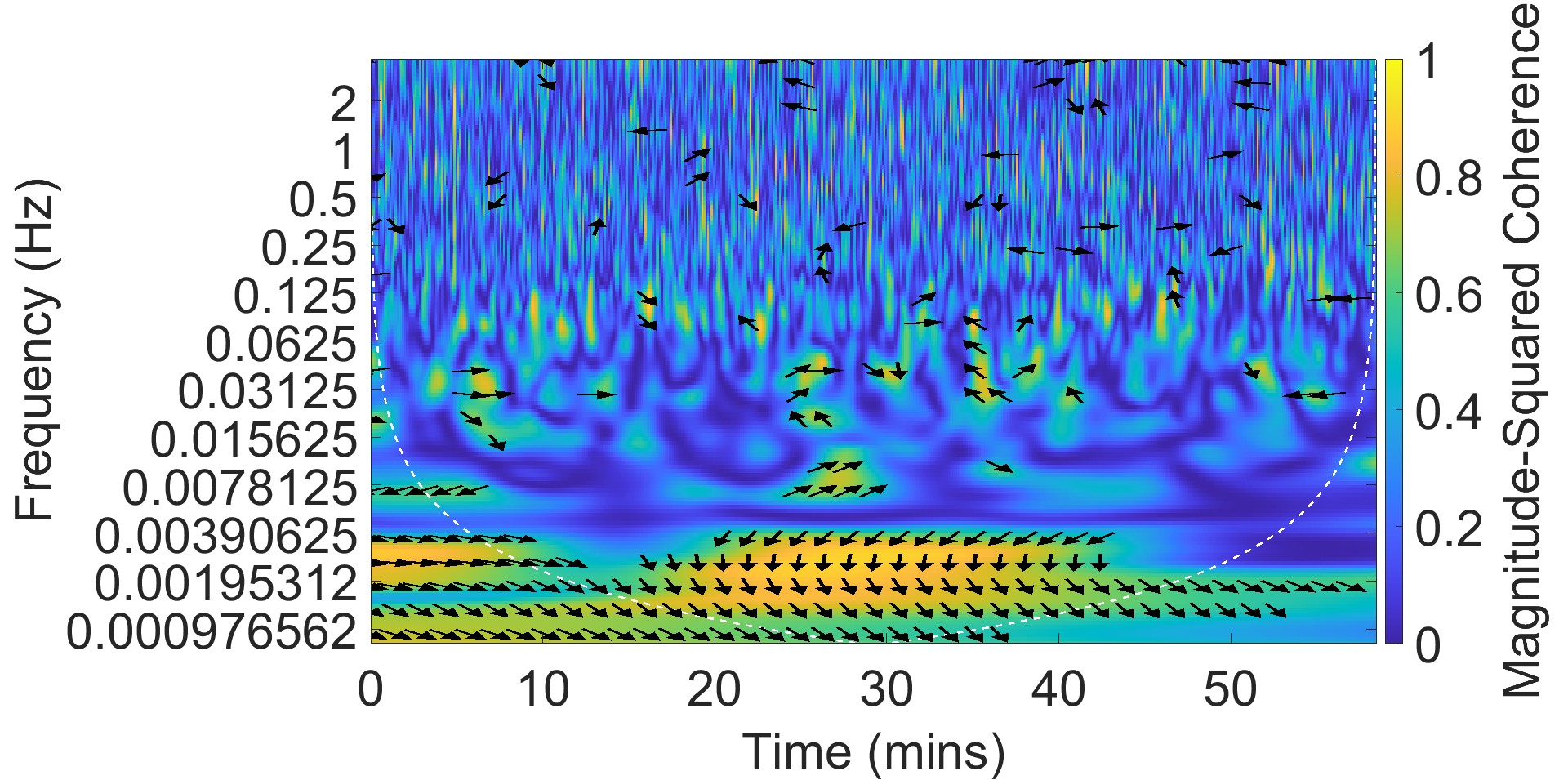}}}
    
   \caption{Impact of location on WC under raw CSI.}
  \label{fig:wc} 
\end{figure*}
\subsubsection{Wavelet Coherence~\cite{grinsted_application_2004}}  
Fig. \ref{fig:wc} shows the Wavelet Coherence (WC) of the raw CSI signals under each of the three studied locations scenarios. The figure shows that WC captures well the impact of the AP's and STA's locations, as the largest yellow zone (highest coherence) among the three scenarios is that of \lab~(Fig. \ref{subfig:wc_room}) with the shortest AP-STA LoS distance. Additionally, larger high coherence zones for raw CSI correspond to higher KGR (Fig. \ref{subfig:KGR_metrics}) and lower BER (Fig. \ref{subfig:BER_metrics}).  
Figs. \ref{subfig:wc_close} and \ref{subfig:wc_far} which capture WC for \lc~and \lo, respectively, show that as the distance between AP and STA increases, the yellow zone of WC shrinks and the phase difference between AP and STA increases, indicating severely impaired channel reciprocity. 
This is attributed to the dominance of NLoS communication between AP and STA in \lc~and \lo. NLoS communication introduces phase differences and noise to the CSI, which appears in WC as a decrease in the high coherence area, an increase in the low coherence area, and an increase in the phase difference.

Fig. \ref{fig:wc} demonstrates that WC provides insights into the CSI signals' correlation in both time and frequency by uncovering the duration of diminished reciprocity, as well as the frequencies that are common in AP's and STA's CSI signals. The yellow zone with the in-phase arrows of WC gives insights about the time duration when the channel is most reciprocal and about the principal, reciprocal frequencies of CSI. Whereas the blue zone captures the channel impairments and noise impact that persists during the entire data collection time at AP but are not reciprocated at STA and vice versa. The figure also shows that high WC values exist at the lower frequencies, indicating that slow variations of the CSI signals are highly correlated at AP and STA compared to the fast CSI signals' variations.
WC also captures the impact of CSI preprocessing effectively as discussed later Sec.~\ref{sec:technique}.

Our findings show that channel reciprocity is severely diminished in practical scenarios. Additionally, our analysis illustrates that channel reciprocity is best assessed using Pearson's correlation and wavelet coherence. Jeffrey's divergence and Wasserstein distance are not reliable for assessing channel reciprocity and do not accurately portray the impact of CSI preprocessing. 
%


%
\subsection{Asynchronous Channel Measurements} 
To obtain their CSIs, STA sends probing signals to AP so that AP can estimate STA-to-AP channel CSI. Reciprocally, AP sends signals for STA so that STA can estimate AP-to-STA channel CSI. The half-duplex WiFi and this process induce a time shift between the CSIs due to the sequential transmissions.
Additionally, the collected CSI data shows that the uplink and downlink of the WiFi exhibit dissimilar patterns of packet losses, specifically in the NLoS scenarios, resulting in time misalignment between AP's CSI and STA's CSI.
This time shift, which varies with the communication channel condition between AP and STA, negatively impacts the channel reciprocity quality. 
In this section, we utilize Time-lagged cross-correlation \cite{paulComparativeStudyMathematical2022} and wavelet coherence to estimate and analyze CSI time shifts. Estimating the time shift will be used to enhance channel reciprocity, as will be shown in 
Sec.~\ref{sec:technique}.   

%
\subsubsection{Time-Lagged Cross-Correlation}
\label{subsec:Cross-corr}
\begin{figure} 
    \centerline{
  \subfloat[\lab\label{subfig:cross_room}]{%
       \includegraphics[ width = \columnwidth]{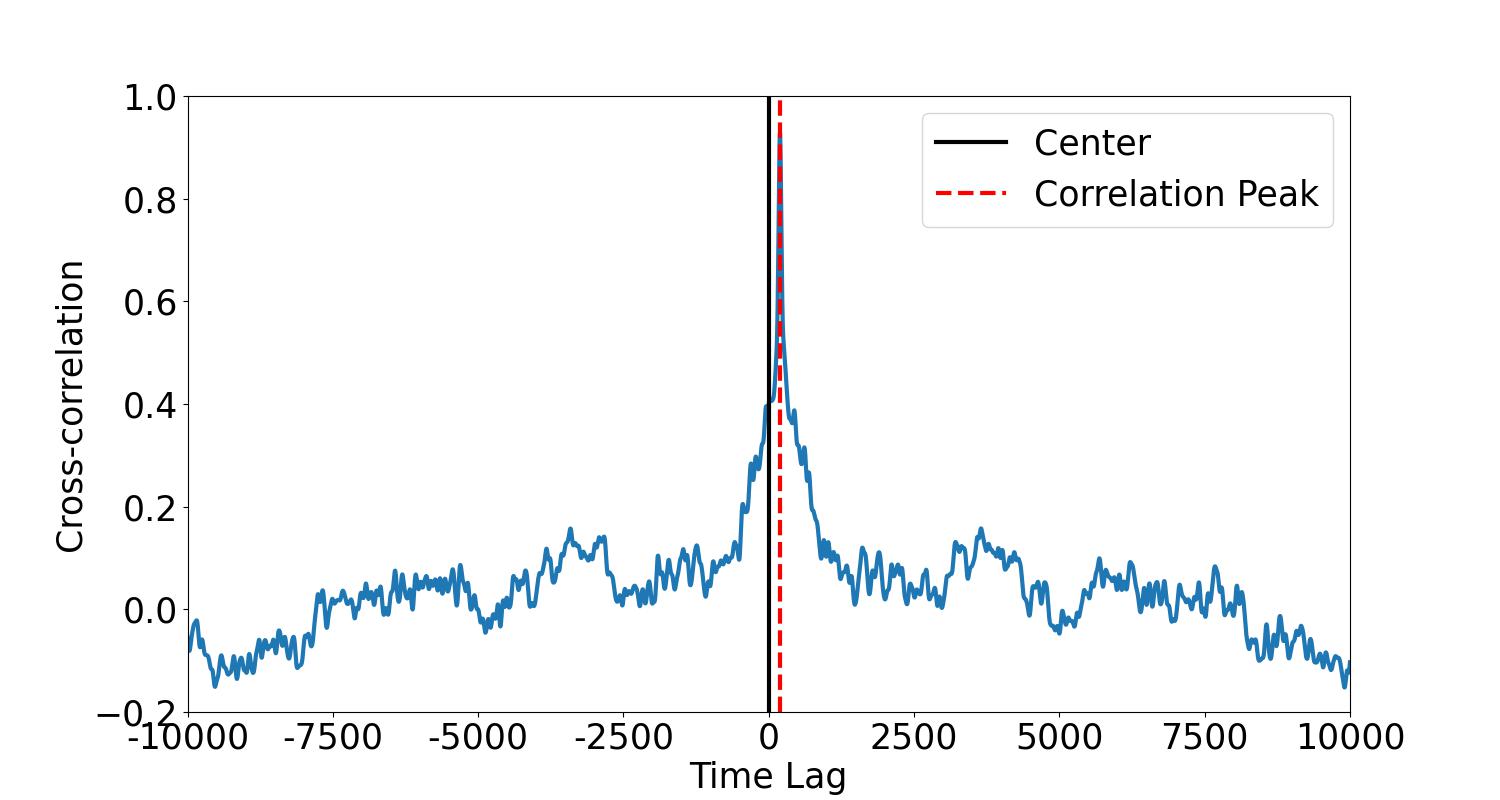}}}
\vspace*{-1mm}
    \centerline{
  \subfloat[\lc \label{subfig:cross_close}]{%
        \includegraphics[width =  \columnwidth]{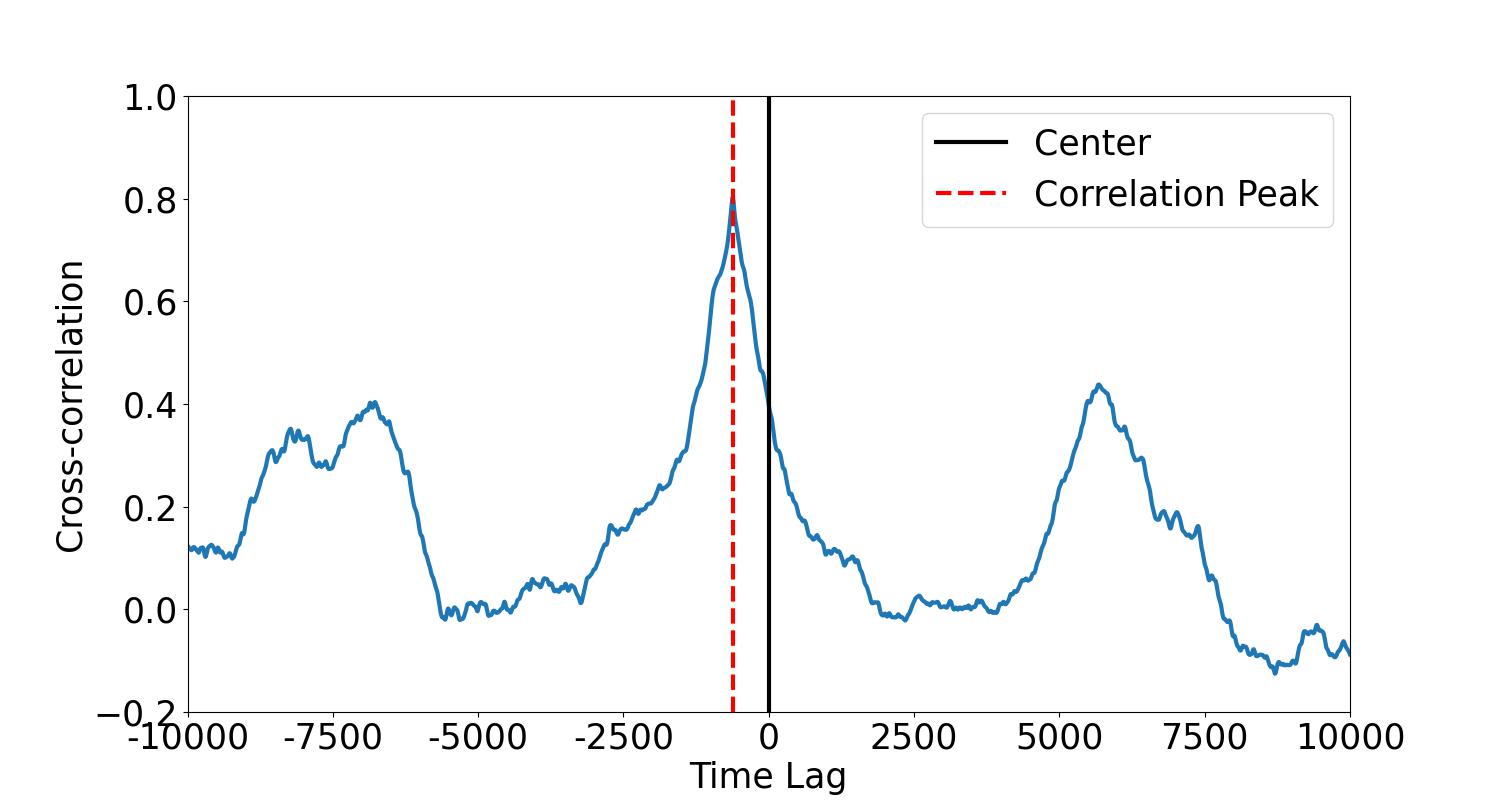}}
    }
\vspace*{-1mm}
    \centerline{
  \subfloat[\lo\label{subfig:cross_far}]{%
        \includegraphics[width =\columnwidth]
        {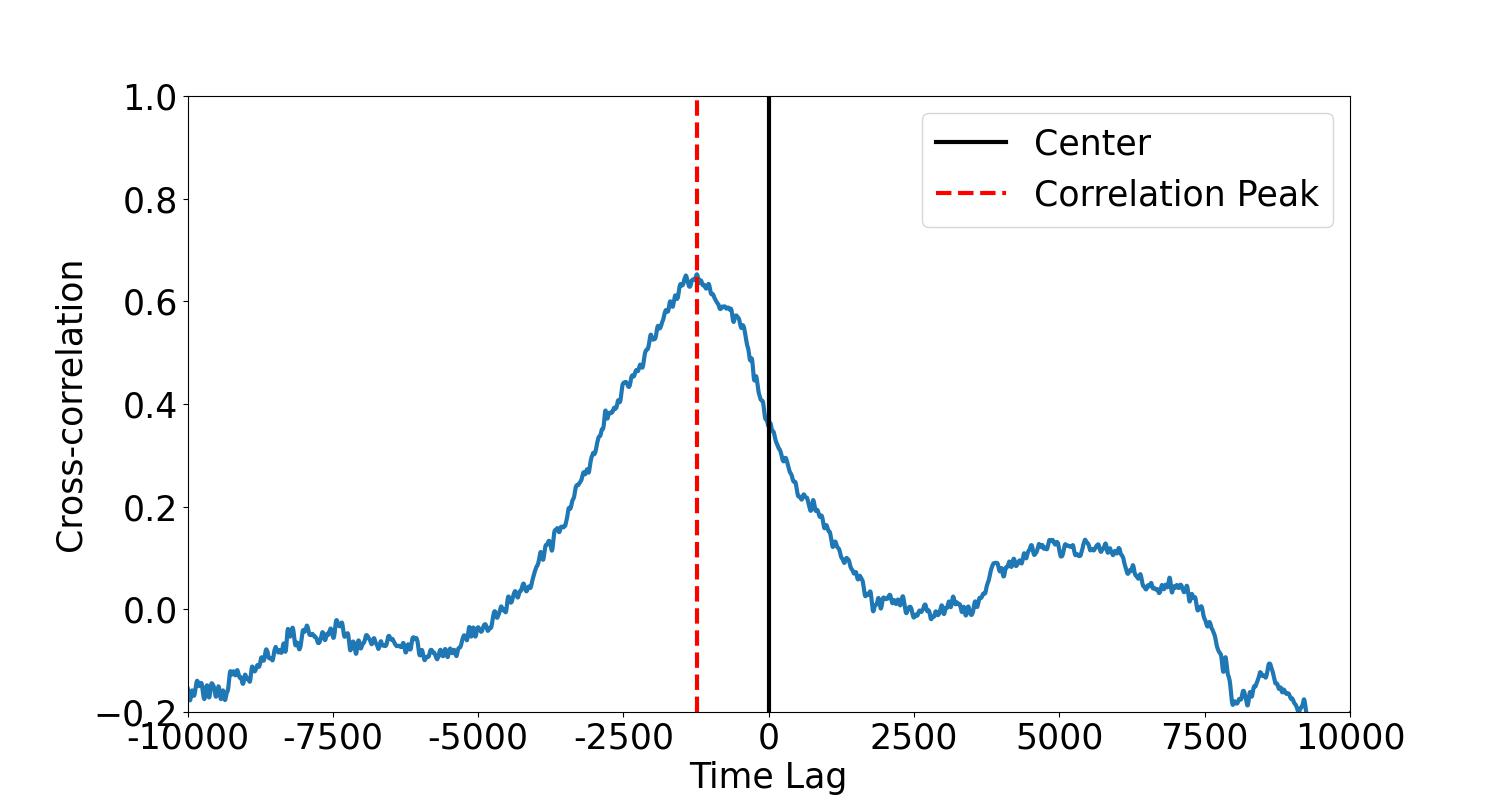}}
    }
    
   \caption{Impact of asynchronous measurements on time-lagged cross-correlation.}
  \label{fig:cross} 
\end{figure}

Fig.~\ref{fig:cross} plots the time-lagged cross-correlation of the CSI signals measured and collected by AP and STA under the three studied location scenarios: \lab~(Fig.~\ref{subfig:cross_room}), \lc~(Fig.~\ref{subfig:cross_close}), and \lo~(Fig.~\ref{subfig:cross_far}).
We draw three observations from this figure. First, we observe that the cross-correlation peaks at a time lag that is different from zero, and this occurs under each of the three studied scenarios. This is due to the time shift in the collected CSIs by AP and STA. The second observation is that the time lag corresponding to the correlation peak increases when going from a LoS connection (Fig.\ref{subfig:cross_room}) to an NLoS connection, as in Figs. \ref{subfig:cross_close} and \ref{subfig:cross_far}. This could be attributed to channel impairments, such as multipath and fading conditions and packet loss. Lastly, we observe that the correlation peak in \lc~(Fig.~\ref{subfig:cross_close}) is greater than the correlation peak in \lo~(Fig.~\ref{subfig:cross_far}), whose channel exhibits higher packet losses and may also exhibit more severe multipath and fading conditions due to AP and STA locations. 
The time lag at the cross-correlation peak provides an estimate for the time shift between the CSI at AP and STA.
\subsubsection{Wavelet Coherence and Inspecting Packet Loss}

In this section, we study the effectiveness of WC in exploring and quantifying CSI packet losses at AP and STA. We perform a large-scale inspection of WC for a long duration of CSI ($2$ hours) and demonstrate that packet losses appear as large low coherence zones and out-of-phase zones all over WC. We also perform a fine-grained small-scale WC inspection using a short duration of CSI ($25$ minutes) and illustrate the efficiency of WC in capturing and quantifying packet losses. 

\paragraph{Large-scale impact of packet loss on wavelet coherence.}

\begin{figure} 
    \centerline{
  \subfloat[No packet loss\label{subfig:wc_pkt_loss_no}]{%
       \includegraphics[width=0.5\columnwidth]{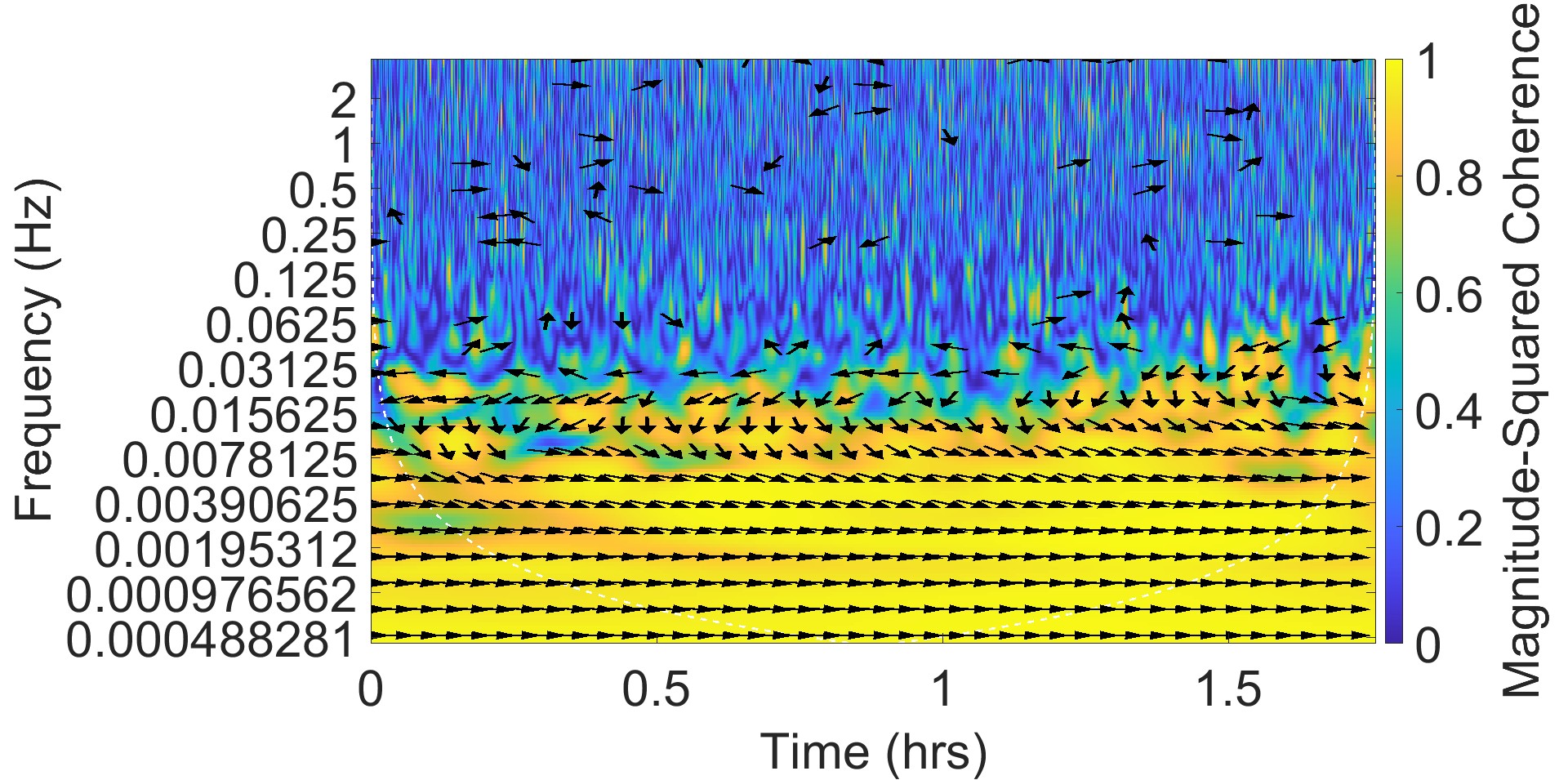}}
  \subfloat[Loss at t = 1 hr\label{subfig:wc_pkt_loss_m}]{%
        \includegraphics[width=0.5\columnwidth]{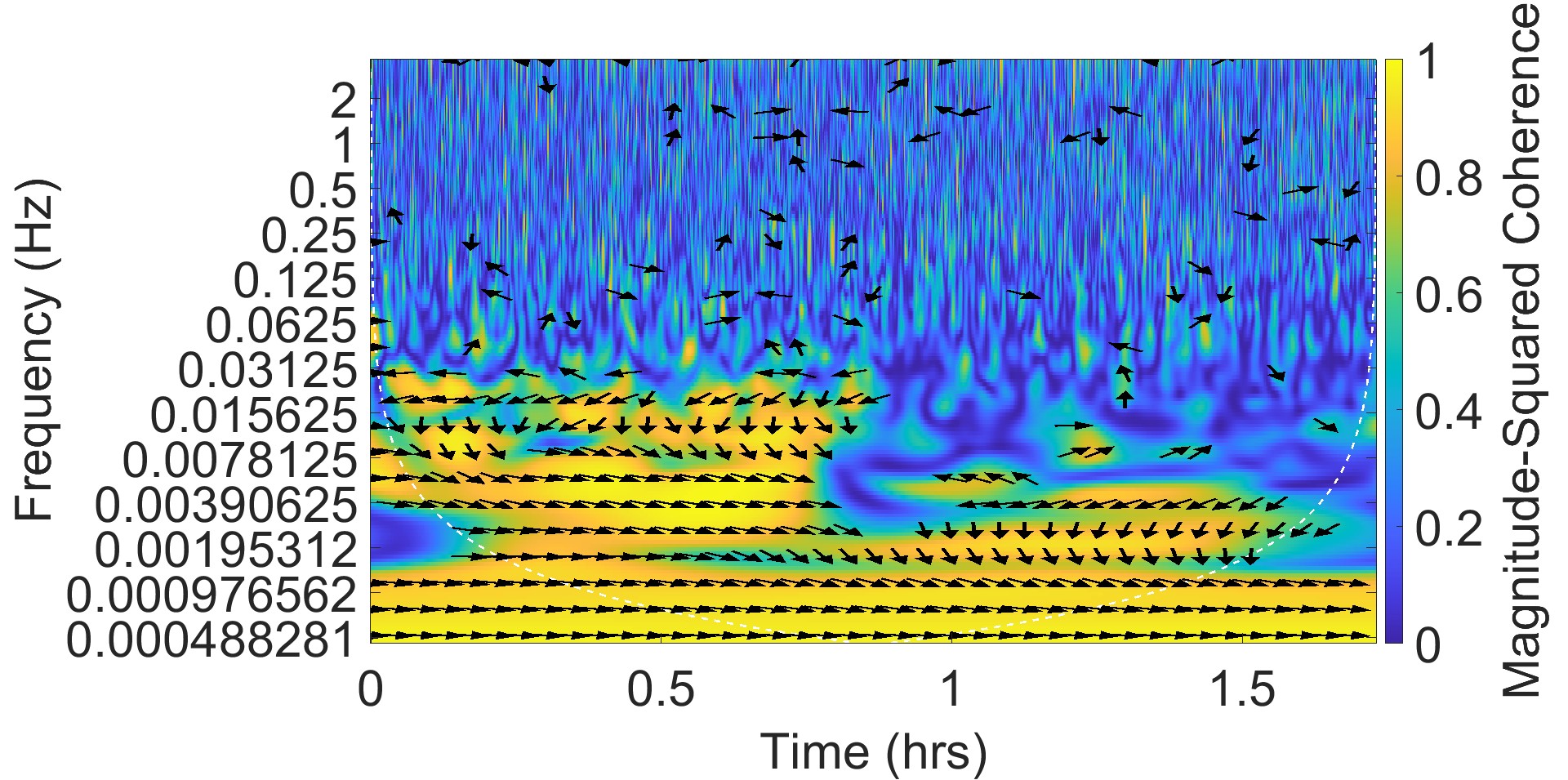}}}
    
    \centerline{
  \subfloat[Loss at t = 0,1 hr\label{subfig:wc_pkt_loss_m_s}]{%
        \includegraphics[width=0.5\columnwidth]{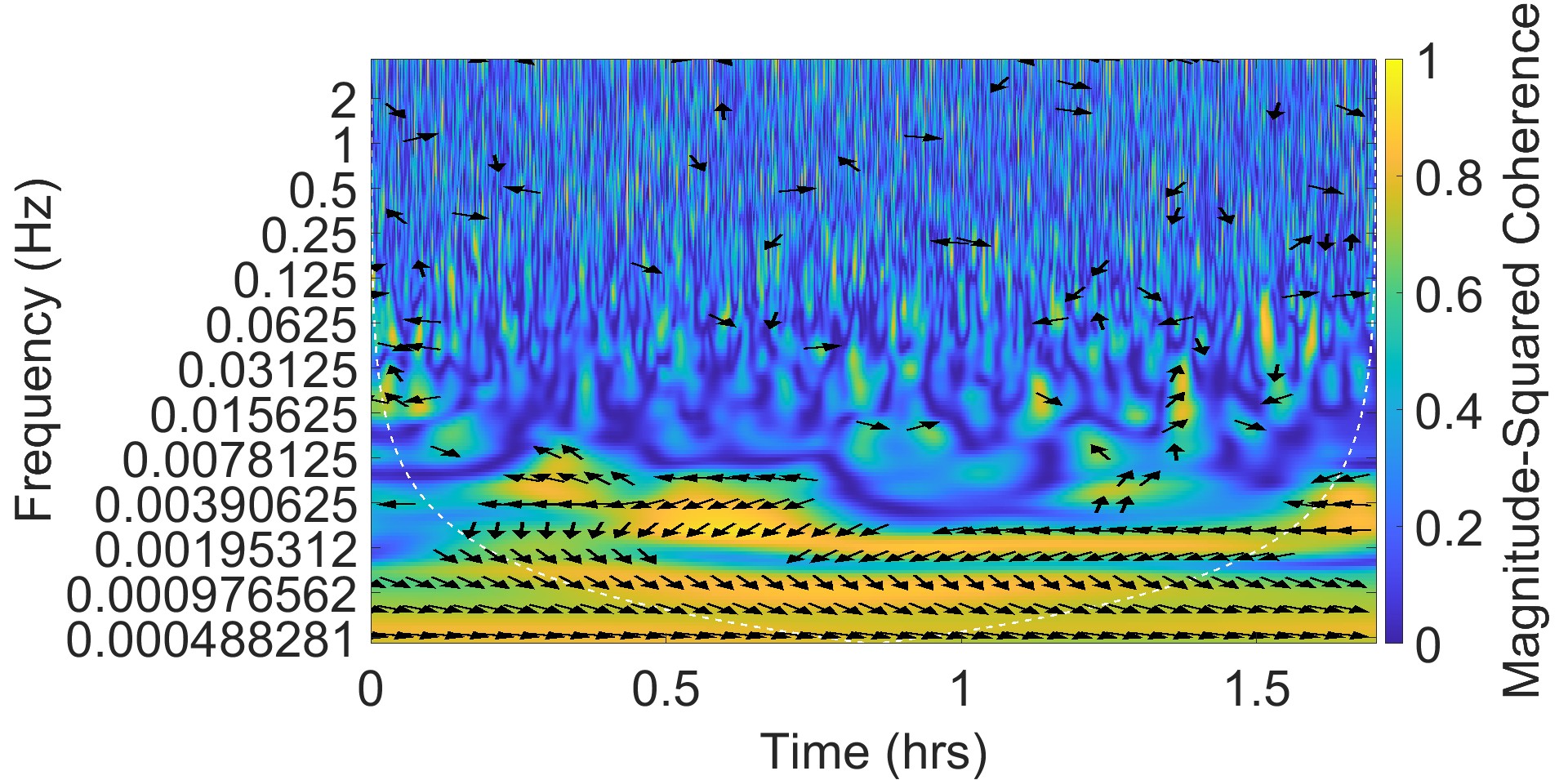}}
        
    \subfloat[Loss at t = 0, 1, 1.6 hr\label{subfig:wc_pkt_loss_m_s_e}]{%
        \includegraphics[width=0.5\columnwidth]{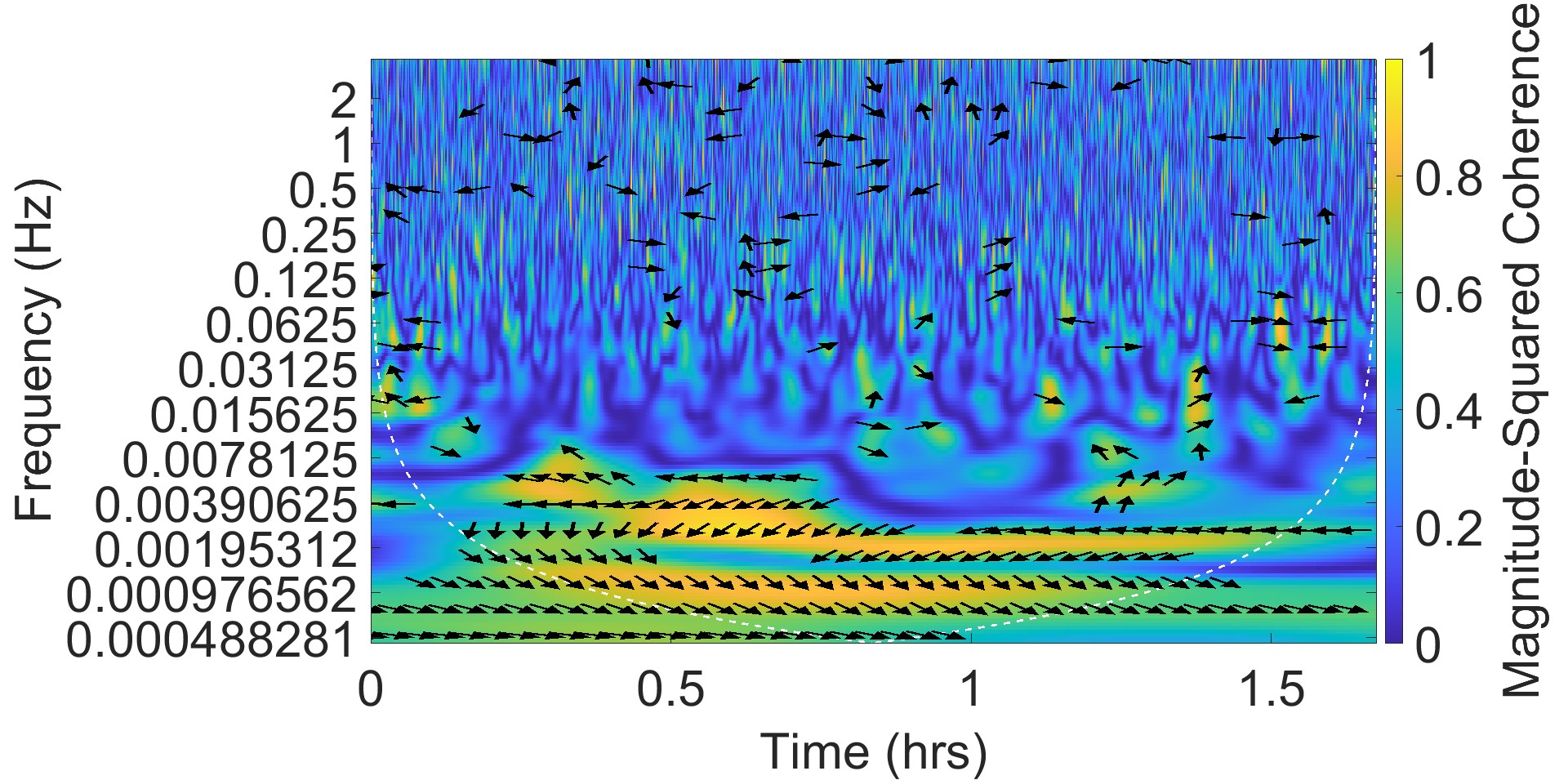}}}
    
   \caption{Large scale impact of induced packet losses on WC metric.}
  \label{fig:wc_loss} 
\end{figure}

Fig.~\ref{fig:wc} illustrates that WC of the NLoS scenarios, which incur high time shifts as previously discussed, exhibit large blue/low-coherence zones and out-of-phase areas when compared to \lab.
To verify that packet losses contribute to the observed high time shift and the low coherence and out-of-phase zones under NLoS scenarios, we take a closer look at the collected CSI data using WC time domain information. For that, intentional packet losses are introduced at specific time instances in a 2-hour CSI dataset collected at AP under the \lab~scenario. 
Specifically, $1000$ packets are dropped after about $1$ hour of data collection, an additional $1000$ packets are dropped at the beginning of data collection, and an additional $1000$ packets are dropped after $1.6$ hours of data collection. Then, WC is computed for each packet loss scenario and presented in Fig.~\ref{fig:wc_loss}.
The figure illustrates that at moments of induced packet losses, blue zones (low coherence) as well as increased phase shifts (black arrows) appear in WC, as compared to the no packet loss scenario depicted in Fig.~\ref{subfig:wc_pkt_loss_no}. For instance, Fig.~\ref{subfig:wc_pkt_loss_m} shows WC when $1000$ packets are dropped after about $1$ hour of AP data collection and illustrates a region of low coherence values around $1$ hour and a change in the phase shift between the CSI at AP and STA due to the dropped packets. These results confirm the effectiveness of WC in capturing variations across the CSIs obtained by AP and STA.
Our findings also show the similarity between WC under the induced packet loss over the entire data collection scenario (Fig.~\ref{subfig:wc_pkt_loss_m_s_e}) and WC under \lo~(Fig.~\ref{subfig:wc_far}), which validates that packet losses in NLoS scenarios produce low coherence and out-of-phase zones as well as high time shifts.

\begin{figure} 
    \centerline{
  \subfloat[No packet loss\label{subfig:clear}]{%
       \includegraphics[width=0.5\columnwidth]{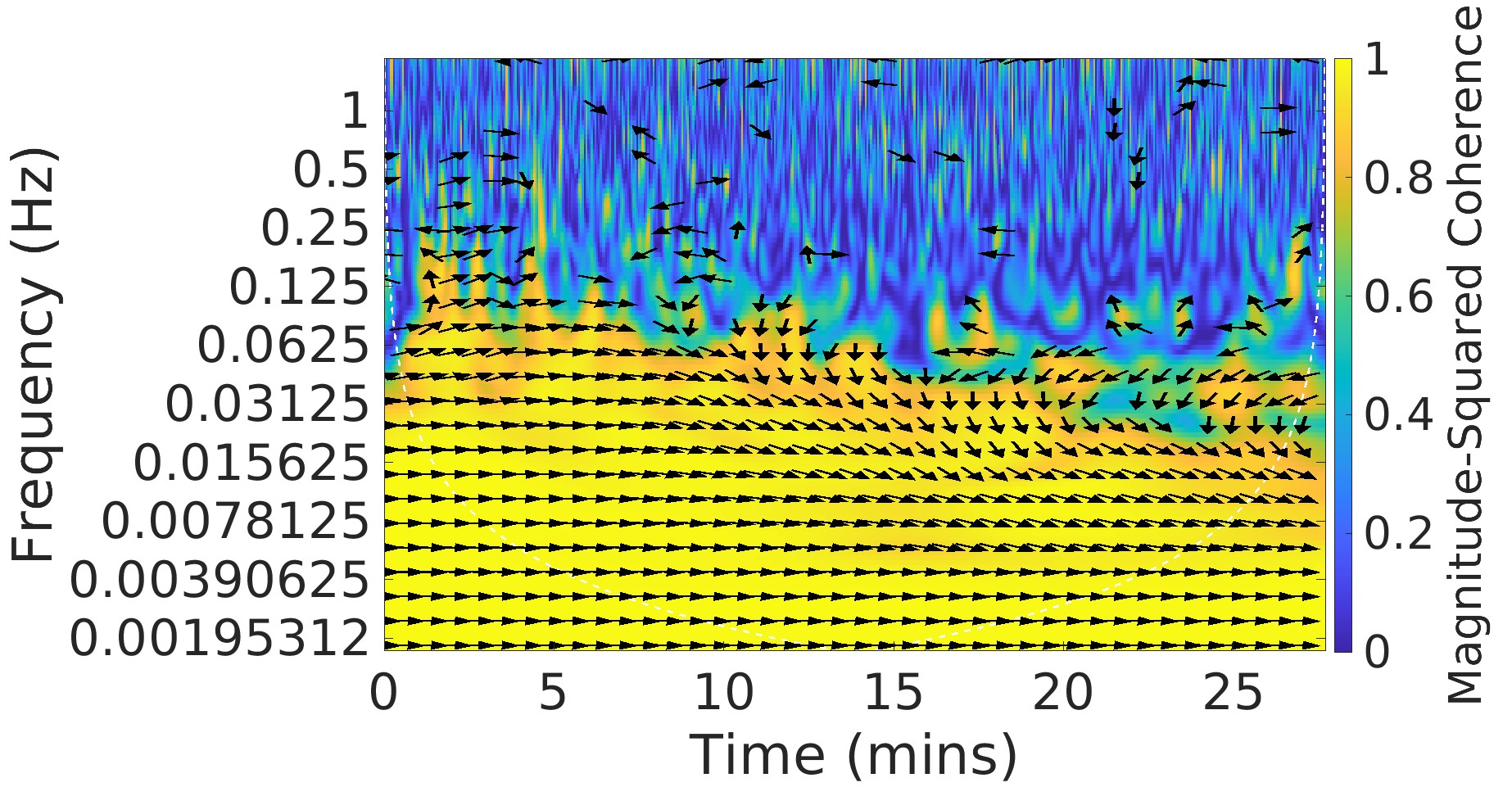}}
  \subfloat[300 packets \label{subfig:1min}]{%
        \includegraphics[width=0.5\columnwidth]{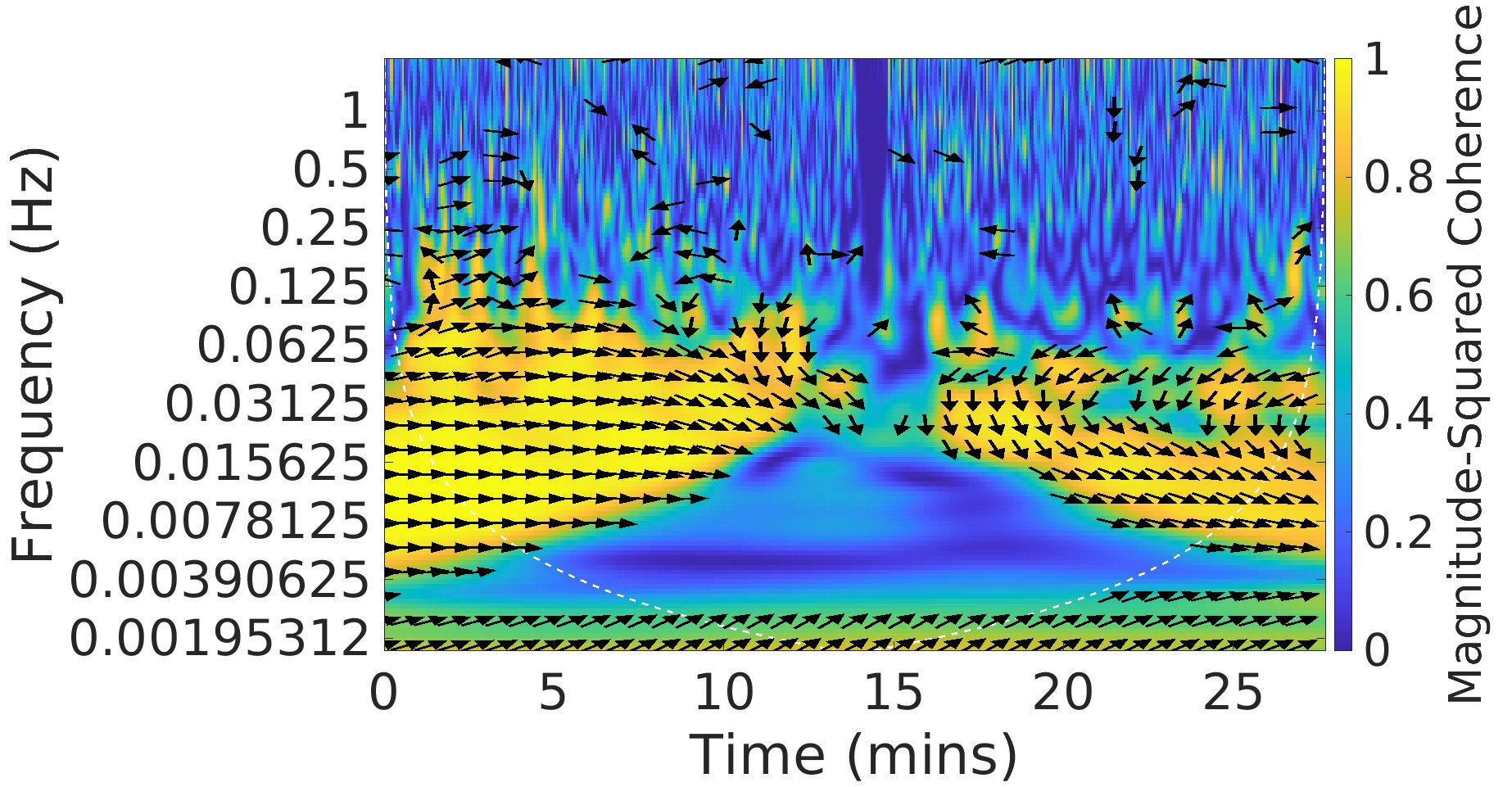}}}
    
    \centerline{
  \subfloat[900 packets \label{subfig:3min}]{%
        \includegraphics[width=0.5\columnwidth]{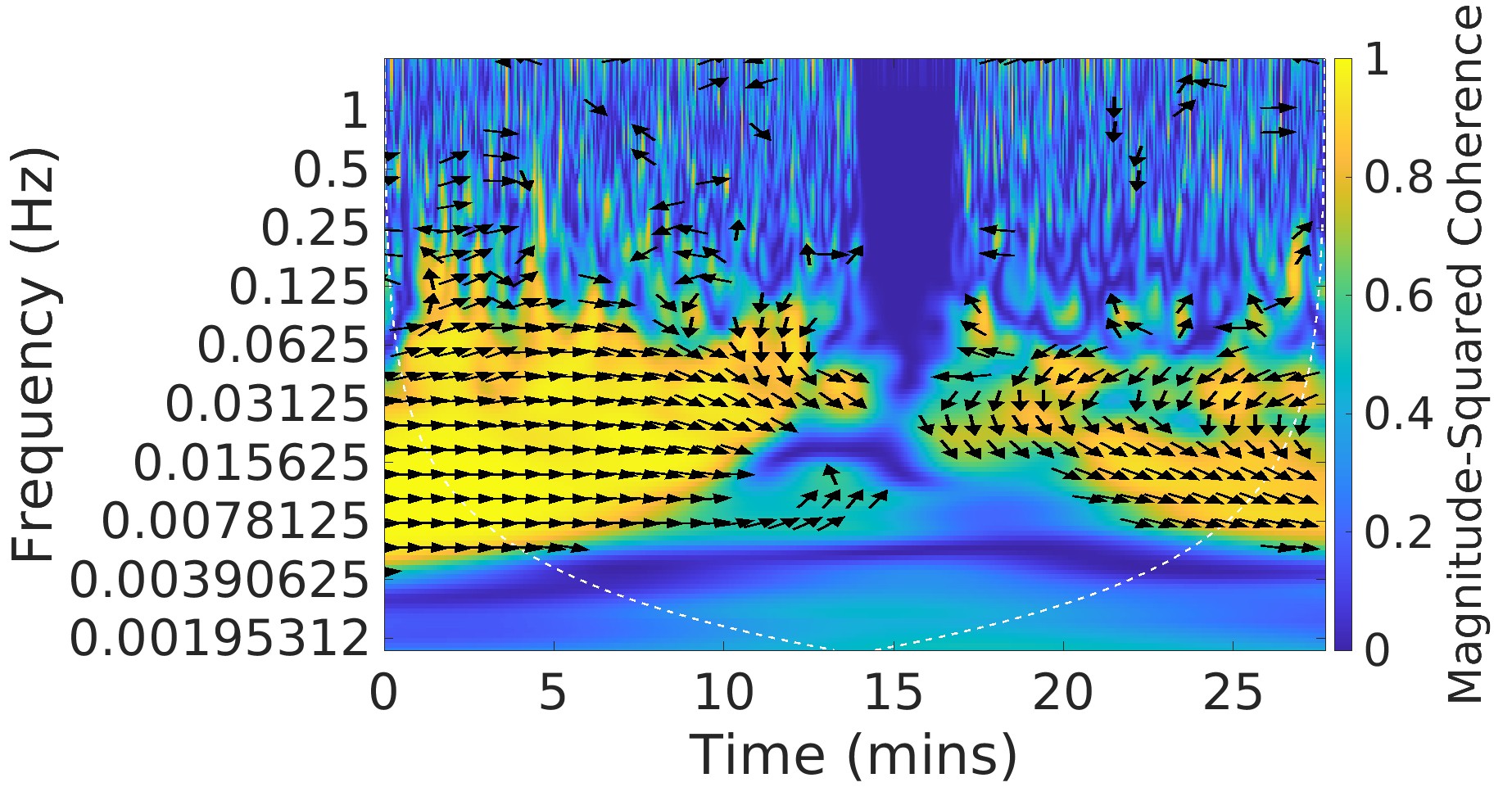}}
        
    \subfloat[1500 packets\label{subfig:5min}]{%
        \includegraphics[width=0.5\columnwidth]{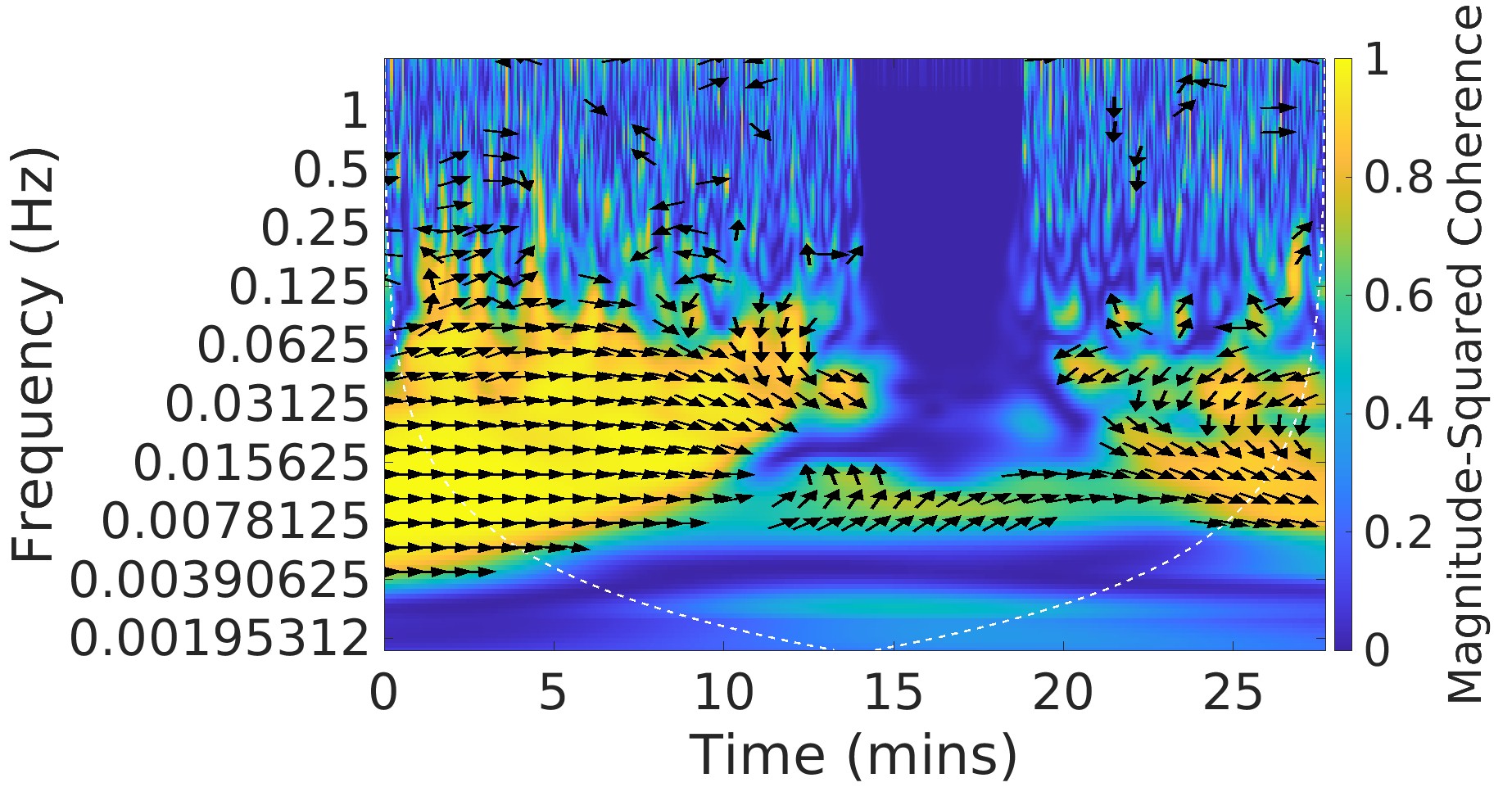}}}

    \centerline{
  \subfloat[600 packets at 5 mins \label{subfig:2min_start}]{%
        \includegraphics[width=0.5\columnwidth]{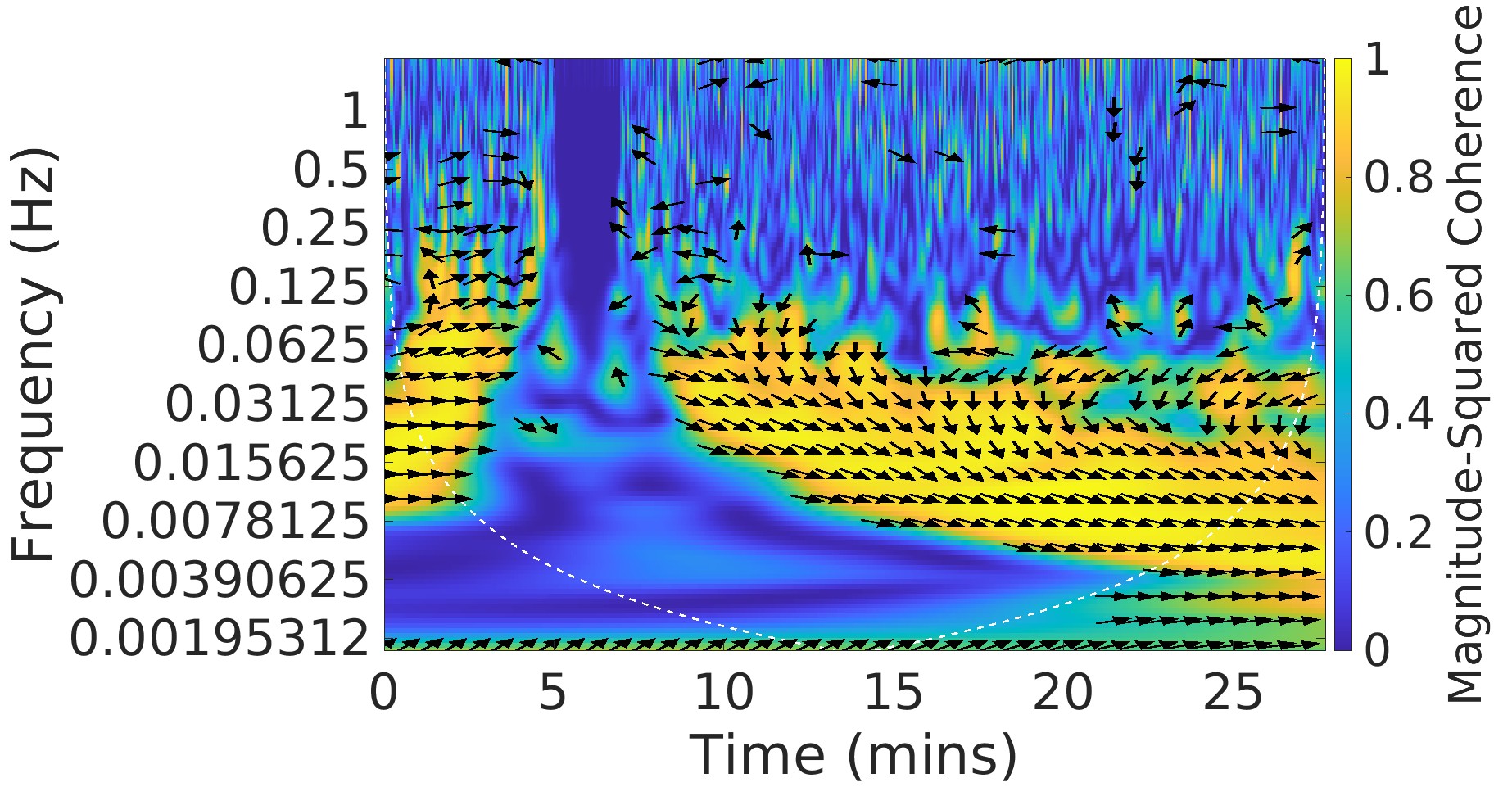}}
        
    \subfloat[600 packets at 5 \& 14 mins \label{subfig:4min_sep}]{%
        \includegraphics[width=0.5\columnwidth]{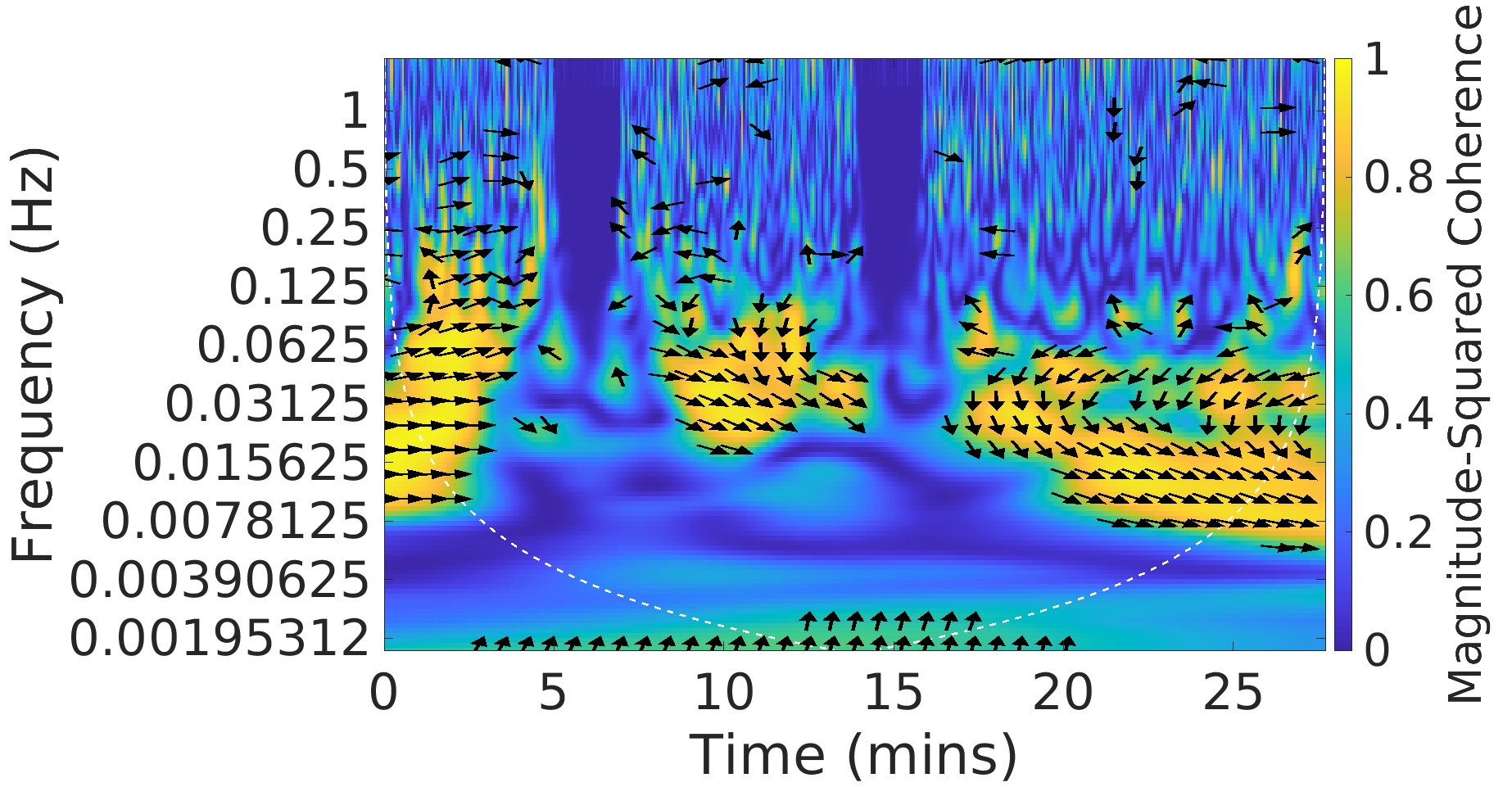}}}
    
   \caption{Small-scale impact of induced packet losses on WC metric.}
  \label{fig:pkt_loss_est} 
\end{figure}

\paragraph{Small-scale impact of packet loss on wavelet coherence}
In this section, we aim at quantifying how much packet loss contributes to low coherence zones. An estimate of packet loss could be obtained using a finer time scale inspection of WC. Therefore, we consider a shorter data collection duration of $25$ minutes at a smaller packet rate of $5$ packets per second.
Fig.~\ref{fig:pkt_loss_est} depicts the impact of packet loss on WC between AP and STA with increasing packet loss values. The figure shows that as the packet loss increases, the width of a low coherence zone in the range of frequencies $[0.06,1.5]$ increases consistently.
Fig.~\ref{subfig:clear} shows WC of CSI at AP and STA without any losses and is commensurate with Fig.~\ref{subfig:wc_pkt_loss_no}. Fig.~\ref{subfig:1min} depicts WC at a packet loss value of $300$ packets, and shows a zero coherence area with a narrow width of about $1$ minute around minute $14$ at frequencies $[0.06, 1.5]$ Hz. 
When the packet loss increases to $900$ after $14$ minutes of data collection in Fig.~\ref{subfig:3min}, we observe a wider low coherence area that has a width proportional to the increased packet loss of about $3$ minutes from $14.1$ minutes to $17.1$ minutes over the frequency range $[0.06,1.5]$. Given the observed duration of packet loss in WC, and the packet exchange rate, we can conclude that about $900$ packets are lost at AP or STA around minute $14$.
The same observation applies to Fig.~\ref{subfig:5min} which depicts WC with a packet loss of $1500$ packets, and shows a wider low coherence zone which confirms that WC effectively captures and quantifies losses. 
Fig.~\ref{subfig:2min_start} shows a case where $600$ packets are dropped from STA's CSI after $5$ minutes of data collection, and depicts the corresponding low coherence zone from $4.6$ minutes to $6.8$ minutes ($2$ minutes) and suggests about $573$ lost packets. Fig.~\ref{subfig:4min_sep} shows WC for a case where packet loss occurs at STA at two separate time instances. The WC demonstrates two low coherence zones of the same width at $4.6$ minutes and $14$ minutes and suggests a total loss of $1200$ packets: $600$ packets dropped after $4.6$ minutes of data collection, and another $600$ packets dropped after $14$ minutes of data collection.
Figs.~\ref{fig:pkt_loss_est} demonstrates that WC effectively quantifies packet losses and shows the time when losses occur.
The figure also highlights the severe impact of packet loss on the coherence of CSI at AP and STA.

%
In conclusion, Pearson's correlation accurately assesses channel reciprocity but overlooks time shifts in CSI at AP and STA caused by asynchronous measurements and packet losses. Combining Pearson's correlation with time-lagged cross-correlation offers a comprehensive assessment of reciprocity. Only WC stands out by providing insights into the similarity between CSI at AP and STA in both time and frequency domains, offering a comprehensive evaluation of the impact of channel impairments and measurements asynchrony.

\section{Channel Reciprocity Enhancement} 
\label{sec:technique}
Motivated by the effectiveness of WC in capturing the impact of the channel impairments and the asynchronous nature of the collected measurements on channel reciprocity (as discussed in Sec. \ref{sec:metrics_analysis}), we introduce a Wavelet Transform (WT)-based CSI reconstruction framework for reciprocity enhancement. 
The proposed framework consists of a WT-based reconstruction of the raw CSI followed by a time synchronization step between AP and STA using the time shift estimated through the cross-correlation as presented in Sec.~\ref{sec:metrics_analysis}. The performance of the proposed framework is then analyzed and compared to Raw CSI, Golay-filtered CSI, and FFT-reconstructed CSI.

%
%
\subsection{WT-Based CSI Reconstruction Framework}
%
The basic idea lies in the fact that if a frequency component is present at both AP's and STA's CSI signals but not for a long enough duration, 
acknowledging it as a common frequency in the reconstructed CSI signals would diminish channel reciprocity.
Driven by this observation, we use WT to analyze and provide insights on where and how the frequency content of the CSI signals observed at AP and STA changes over time, and use this WT analysis to locate frequency components that are common to both CSI signals. Specifically, we propose to use WC to determine the frequencies that are present in both of the CSI signals, collected by AP and STA, for some substantial duration, and we do so by considering the frequency components whose WC values are higher than some predefined threshold, $\alpha$, and that are present over some predefined duration determined by a tunable threshold number of samples, $\beta$.
%
The set of these frequencies, $F_{rec}$, is then used to reconstruct a less noisy, more reciprocal version of the CSI using Inverse WT (IWT) \cite{doi:10.1137/0515056}. The proposed WT-based reconstruction of CSI signals is depicted in Algorithm \ref{Al1}. 
%
\begin{algorithm}\small
\algsetup{linenosize=\smallsize}
\caption{WT-based CSI Reconstruction}\label{Al1}
 \begin{algorithmic}
 \renewcommand{\algorithmicrequire}{\textbf{Input:}}
 \renewcommand{\algorithmicensure}{\textbf{Output:}}
 \REQUIRE ~
 $\text{WT}_{\text{AP/STA}}$: WT for AP's STA's CSI signals\\
 F: A set of WC frequency components\\
 T: A set of WC time instances \\
 WC: A set of AP's and STA's WC values = $\{WC_{f,t} |f \in F, t \in T\}$ \\
 $\alpha$, $\beta$: Thresholds of WC value and number of samples 
 \ENSURE  
 $\text{Processed CSI}_{\text{AP}}$, $\text{Processed CSI}_{\text{STA}}$
  \STATE  $WC_{\text{High}} =  \{ WC_{f,t} | WC_{f,t} \geq \alpha, f \in F, t \in T \}$ 
  \FOR {\text{each} $ f_i \in F$} 
  \STATE $WC_{rec}(f_i) = \{WC_{\text{High}_{f,t}}| f = f_{i}, t \in T\}$ 
  \IF {$|WC_{rec}(f_i)| \geq \beta$ }
  \STATE $\{F_{rec}\} \gets f_i$
  \ENDIF
  \ENDFOR
  \STATE $\text{Processed CSI}_{\text{AP}} = \text{IWT}(WT_{\text{AP}}, [min(\{F_{rec}\}), max(\{F_{rec}\})])$
\STATE $\text{Processed CSI}_{\text{STA}} =  \text{IWT}(WT_{\text{STA}}, [min(\{F_{rec}\}), max(\{F_{rec}\})])$
  \RETURN 
  $\text{Processed CSI}_{\text{AP}}$, $\text{Processed CSI}_{\text{STA}}$
\end{algorithmic} 
 \end{algorithm}
\begin{figure}
\includegraphics[ width=\columnwidth]{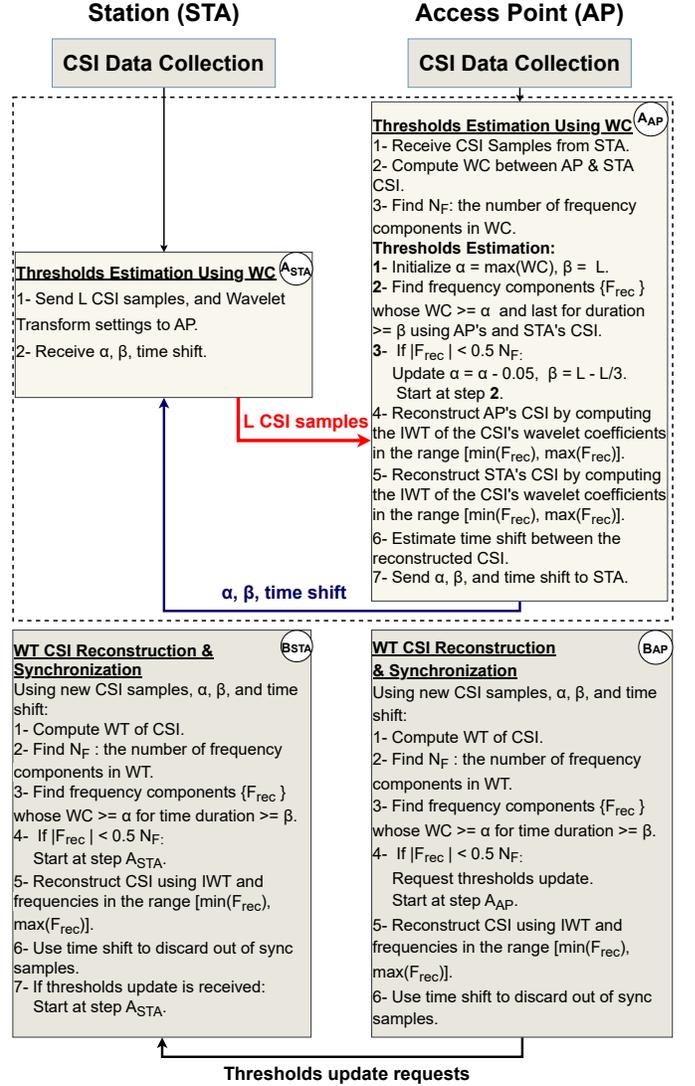}
\caption{WT-Based CSI Reconstruction \& Synchronization. }
\label{fig:wt_reconst}
\end{figure}
Because of the half-duplex WiFi, the inability of AP and STA to collect their CSI signals simultaneously, and packet loss, a time shift is present across the two CSI signals collected by AP and STA as discussed in Sec.~\ref{sec:metrics_analysis}. To make up for this synchronization issue, we propose to use cross-correlation to estimate the time shift between AP's and STA's WT-reconstructed data and then use the estimated time shift to align the two CSI signals.
Different delays upon different periods between AP and STA are observed in the NLoS scenarios, which is similar to time warping. Dynamic time warping (DTW) algorithm \cite{1163055} could be used to align CSI and mitigate the impact of packet loss by stretching and compressing certain segments to achieve the best possible match. Since DTW requires STA to send its entire CSI to AP over the public channel for alignment which is neither secure nor realistic since the delays keep changing, we opt for time-lagged cross-correlation to estimate a value of the overall time shift between CSI data at AP and STA aiming for a “best effort” CSI alignment.

Fig. \ref{fig:wt_reconst} depicts the steps of the proposed WT-based CSI reconstruction and synchronization at AP and STA. After CSI collection, STA starts step $A_{STA}$ and sends $L$ CSI samples to AP. Then, in step $A_{AP}$, AP receives CSI samples from STA and finds the reciprocal frequency components $\{F_{rec}\}$ within AP and STA CSI samples using the wavelet coherence (WC). To find $\{F_{rec}\}$, first, AP computes WC between the received $L$ CSI samples from STA and the corresponding $L$ CSI samples collected by AP. Second, AP searches for frequencies that exist through long duration in the CSI defined by the number of samples threshold $\beta$ and where the wavelet coherence values are greater than the wavelet coherence threshold $\alpha$. AP initializes $\beta$ with the value $L$ and initializes $\alpha$ with the maximum coherence value in the computed WC. By this, AP attempts to find highly correlated frequency components that exist during the entire CSI duration. These initial values of $\alpha$ and $\beta$ usually lead to a smooth reconstructed CSI with limited randomness. Therefore, we reduce the threshold value gradually until the set of reciprocal frequencies includes at least half of the available frequency components in the WC. After finding $\{F_{rec}\}$, AP finds the reconstructed CSI for AP and STA by computing the IWT using the wavelet coefficients and the frequency components in the range $[min(\{F_{rec}\}), max(\{F_{rec}\})]$ \cite{doi:10.1137/0515056}. AP then estimates the time shift between the reconstructed AP and STA CSIs using the time-lagged cross-correlation as explained in Sec. \ref{subsec:Cross-corr}. 
Lastly, AP sends $\alpha$, $\beta$, and the estimated time shift to STA.
In step $B_{STA}$, STA uses new STA CSI samples, $\alpha$, and $\beta$ to find its set of reciprocal frequencies $\{F_{rec}\}$, then reconstructs the CSI using IWT and the wavelet coefficients of the CSI for the frequency components in the range $[min(\{F_{rec}\}), max(\{F_{rec}\})]$. At the same time, in step $B_{AP}$, AP uses new AP CSI samples, $\alpha$, and $\beta$ to find its new set of reciprocal frequencies $\{F_{rec}\}$, then reconstructs the CSI using IWT and the wavelet coefficients of the CSI in the frequency range $[min(\{F_{rec}\}), max(\{F_{rec}\})]$. Our study shows that setting $\alpha$ to a value slightly below the maximum coherence value obtained in step $A_{AP}$ and setting $\beta$ to around $L/3$ produce highly reciprocal CSI at AP and STA.
After CSI reconstruction by IWT, STA and AP use the estimated time shift to discard out-of-sync samples.
Due to channel variations, AP and STA may require to update their threshold values $\alpha$ and $\beta$. This happen when new sets of the reciprocal frequencies at AP or STA have few frequencies less than at least half of the available frequency components in the wavelet transform of the CSI. When thresholds updates are needed at STA, it starts at step $A_{STA}$. When thresholds updates are needed at AP, it requests new $L$ CSI samples from STA, and then STA starts step $A_{STA}$.

%

\begin{figure}
    \centering
    \includegraphics[width=1\linewidth]{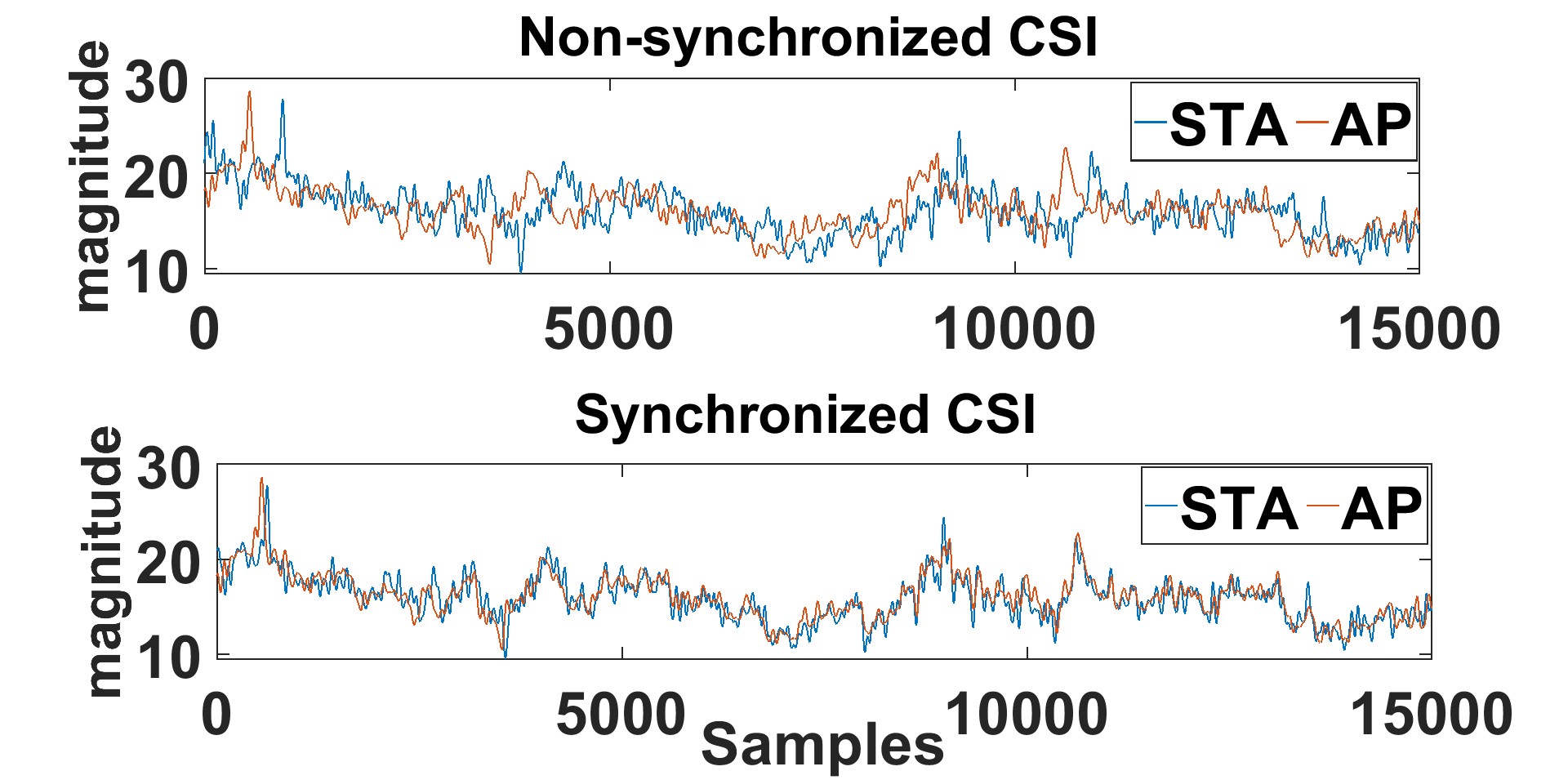}
    \vspace*{-4mm}
    \caption{Synchronized CSI vs. non-synchronized CSI: CSI obtained under the \lo~scenario and preprocessed using WT-based CSI reconstruction.}
    \label{fig:sync}
\end{figure}

\subsection{Result Analysis}

\subsubsection{Impact of synchronization on CSI}   
To illustrate the benefit of synchronization, we plot in Fig.~\ref{fig:sync} the CSI signals obtained at AP and STA and reconstructed using the proposed WT-based CSI reconstruction, with (bottom figure) and without (top figure) CSI synchronization.  
%
The figure clearly shows that synchronizing the two CSI signals by compensating for the time shift aligns well the CSI signals, and thereby enhances the reciprocity of the channel between AP and STA, as will be shown later when using it for secret key generation.

\begin{figure} 
    \centerline{
  \subfloat[Raw CSI \label{subfig:raw_csi}]{%
       \includegraphics[width=0.25\textwidth]{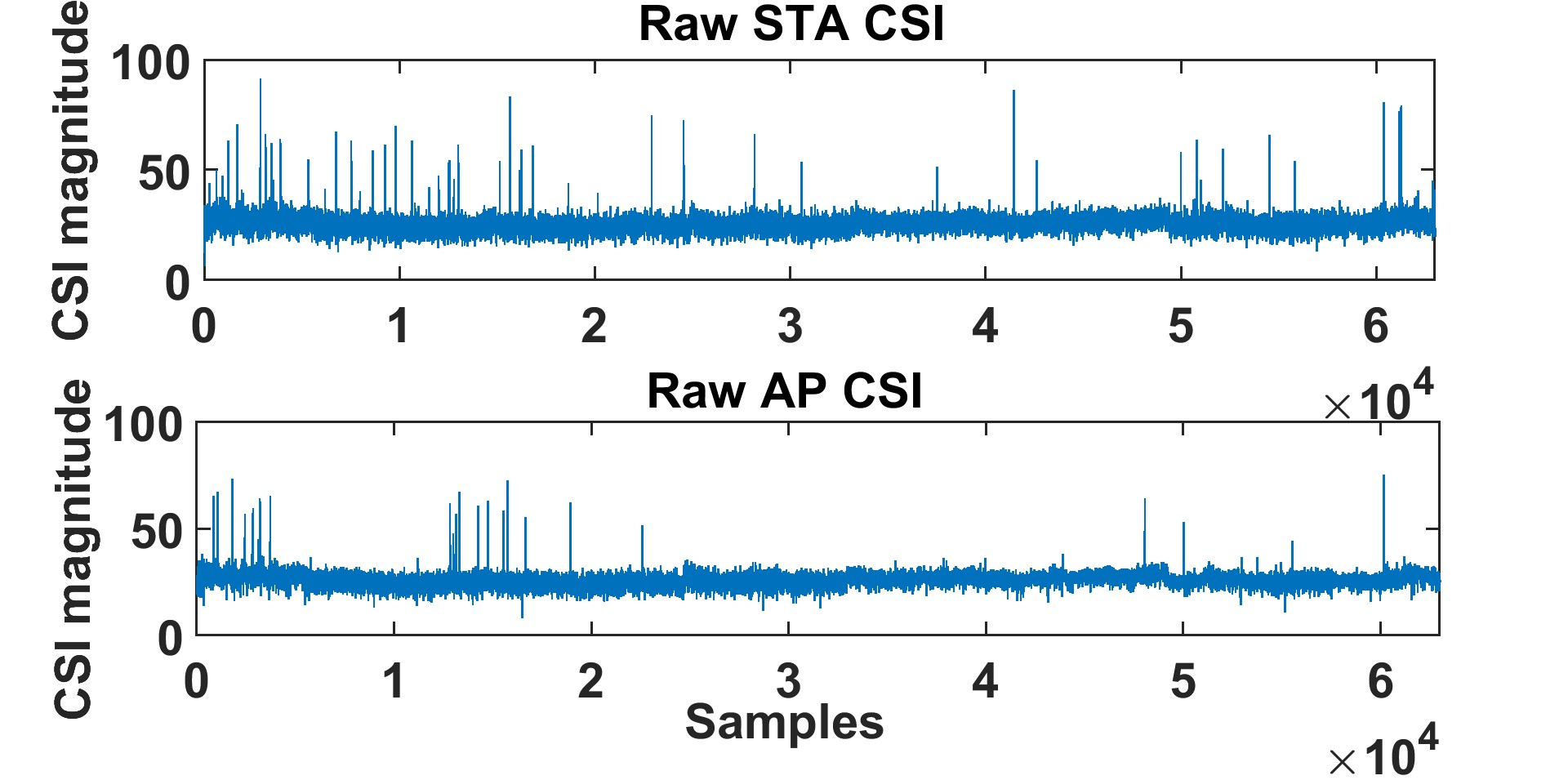}}
    \subfloat[Golay-filterd CSI \label{subfig:golay_csi}]{%
       \includegraphics[width=0.25\textwidth]{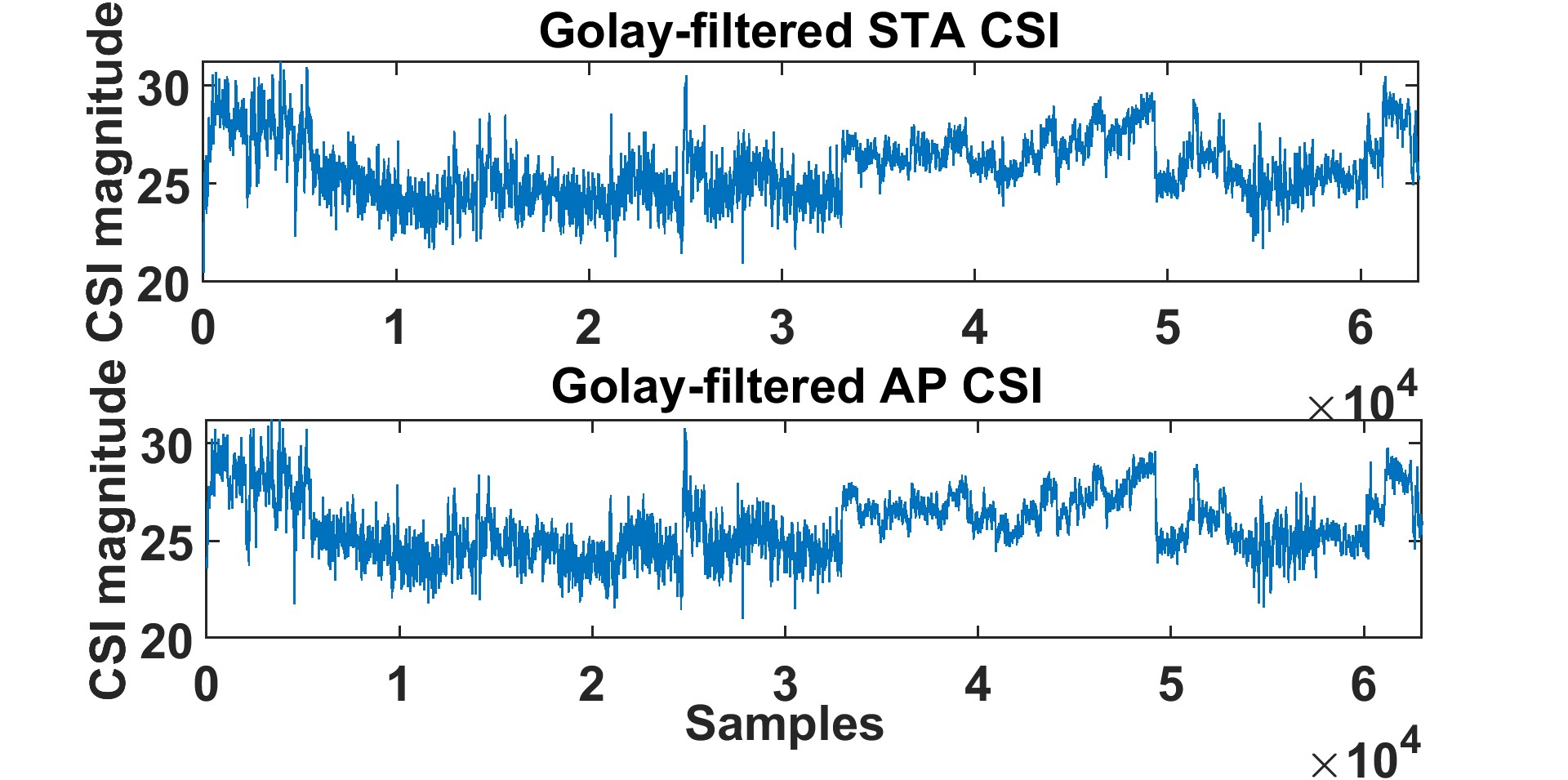}}
    }
    \centerline{
    
     \subfloat[FFT-reconstructed CSI \label{subfig:IFFT_csi}]{%
       \includegraphics[width=0.25\textwidth]{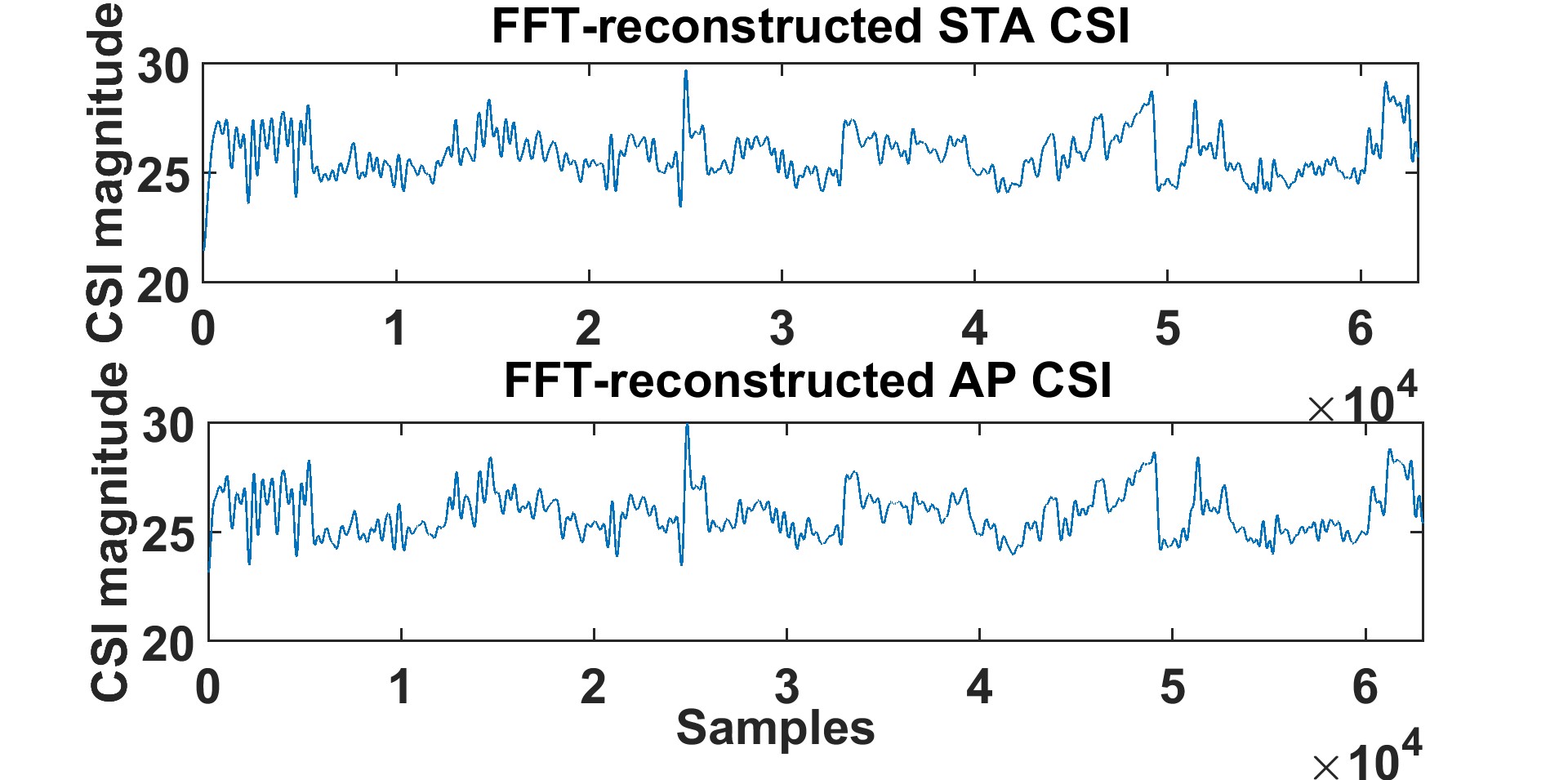}
      }
      \subfloat[WT-reconstructed CSI\label{subfig:wt_csi}]{%
        \includegraphics[width=0.25\textwidth]{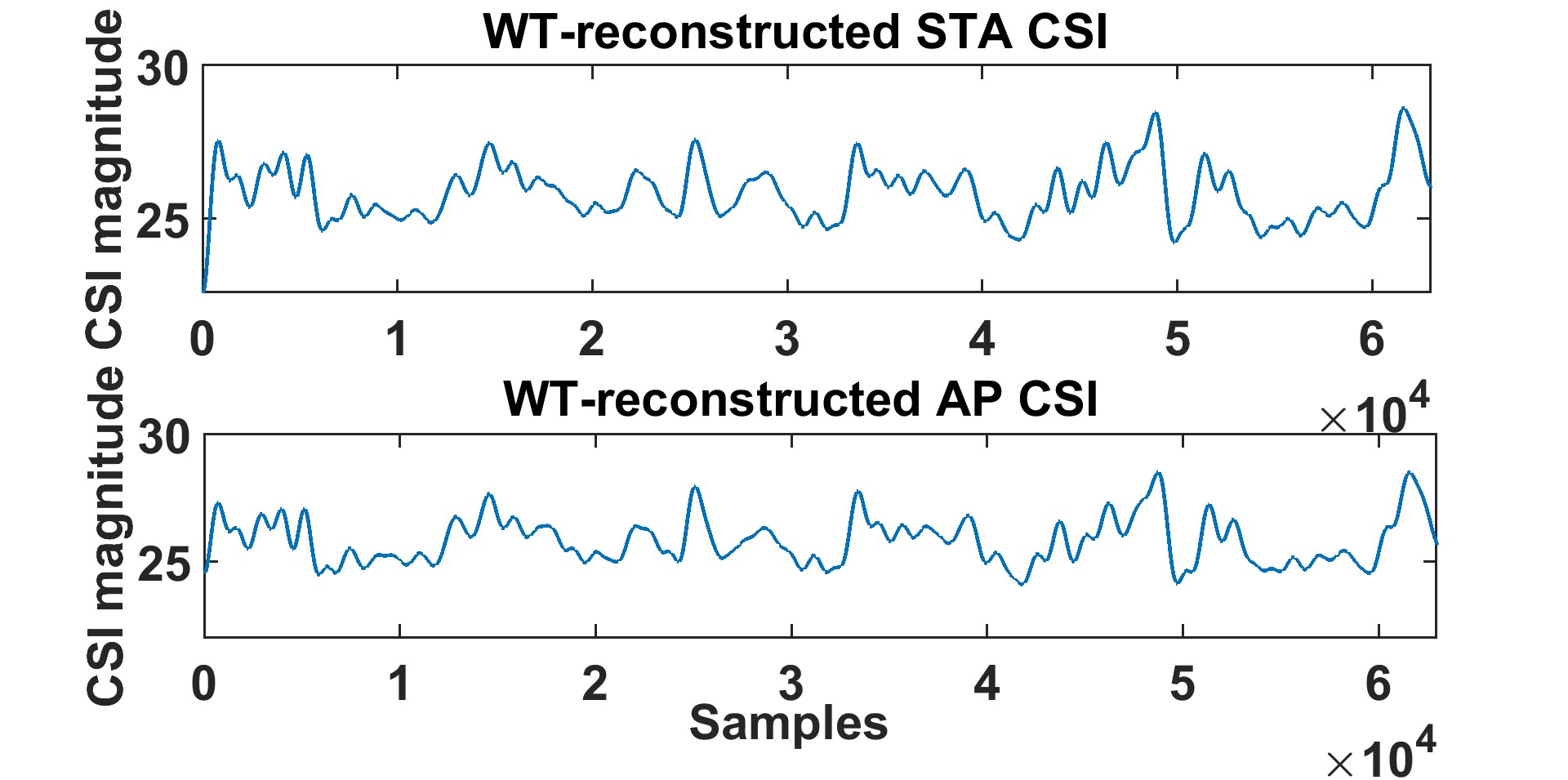}}}
    
   \caption{Impact of preprocessing on CSI reciprocity.}
  \label{fig:processing_csi} 
\end{figure}
\subsubsection{Comparison of Different CSI Reconstructions}
Fig.~\ref{fig:processing_csi} compares the proposed WT-based CSI reconstruction with the FFT-based and Golary filtering-based CSI reconstructions vis-a-vis their ability to improve reciprocity and CSI consistency across AP and STA. 
%
AP's and STA's raw CSI signals (without any preprocessing) are also presented to use as a baseline comparison. 
The figure illustrates the significant improvement achieved by the WT-reconstructed CSI (Fig.~\ref{subfig:wt_csi}), in comparison with raw CSI (Fig.~\ref{subfig:raw_csi}), Golay-filtered CSI (Fig.~\ref{subfig:golay_csi}), and FFT-reconstructed CSI (Fig.~\ref{subfig:IFFT_csi}), all conducted under the \lab~scenario.
The figure highlights the enhancement in CSI obtained through Golay filtering, which eliminates non-reciprocal noise spikes present in the raw CSI. It also demonstrates the smoothing effect of the FFT-reconstructed CSI when compared to raw CSI, achieved by removing the high frequency components with low power contribution to the CSI.
Additionally, the figure suggests that both the proposed WT reconstruction and FFT reconstruction may exhibit comparable improvements in CSI. However, as shown later in Sec.~\ref{sec:skg}, we demonstrate that the performance of our proposed WT-based CSI reconstruction is more stable and consistent across different experimental scenarios when compared to FFT reconstruction. 

\subsubsection{Wavelet Coherence of Different CSI Reconstructions}
\begin{figure*} 
    \centerline{
  \subfloat[Raw CSI \label{subfig:wc_raw}]{%
       \includegraphics[width=0.3\textwidth]{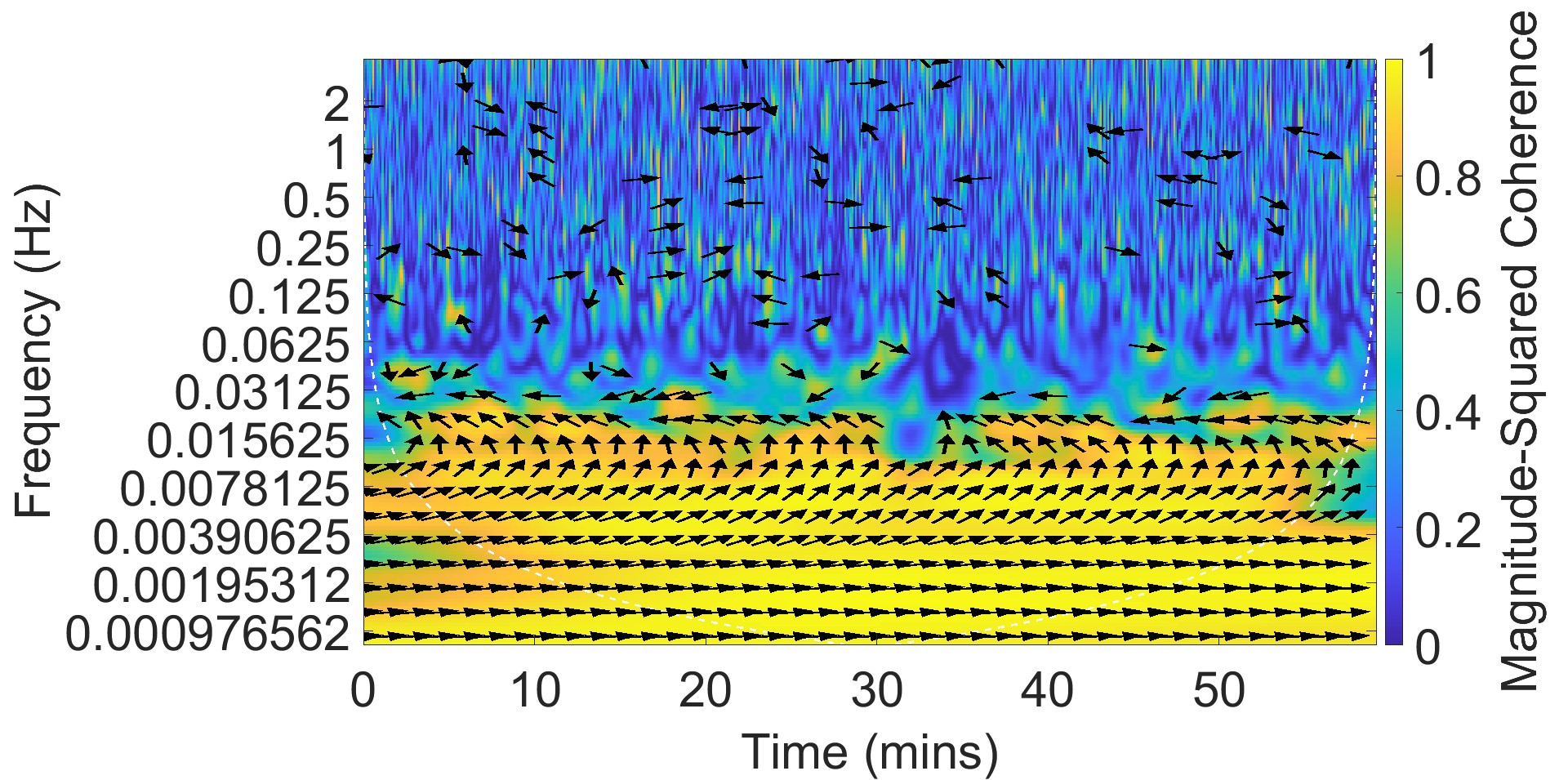}}
  \subfloat[Golay filtered CSI\label{subfig:wc_golay}]{%
        \includegraphics[width=0.3\textwidth]{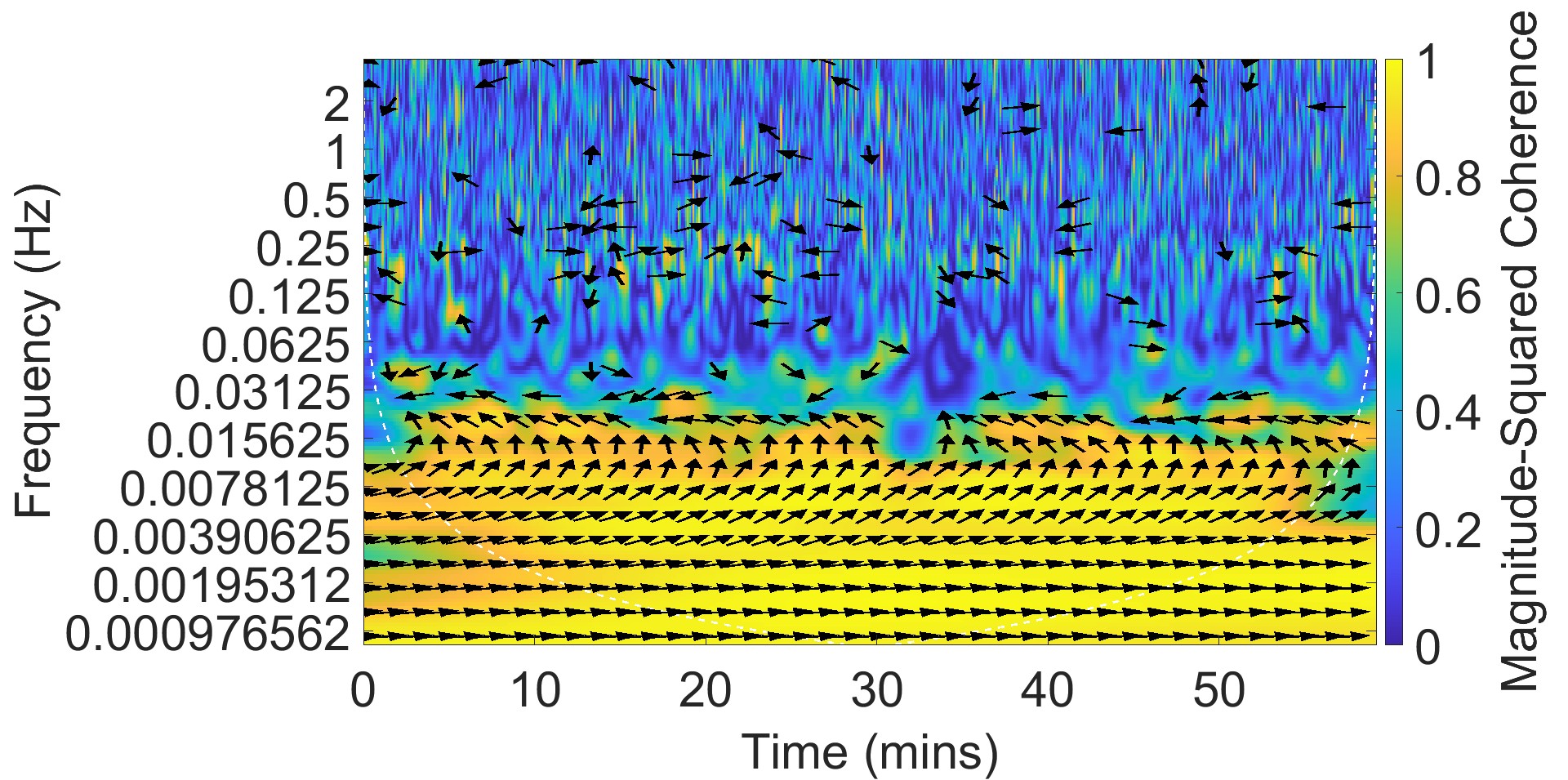}}
  \subfloat[Filtered, synced CSI\label{subfig:wc_golay_sync}]{%
        \includegraphics[width=0.3\textwidth]{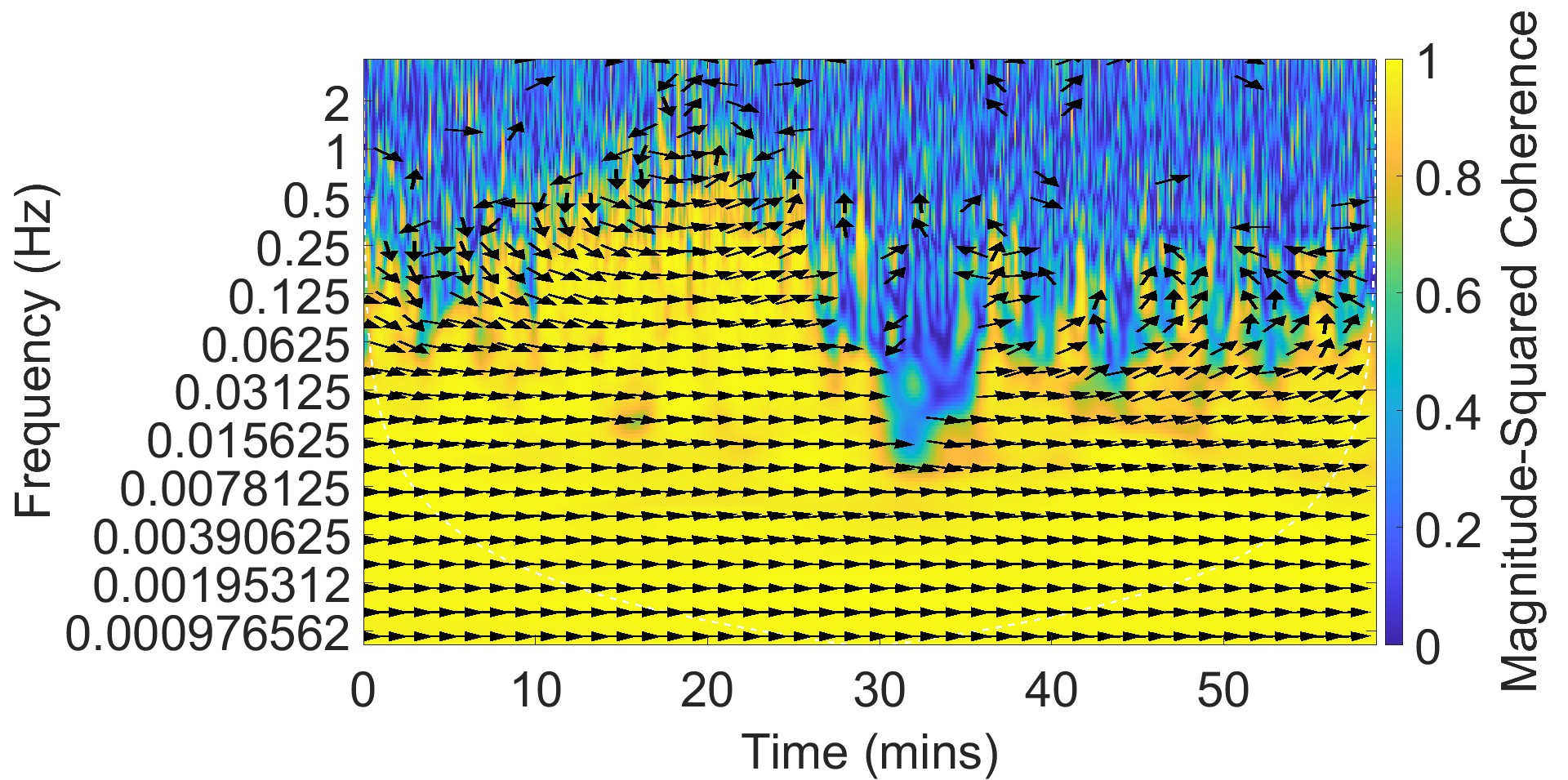}}}
    \centerline{
  \subfloat[FFT-reconstucted CSI\label{subfig:wc_ifft}]{%
       \includegraphics[width=0.3\textwidth]{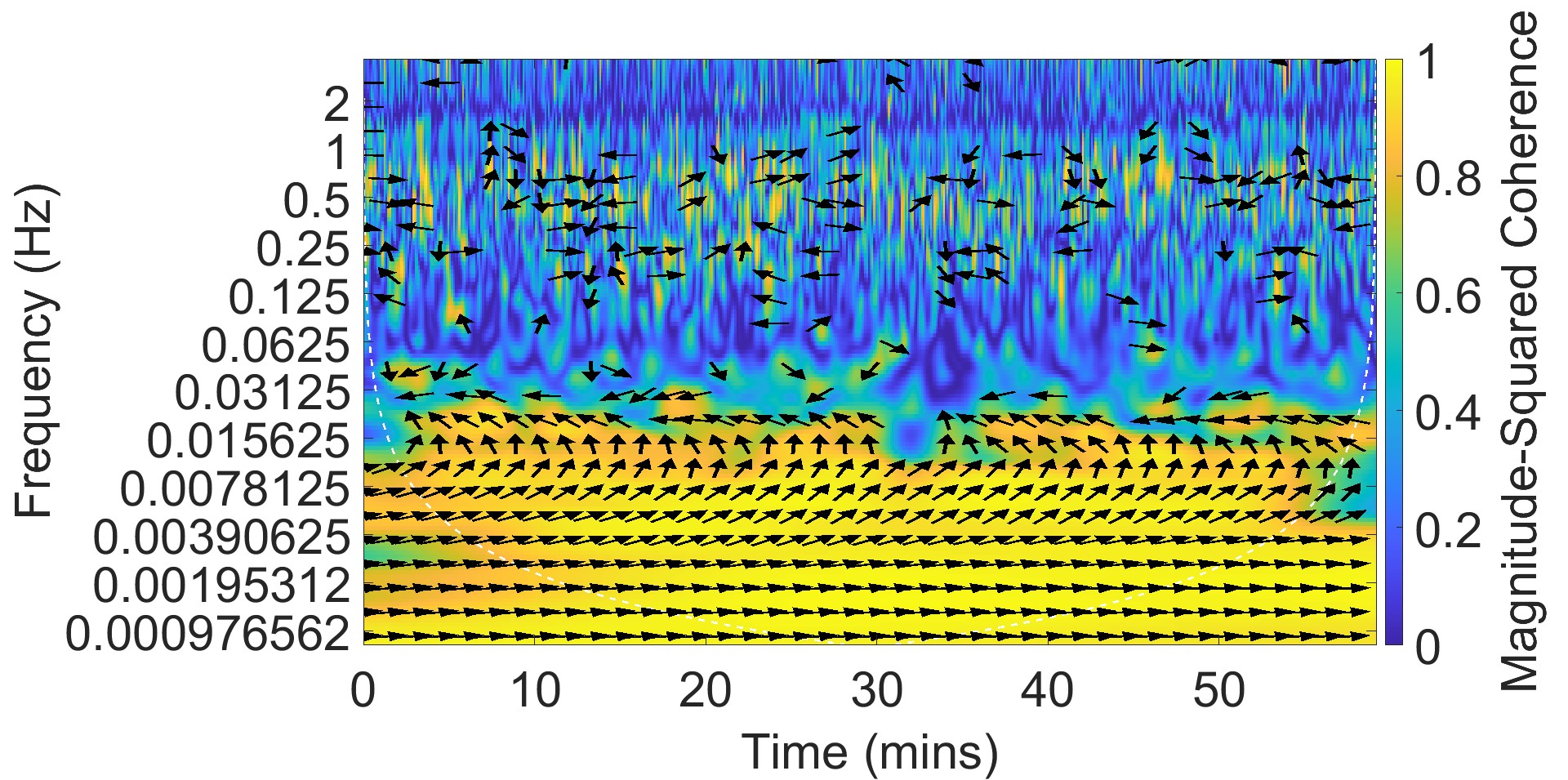}}
  \subfloat[WT-reconstructed CSI\label{subfig:wc_wt}]{%
        \includegraphics[width=0.3\textwidth]{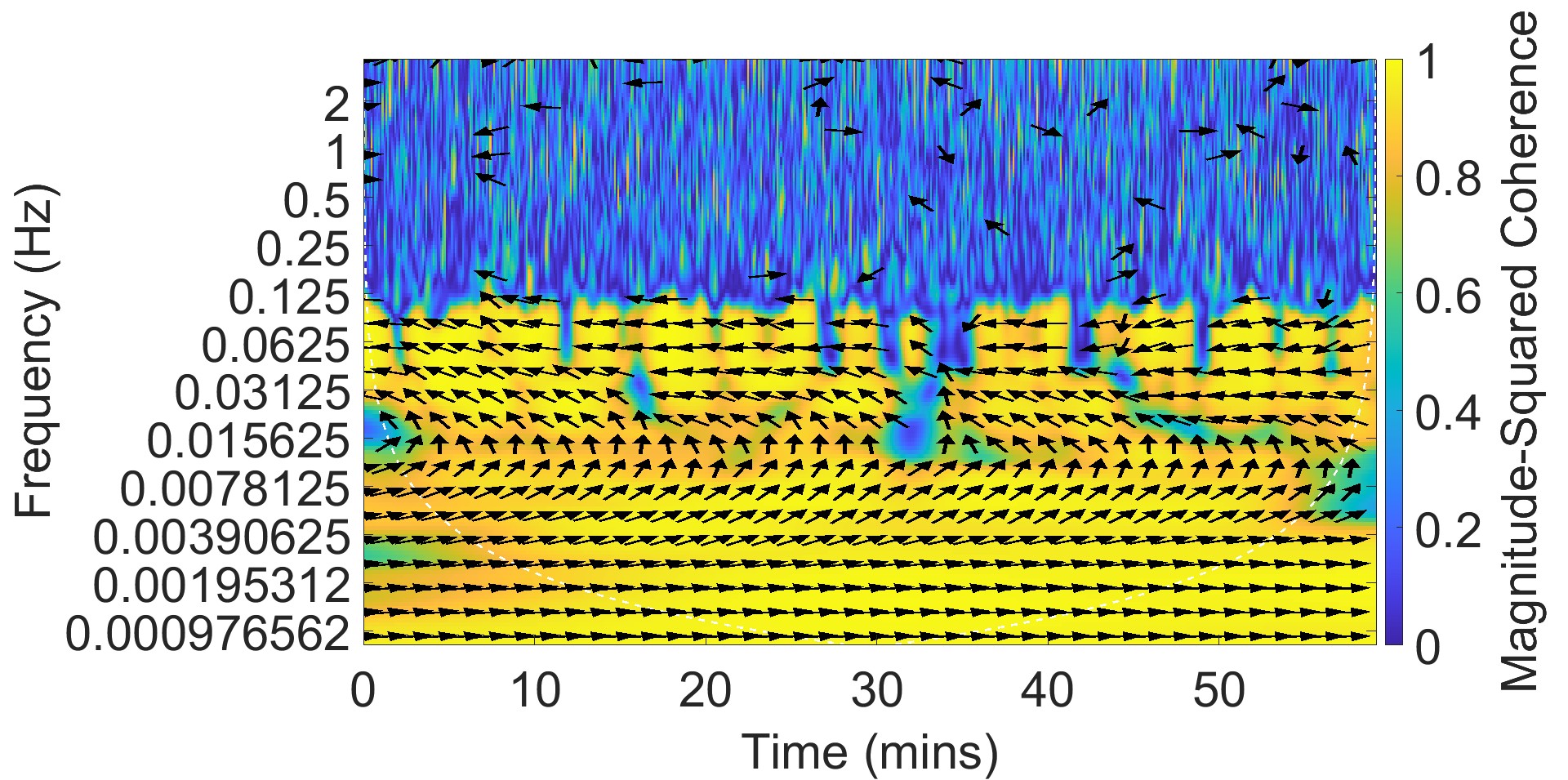}}
  \subfloat[WT-reconstrucetd, synced CSI\label{subfig:wc_wt_sync}]{%
        \includegraphics[width=0.3\textwidth]{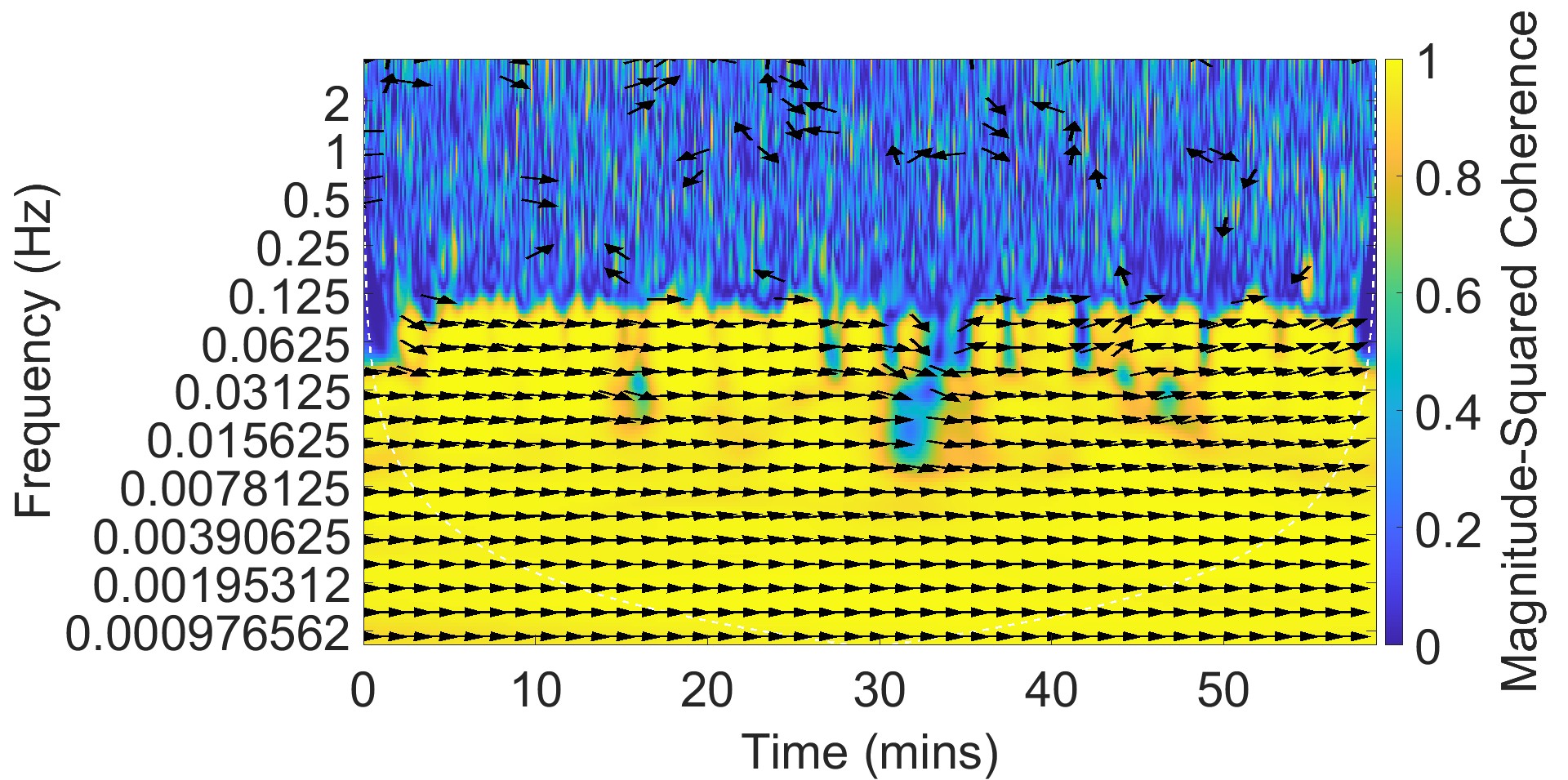}}}
        
    
   \caption{Wavelet coherence of the different CSI reconstructing approaches under the \lab~scenario.}
  \label{fig:wc_lab} 
\end{figure*}
%
\begin{figure*} 
    \centerline{
  \subfloat[Raw CSI \label{subfig:wc_raw_lc}]{%
       \includegraphics[width=0.3\textwidth]{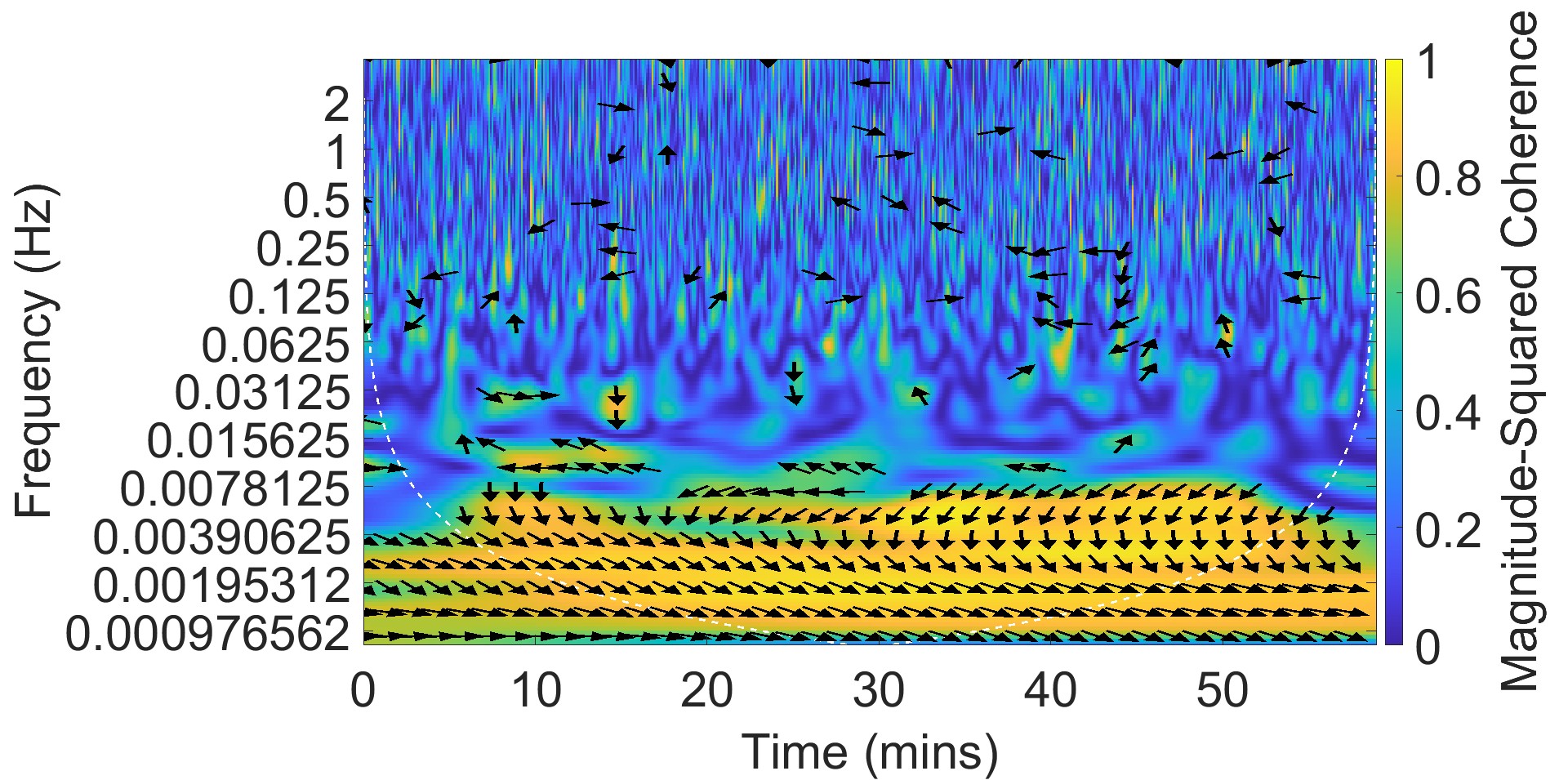}}
  \subfloat[Golay filtered CSI\label{subfig:wc_golay_lc}]{%
        \includegraphics[width=0.3\textwidth]{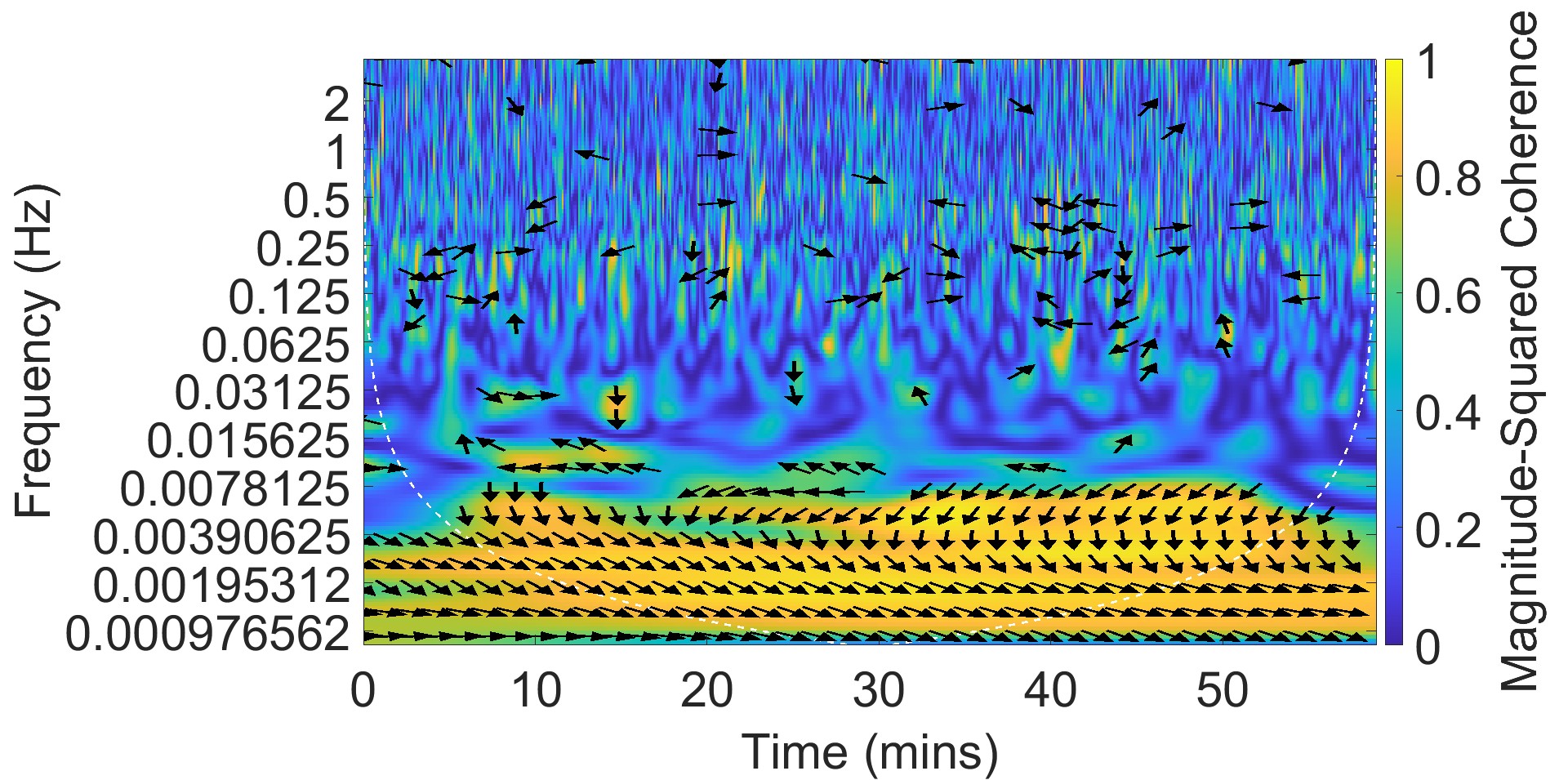}}
  \subfloat[Filtered, synced CSI\label{subfig:wc_golay_sync_lc}]{%
        \includegraphics[width=0.3\textwidth]{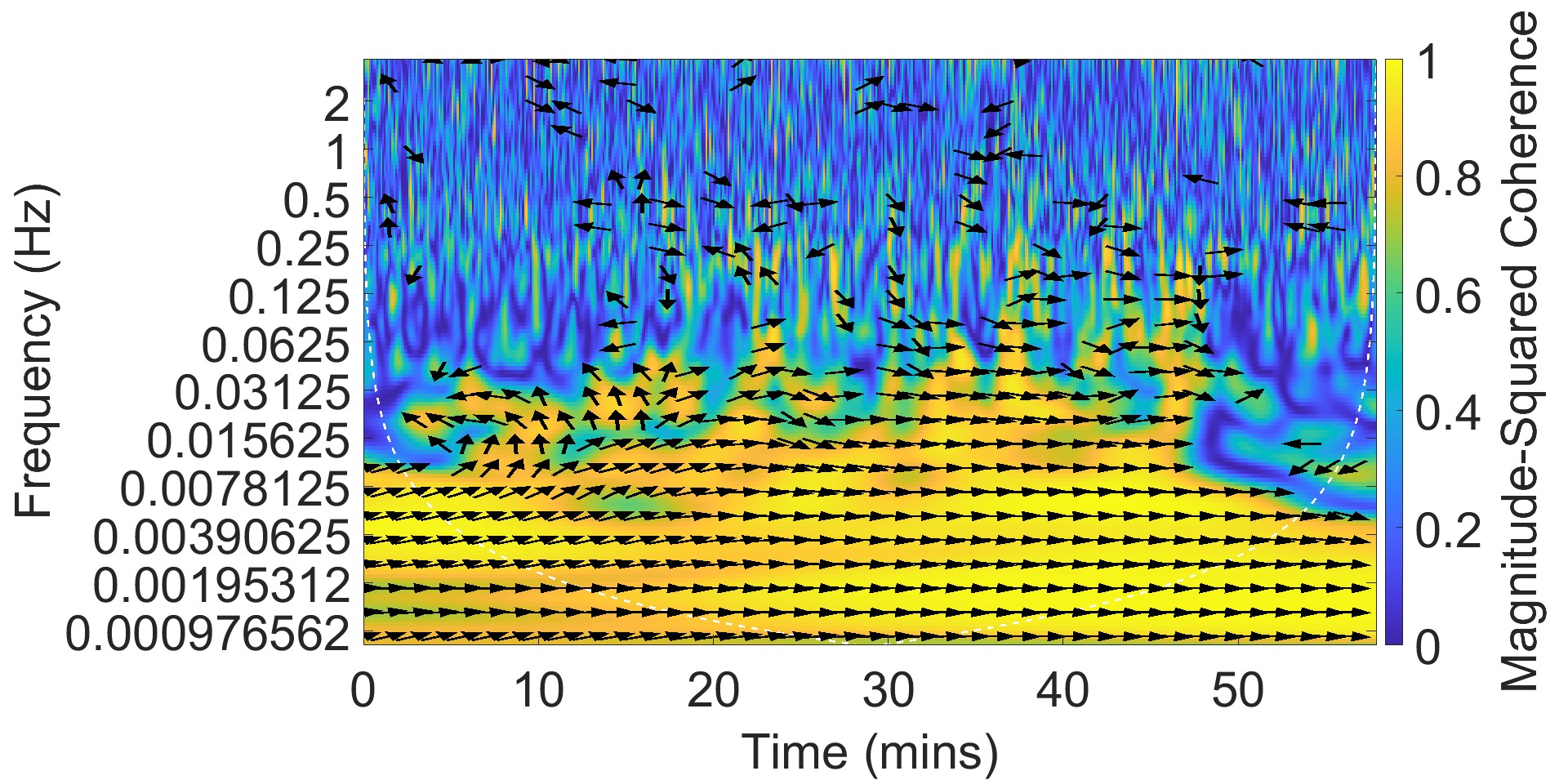}}}
    \centerline{
  \subfloat[FFT-reconstructed CSI\label{subfig:wc_ifft_lc}]{%
       \includegraphics[width=0.3\textwidth]{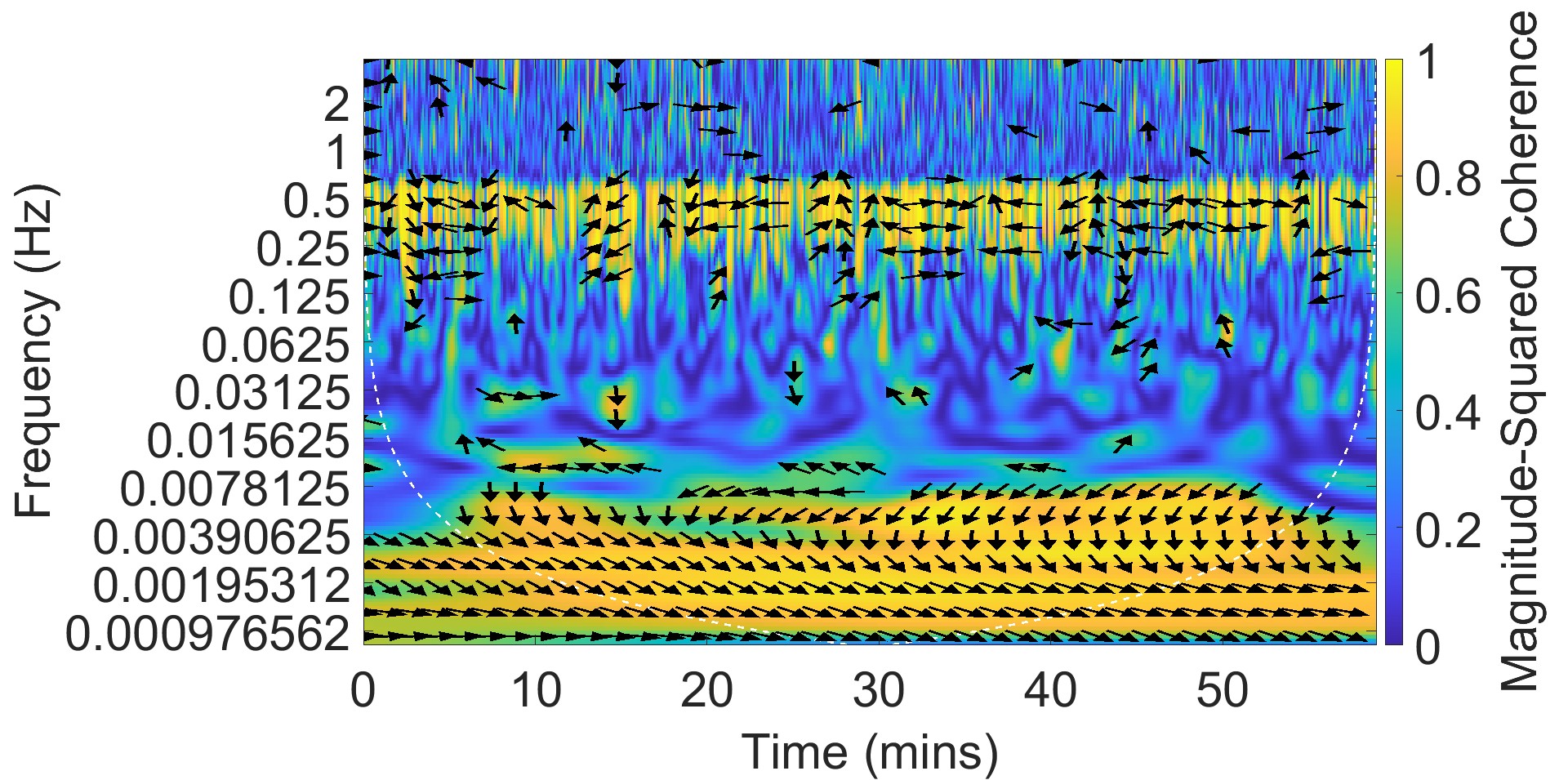}}
  \subfloat[WT-reconstructed CSI\label{subfig:wc_wt_lc}]{%
        \includegraphics[width=0.3\textwidth]{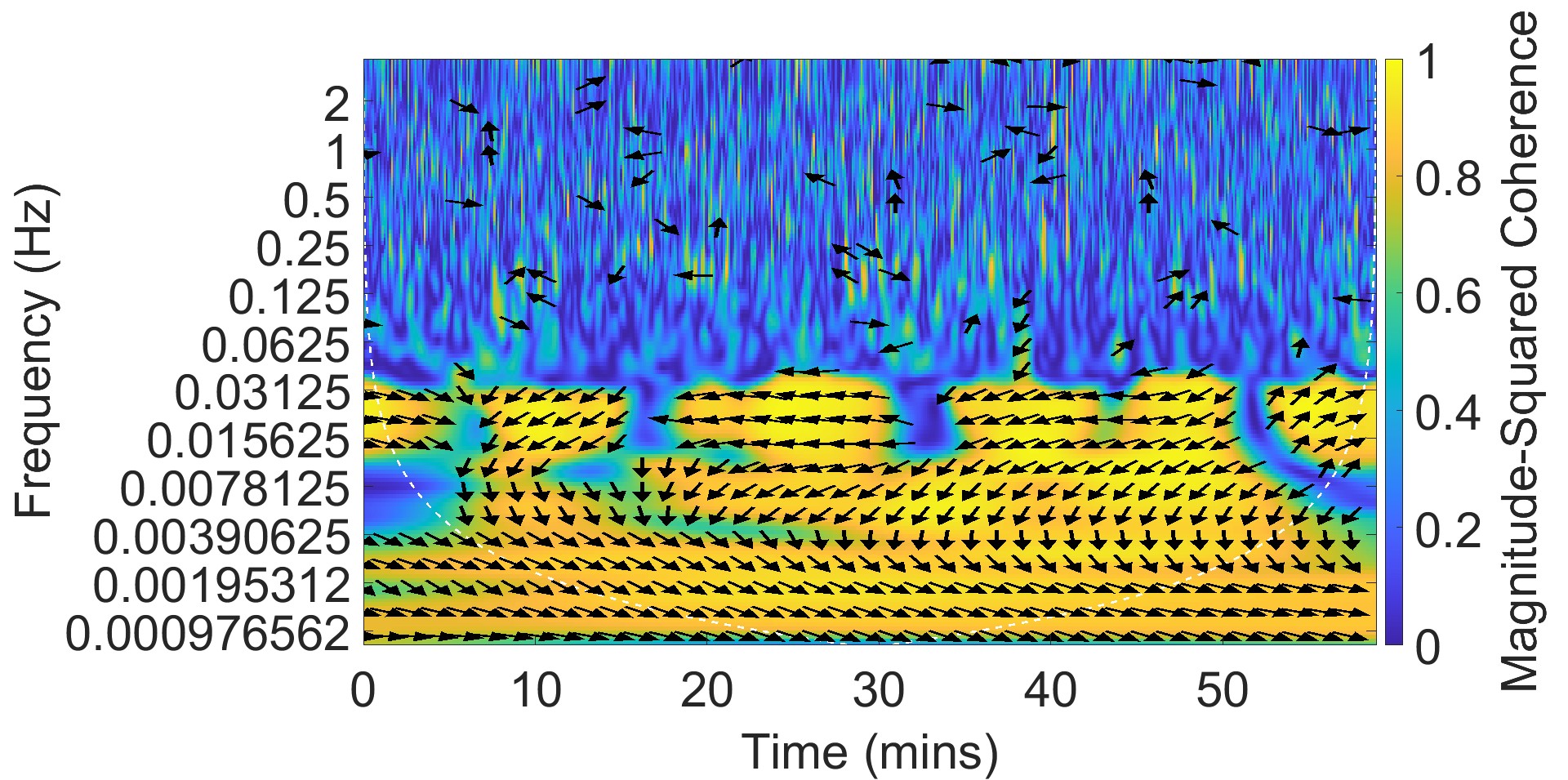}}
  \subfloat[WT-reconstructed, synced CSI\label{subfig:wc_wt_sync_lc}]{%
        \includegraphics[width=0.3\textwidth]{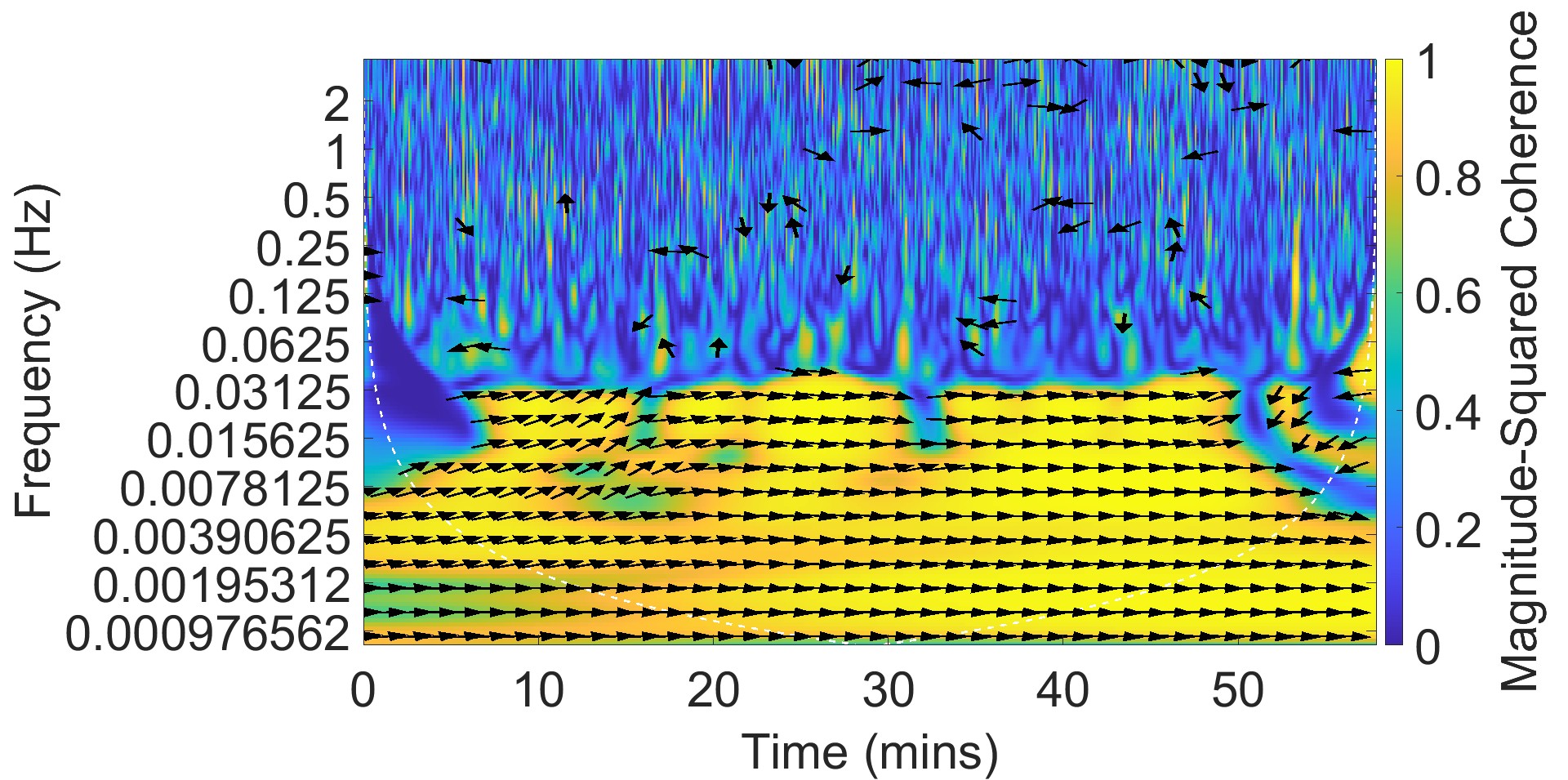}}}
        
   
   \caption{Wavelet coherence of the different CSI reconstructing approaches under the \lc~scenario.}
  \label{fig:wc_lc} 
\end{figure*}
%
\begin{figure*} 
    \centerline{
  \subfloat[Raw CSI \label{subfig:wc_raw_lo}]{%
       \includegraphics[width=0.3\textwidth]{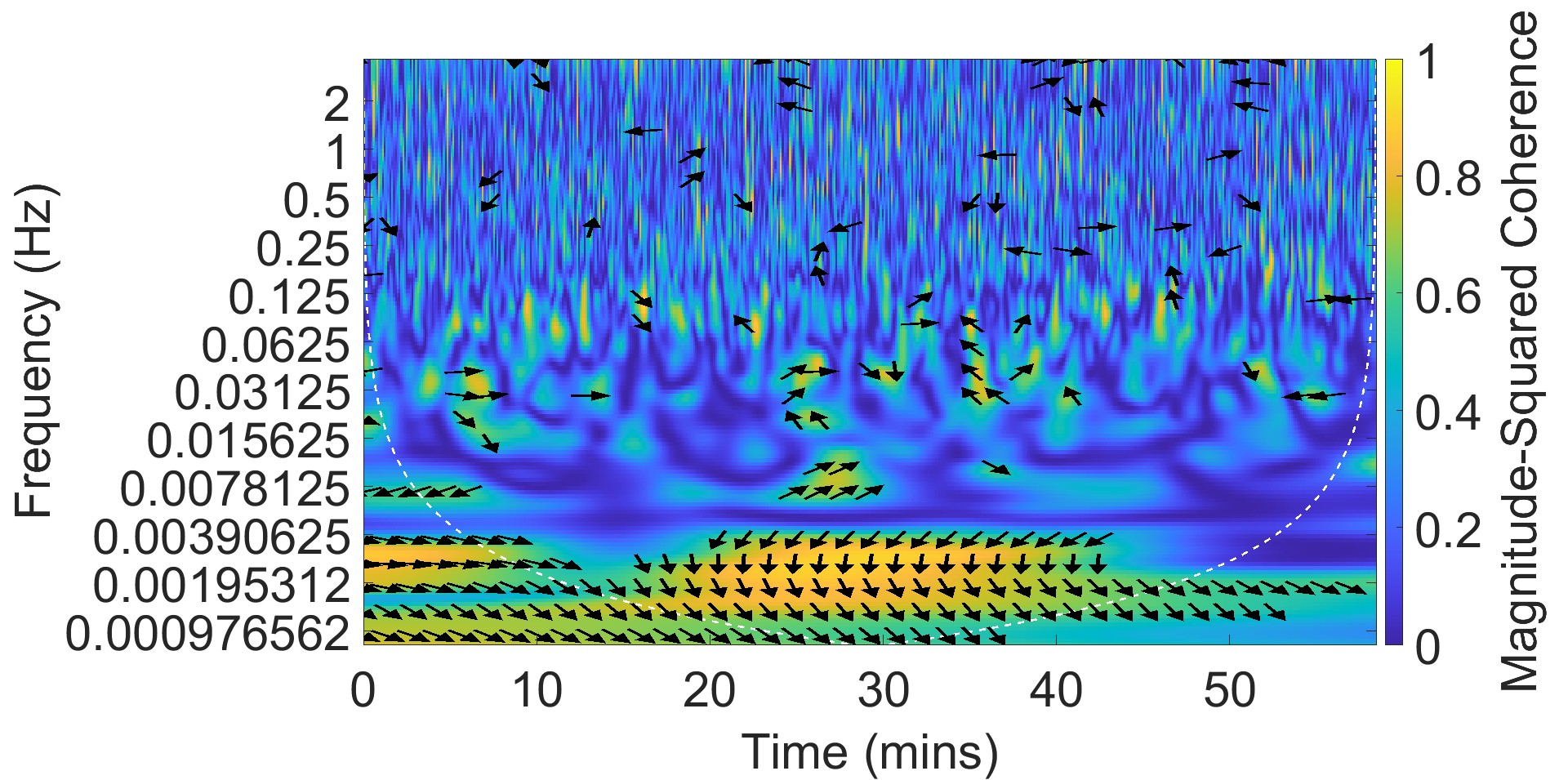}}
  \subfloat[Golay filtered CSI\label{subfig:wc_golay_lo}]{%
        \includegraphics[width=0.3\textwidth]{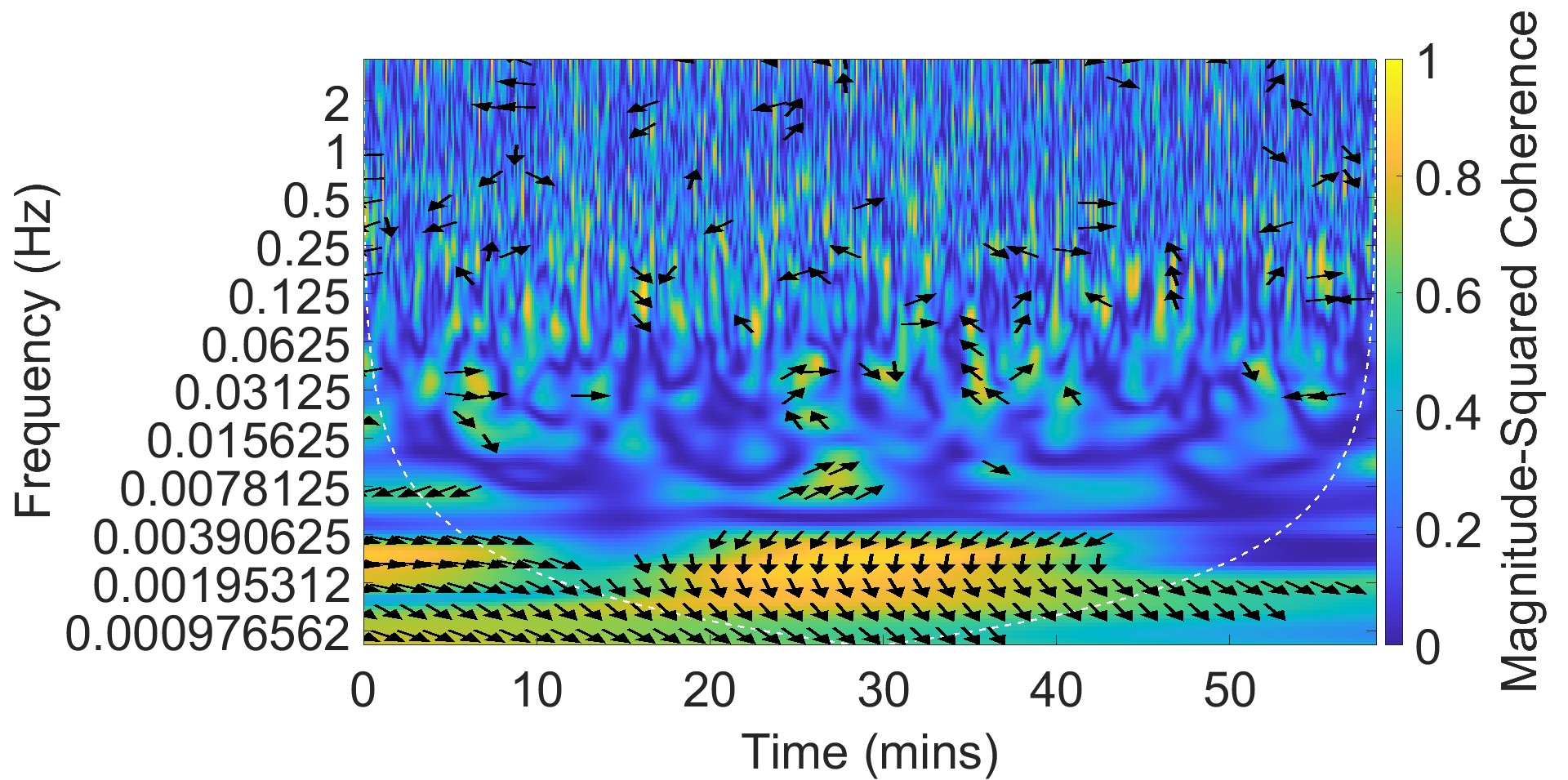}}
  \subfloat[Filtered, synced CSI\label{subfig:wc_golay_sync_lo}]{%
        \includegraphics[width=0.3\textwidth]{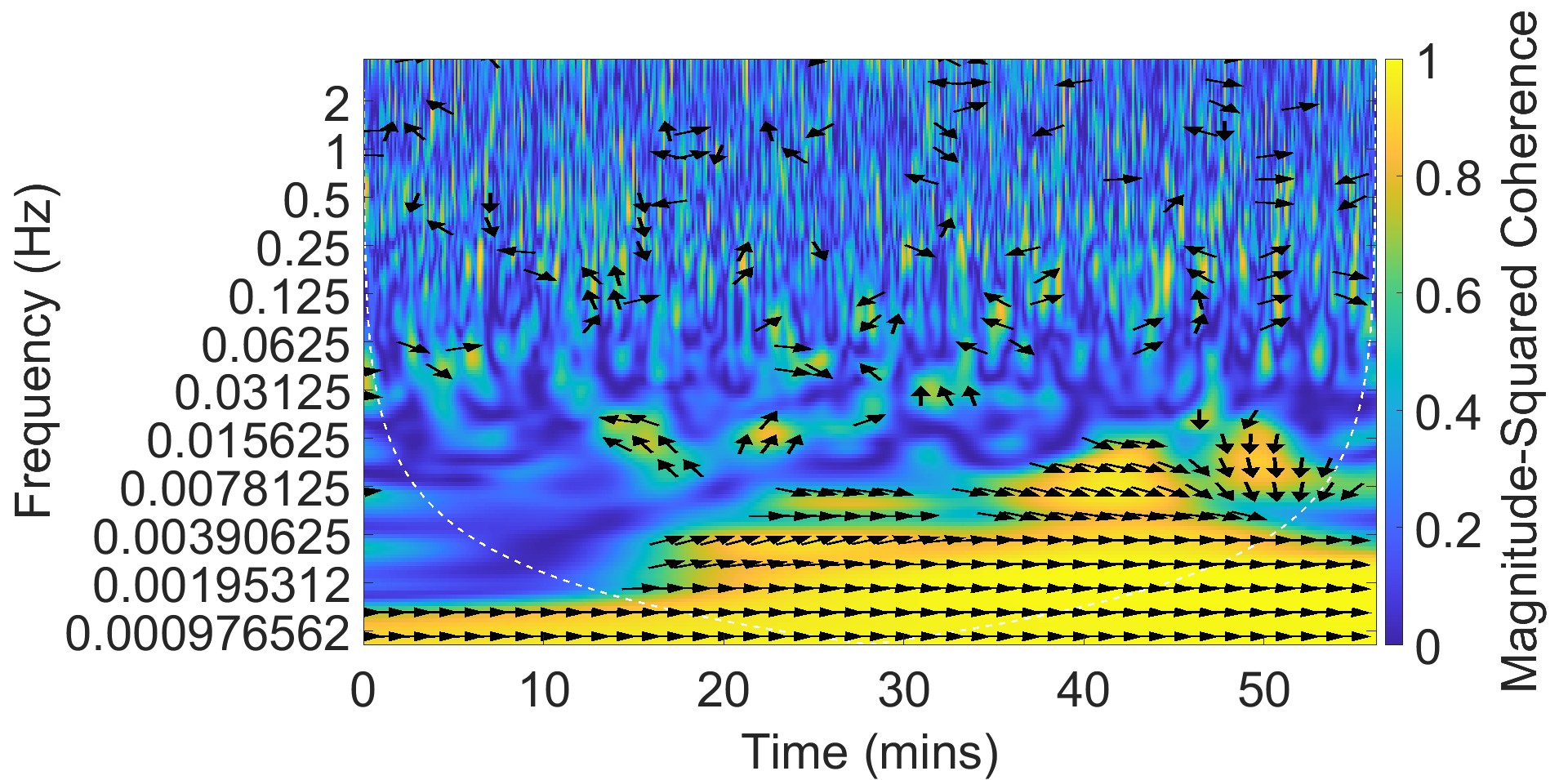}}}
    \centerline{
  \subfloat[FFT-reconstructed CSI\label{subfig:wc_ifft_lo}]{%
       \includegraphics[width=0.3\textwidth]{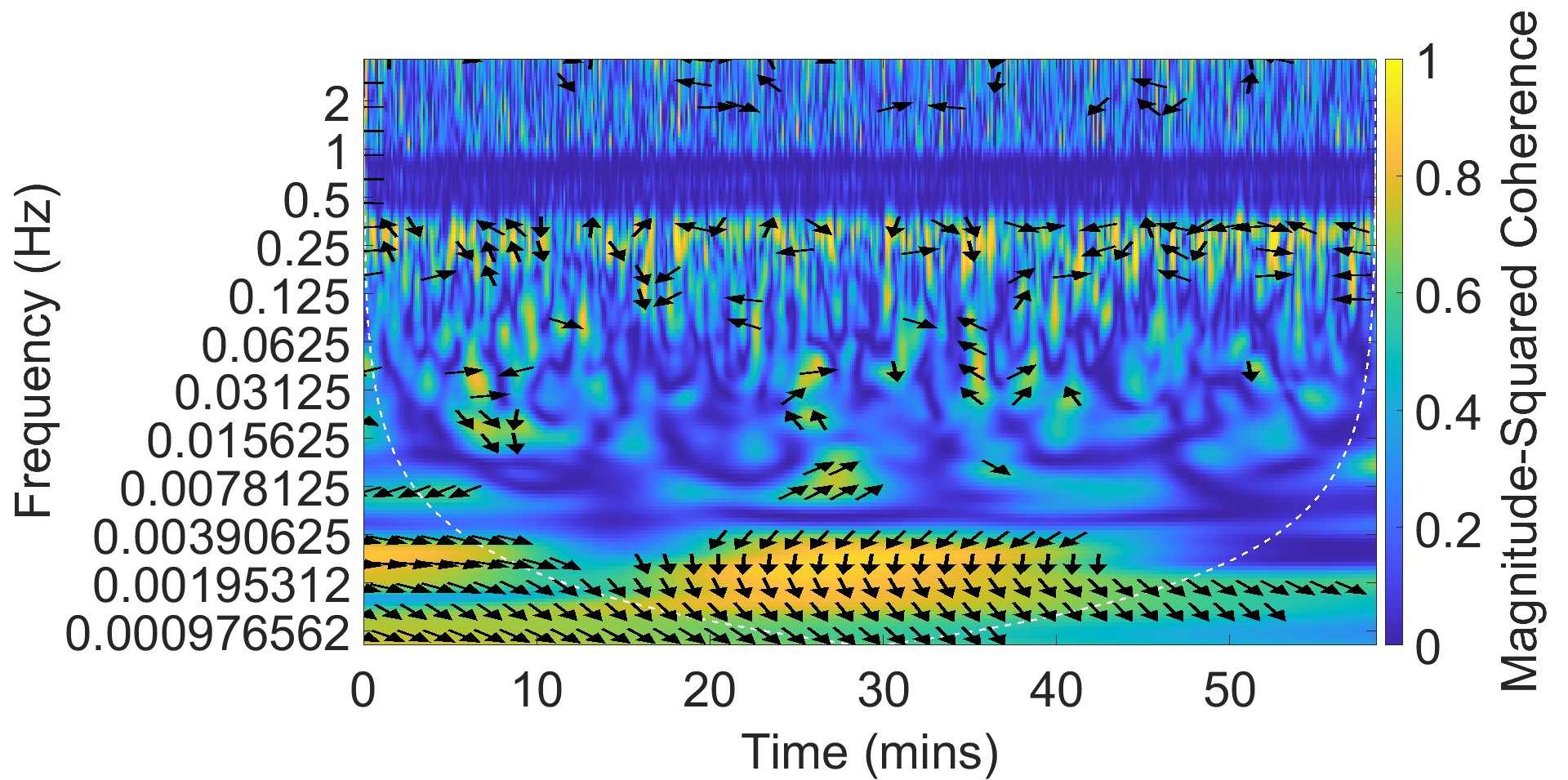}}
  \subfloat[WT-reconstructed CSI\label{subfig:wc_wt_lo}]{%
        \includegraphics[width=0.3\textwidth]{ 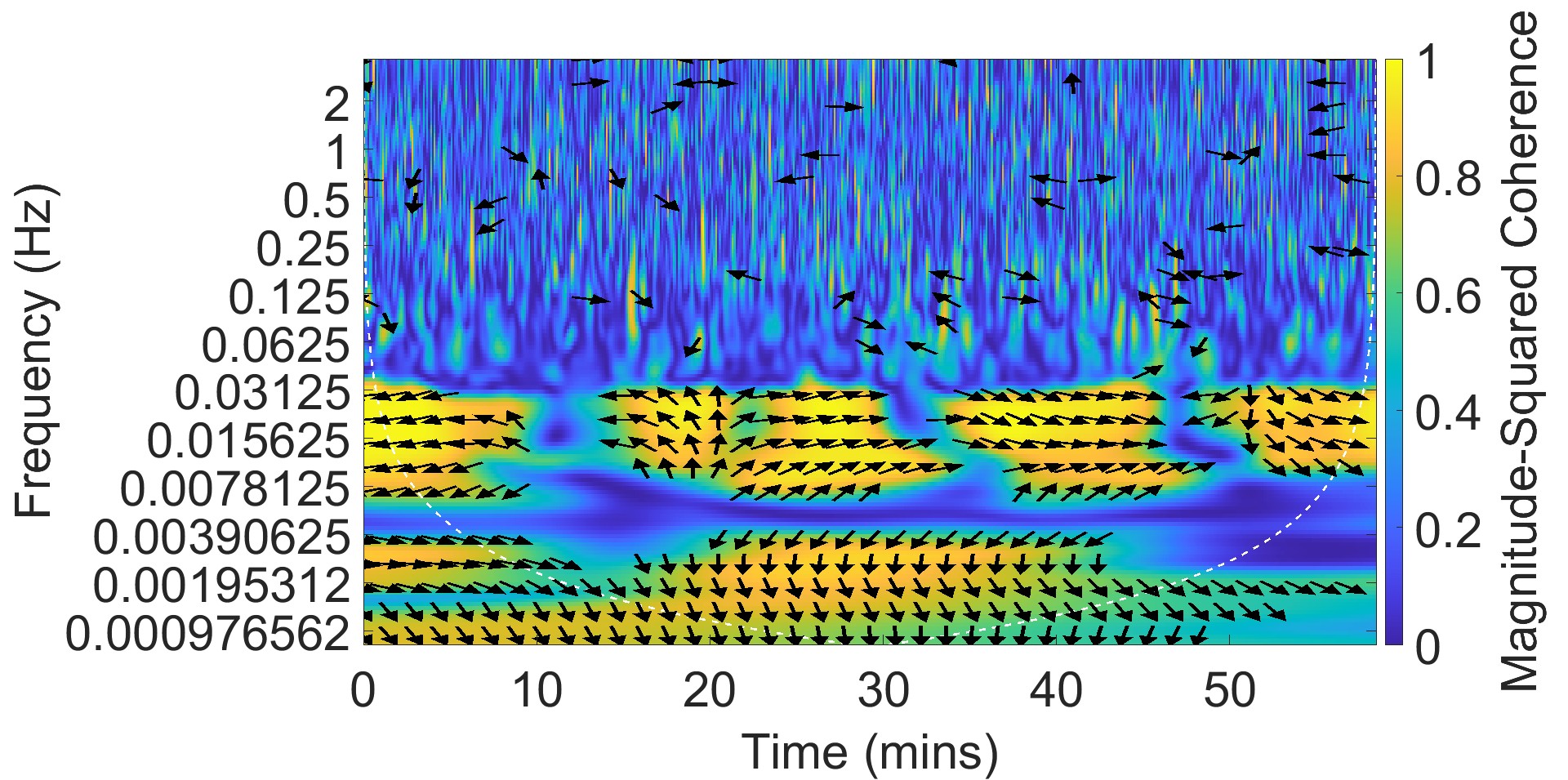}}
  \subfloat[WT-reconstructed, synced CSI\label{subfig:wc_wt_sync_lo}]{%
        \includegraphics[width=0.3\textwidth]{ 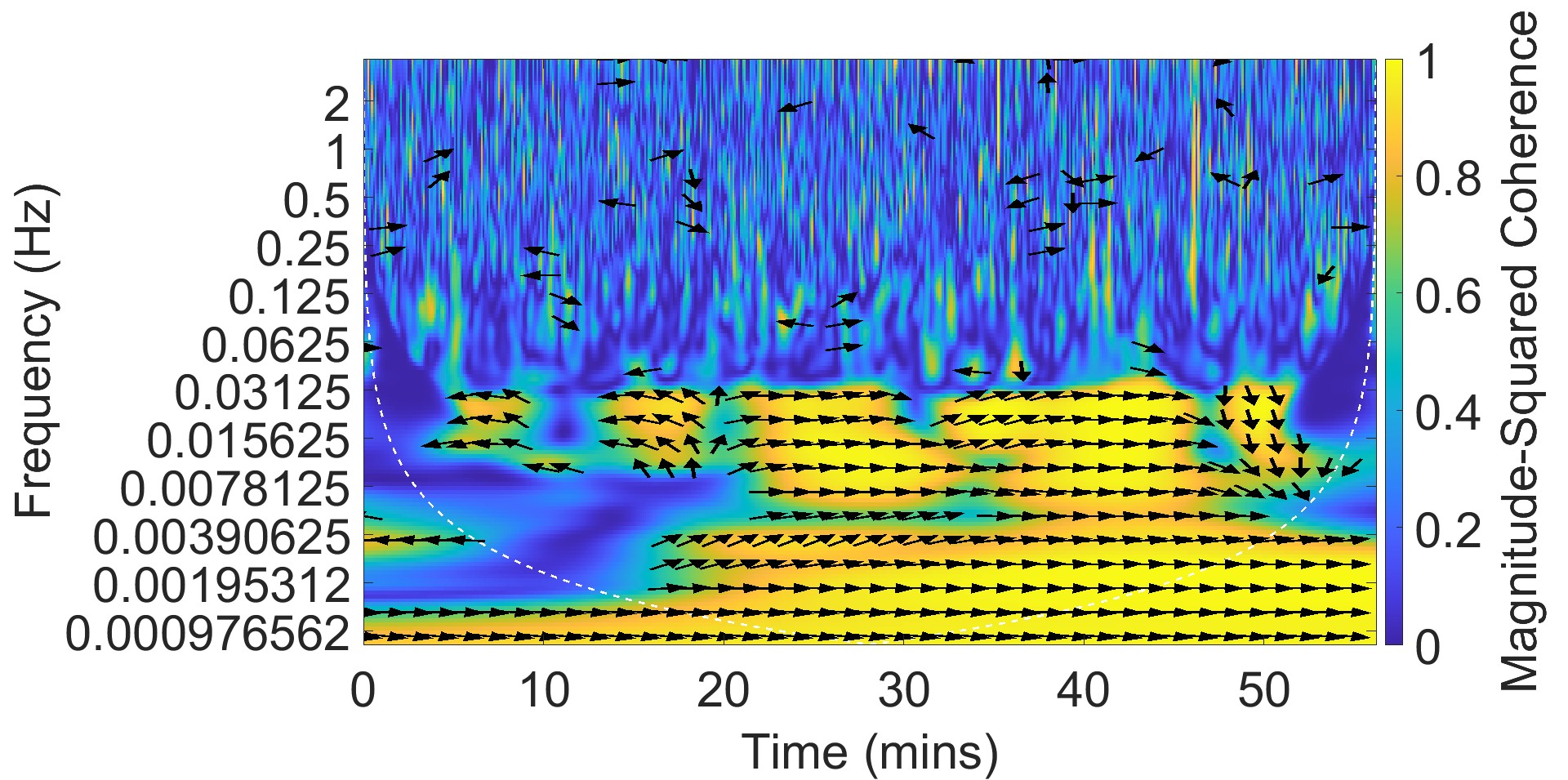}}}
        
    
   \caption{Wavelet coherence of the different CSI reconstructing approaches under the \lo~scenario.}
  \label{fig:wc_lo} 
\end{figure*}
Figs.~\ref{fig:wc_lab},~\ref{fig:wc_lc}, and \ref{fig:wc_lo} depict the impact of Golay filtering, FFT reconstruction, WT reconstruction, and CSI Synchronization on the WC metric for the three studied location scenarios.
The figures show commensurate results with Sec.~\ref{sec:metrics_analysis} and demonstrate WC's effectiveness in quantifying channel reciprocity and capturing the impact of the measurement asynchrony and data preprocessing on the CSI coherence and phase difference. 
%
%
Golay filtering (Fig.~\ref{subfig:wc_golay}) and FFT reconstruction (Fig.~\ref{subfig:wc_ifft}) exhibit similar effects on WC, introducing additional high coherence instances over the frequency range of $0.03-2$ Hz compared to raw CSI (Fig.~\ref{subfig:wc_raw}) which is attributed to the smoothing effect of Gloay filtering and FFT-reconstruction. The proposed WT-based reconstruction technique (Fig.~\ref{subfig:wc_wt}) enhances channel reciprocity by extending the high-coherence area to cover the frequency range from $0-0.125$ Hz, an improvement over raw CSI.
The figure also demonstrates the impact of CSI synchronization on WC. Synchronization widens the range of high coherence values and adjusts the phase difference between the CSI at AP and STA, indicated by the black arrows at zero angle. Combining synchronization with the WT reconstruction technique results in the best WC performance (Fig.~\ref{subfig:wc_wt_sync}), characterized by a wide high-coherence, in-phase area.
It is worth noting that combining synchronization with Golay filtering (Fig.~\ref{subfig:wc_golay_sync}) achieves a wider high-coherence area compared to the synchronized, WT-reconstructed CSI. However, the additional high-frequency components in the Golay-filtered WC are coherent during a limited time duration of about 10 minutes and will adversely impact the key-generation performance.
Similar observations hold for the NLoS scenarios, \lc~and \lo~(Figs.~\ref{fig:wc_lc}, and \ref{fig:wc_lo}), where channel impairments and asynchronous measurements limit the frequency range and time durations of high coherence values for the raw CSI and the preprocessing techniques. Under \lo~(Fig.~\ref{fig:wc_lo}), the proposed Wavelet technique has a significant positive impact on WC values, mitigating the impact of the lack of synchronous measurements across AP and STA. 

\section{Secret-Key Generation}
\label{sec:skg}
In this section, the WT-based CSI reconstruction and synchronization, described in Sec.~\ref{sec:technique}, are utilized to generate secret keys between AP and STA. The secret key generation process initiates with channel probing, where STA and AP exchange signals to estimate their CSIs. Subsequently, both devices convert their CSIs into bit streams using a quantization scheme and Gray coding, similar to an analog-to-digital converter. Quantization employs thresholds determined by the Cumulative Distribution Function (CDF) of the CSI samples. Error correction coding (ECC) is then applied to rectify key mismatches. If the error exceeds ECC capacity, key agreement failure prompts a restart with channel probing. Repeated failures result in low KGR. To mitigate eavesdropper information, cryptographic hash functions are employed in privacy amplification, producing an output closer to a uniformly random key~\cite{zhangKeyGenerationWireless2016c}.
%
\subsection{Wavelet-based Secret Key Generation}
\begin{figure}
\includegraphics[ width=\columnwidth]{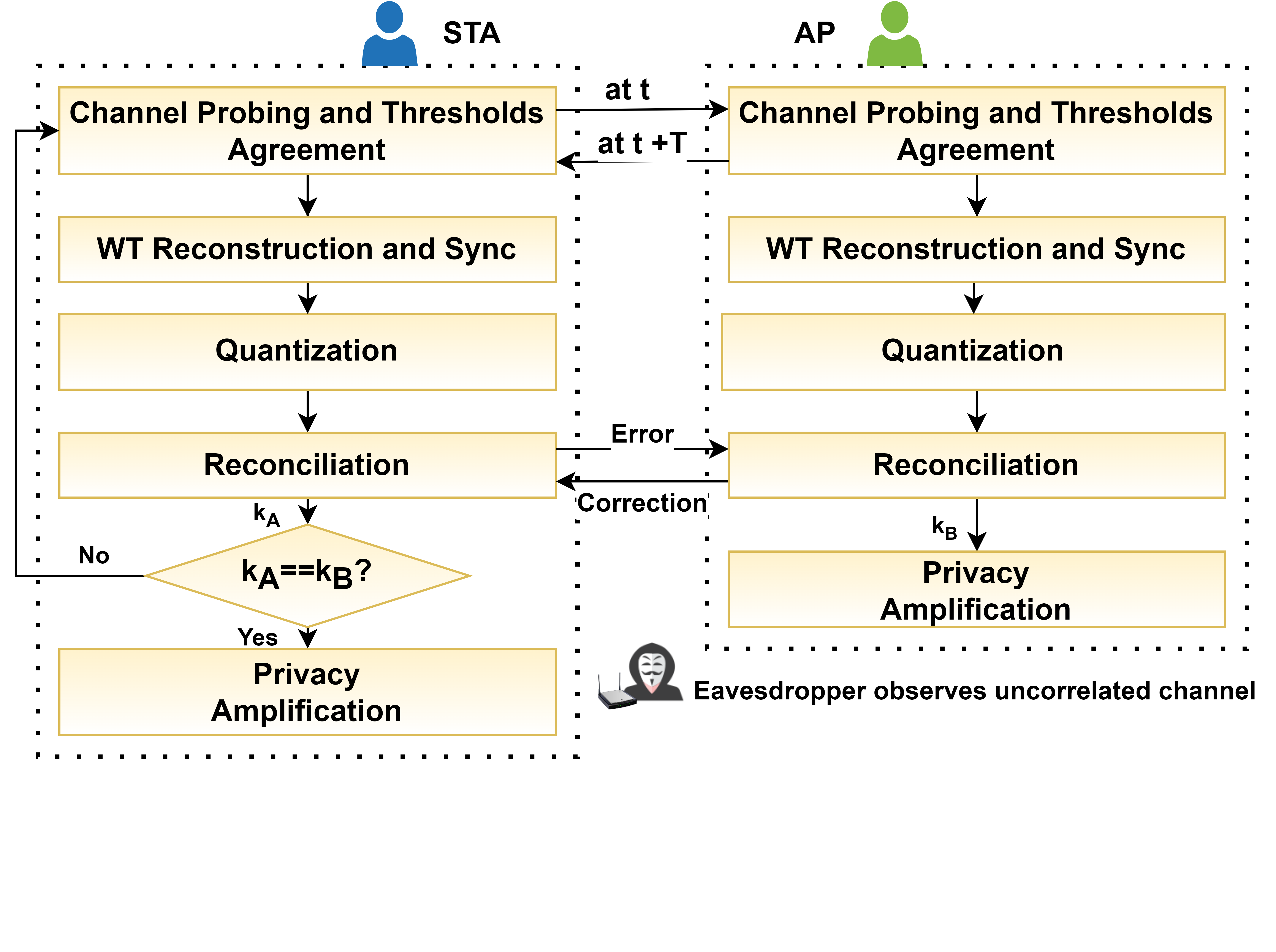}
\vspace*{-15mm}
\caption{Wavelet-based Key Generation (\WSKG) scheme.}
\label{steps}
\end{figure}

Fig.~\ref{steps} illustrates the proposed Wavelet-based Secret Key Generation (\WSKG) scheme, with emphasis on the initial three steps for improving channel reciprocity. Information reconciliation and privacy amplification, not central to reciprocity, are excluded from consideration in this study.

\myitemizebegin

\item \textbf{Step 1: Channel Probing and Thresholds Agreement.} First, STA and AP exchange probing signals to estimate CSIs. Second, they agree on coherence and sample thresholds, $\alpha$ and $\beta$, and estimate CSI time shift. To do so, STA sends CSI samples to AP, which computes WC and determines $\alpha$ and $\beta$. AP estimates time shift using time-lagged cross-correlation and communicates $\alpha$, $\beta$, and time shift to STA publicly.
 
\item \textbf{Step 2: WT Reconstruction and Synchronization.} Now that AP and STA know the threshold values, they use the WT-based CSI reconstruction in Algorithm \ref{Al1} with new CSI samples that are not exchanged over the public channel to reconstruct the CSI at AP and STA. AP and STA also sync their CSIs using the estimated time shift. 
Channel patial decorrelation prevents an attacker located at a half a wavelength distance ($6.25$ cm) away from AP and STA from reconstructing reciprocal CSI signals even if it has access to $\alpha$, $\beta$, and the CSI time shift~\cite{zhangKeyGenerationWireless2016c}.

\item \textbf{Step 3: Quantization.} A $4$-level lossless, uniform cumulative distribution function (CDF) based quantization is used to convert the CSI samples collected as described in Sec.~\ref{sec:testbed} into sequences of bits. Gray coding is later used to encode the quantization levels into bits. We used $100$ CSI samples to create $200$-bit key sequences.
Information reconciliation has a limited error correction capability. For that, we studied the performance of the proposed \WSKG~scheme at three error-threshold values: $5$, $15$, and $20$ bits. The error thresholds refer to the maximum acceptable number of bit errors (or mismatches) between the generated secret keys at AP and STA.  

\myitemizeend
%
\subsection{Result Analysis}
To evaluate the proposed WSKG approach, we compare its performance against \textbf{Golay Filtering} \cite{junejoLoRaLiSKLightweightShared2022a}, \textbf{FFT- reconstruction} \cite{zhang_efficient_2016}, Wavelet Packet Transform (WPT) key generation \cite{kumar_physical_2021} as well as two DL-based key generation benchmarks \textbf{Denoising Autoencoder (AE)} \cite{zhouPhysicalLayerSecret2022}, and \textbf{Bidirectional Convolutional Feature Learning (BCFL)} \cite{chenPhysicalLayerSecretKey2023}.
We also evaluate the importance of CSI synchronization by comparing the performance of FFT-based and Golay filtering-based CSI constructions with and without synchronization. The performance of the proposed \WSKG~ scheme is assessed using KGR and BER. 

\begin{figure*} 
    \centerline{
  \subfloat[Error = 5 bits\label{subfig:KGR_5}]{%
       \includegraphics[width =0.3\textwidth]{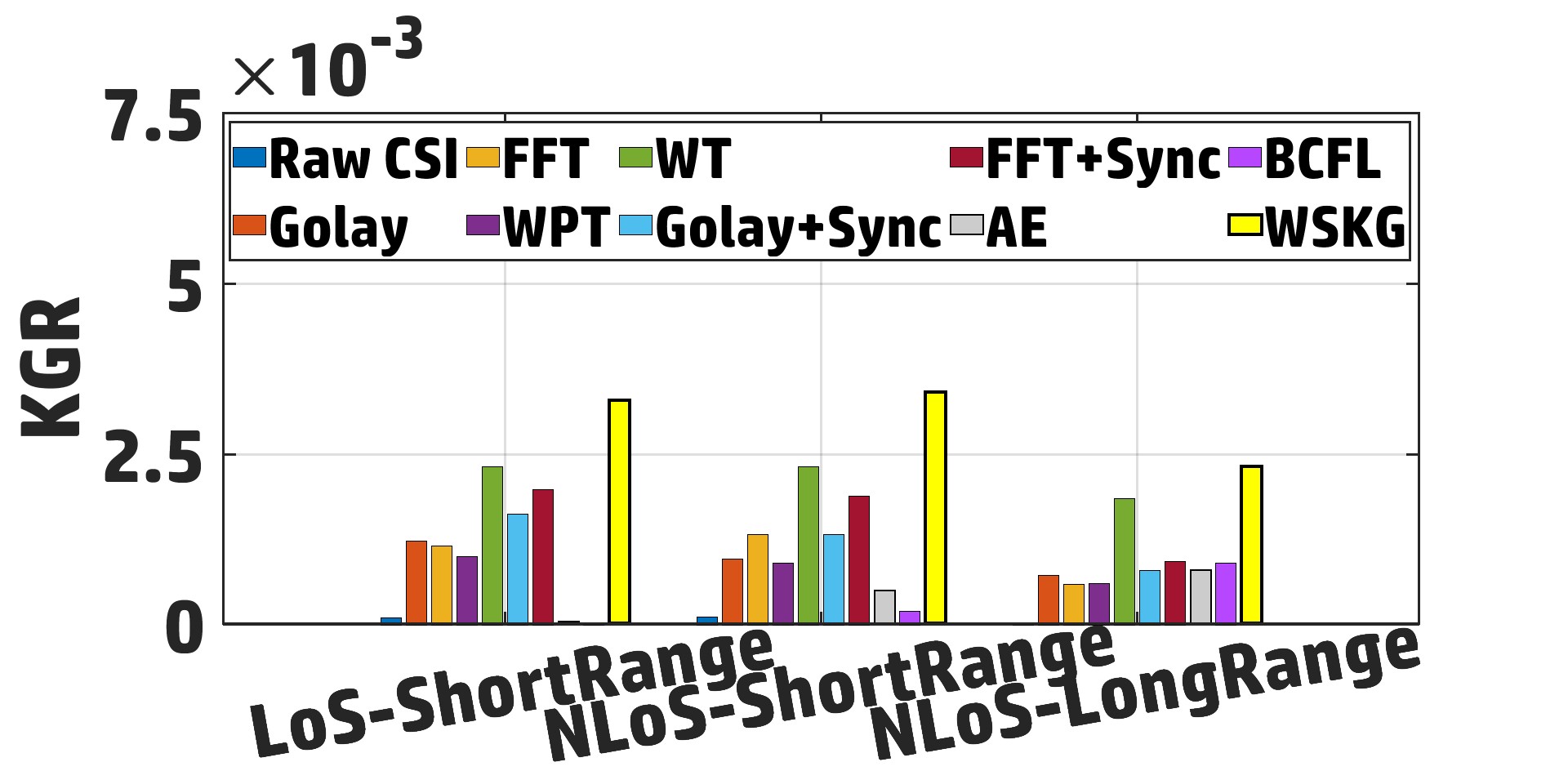}}
    \hspace{-0.1in}
  \subfloat[Error = 15 bits\label{subfig:KGR_15}]{%
        \includegraphics[width =0.3\textwidth]{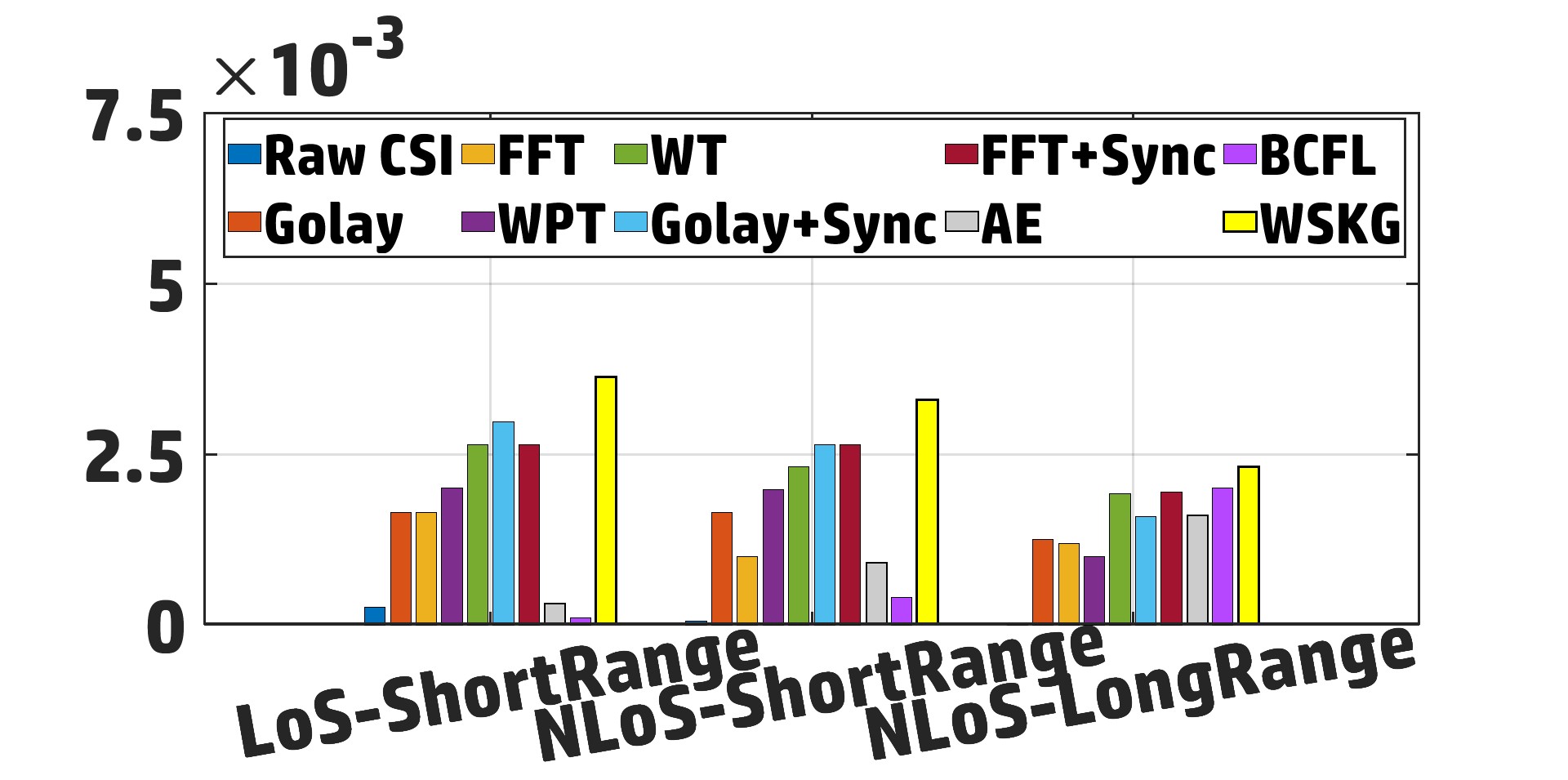}}
    \hspace{-0.1in}
  \subfloat[Error = 20 bits\label{subfig:KGR_20}]{%
        \includegraphics[width =0.3\textwidth]{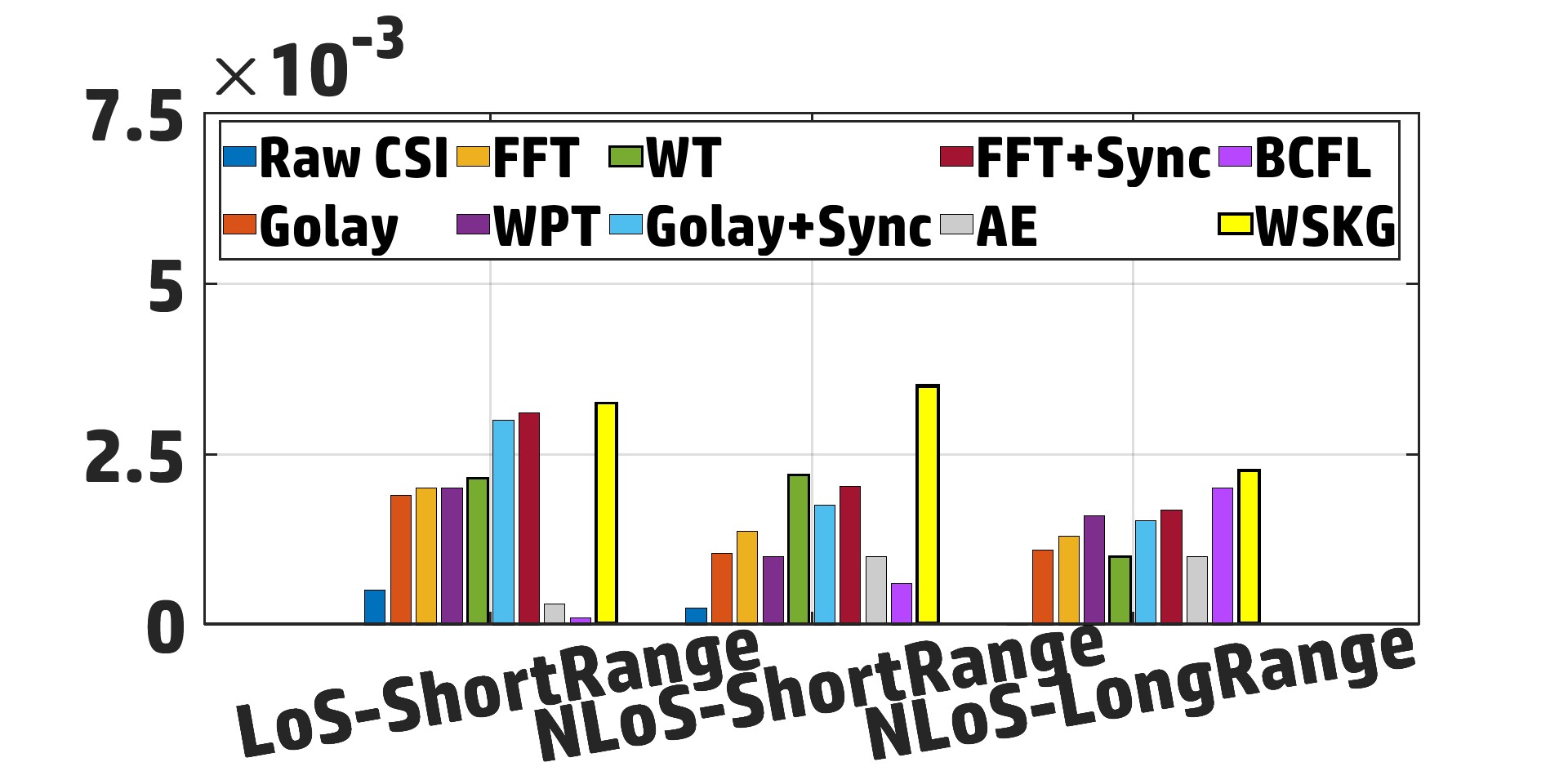}}}
    
   \caption{  Comparison of KGR performances. }
  \label{fig:KGR} 
\end{figure*}

\begin{figure*} 
    \centerline{
  \subfloat[Error = 5 bits\label{subfig:BER_5}]{%
       \includegraphics[width = 0.3\textwidth]{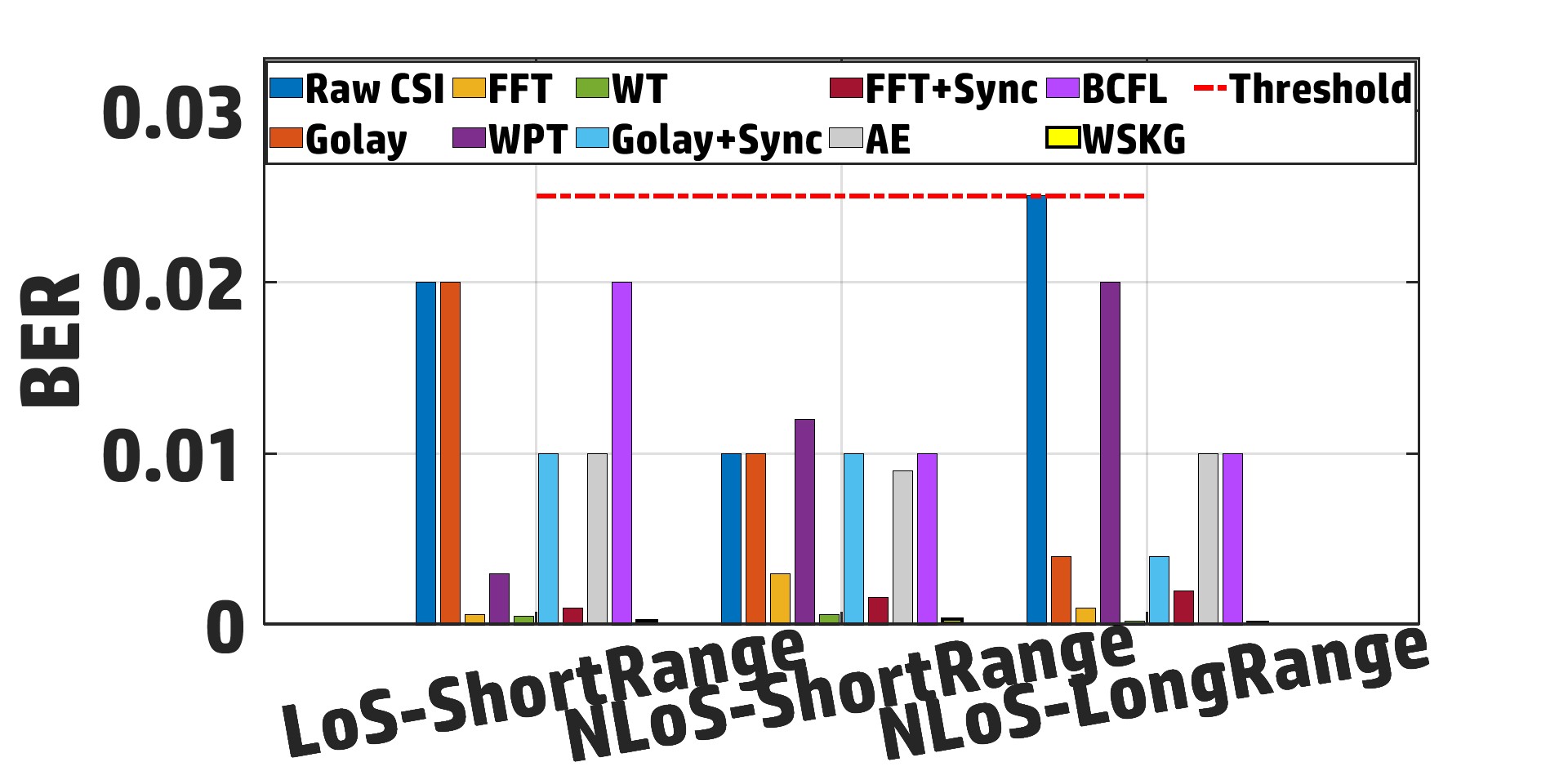}}
  \subfloat[Error = 15 bits\label{subfig:BER_15}]{%
        \includegraphics[width =0.3\textwidth]{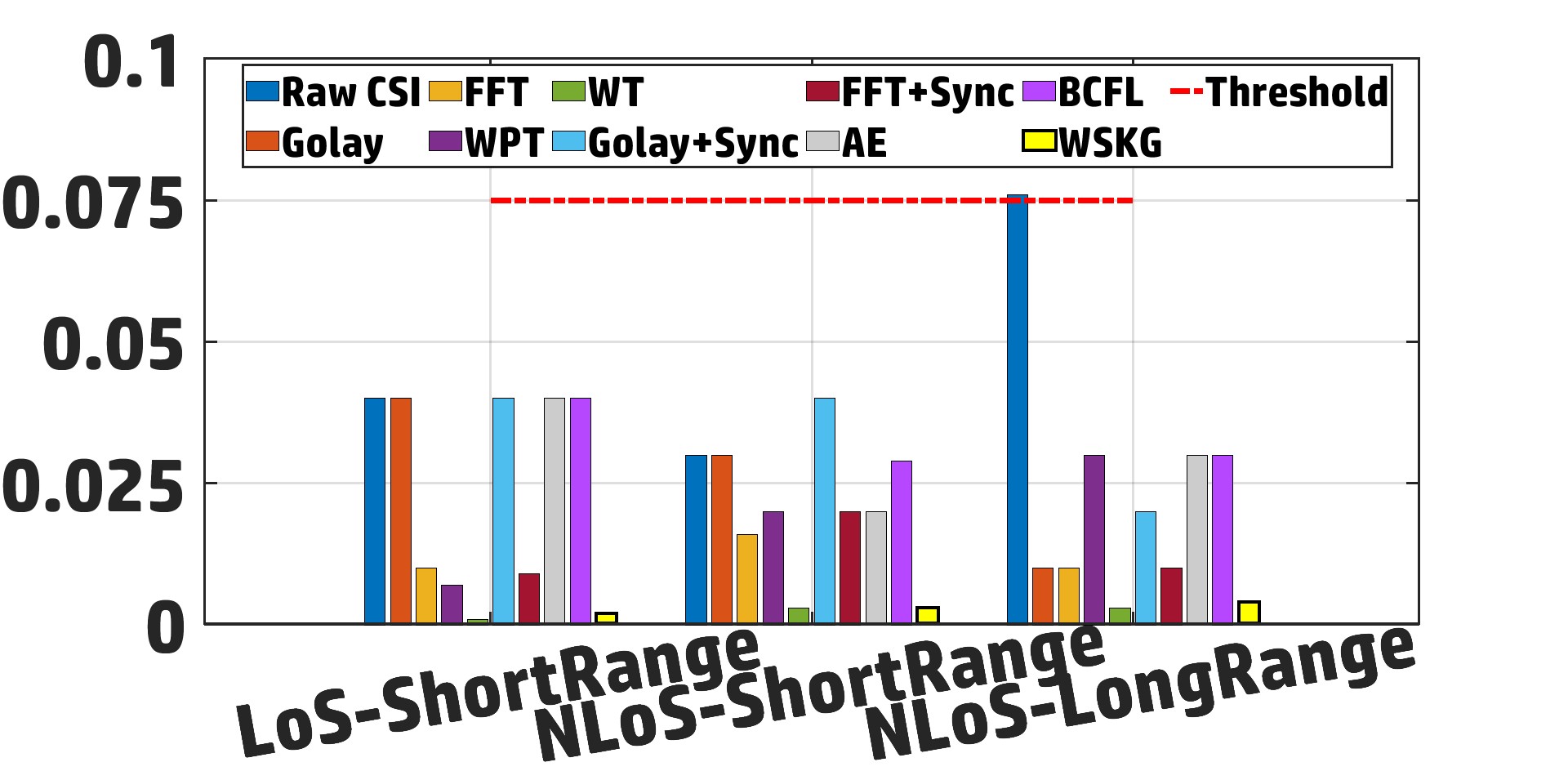}}
  \subfloat[Error = 20 bits\label{subfig:BER_20}]{%
        \includegraphics[width =0.3\textwidth]{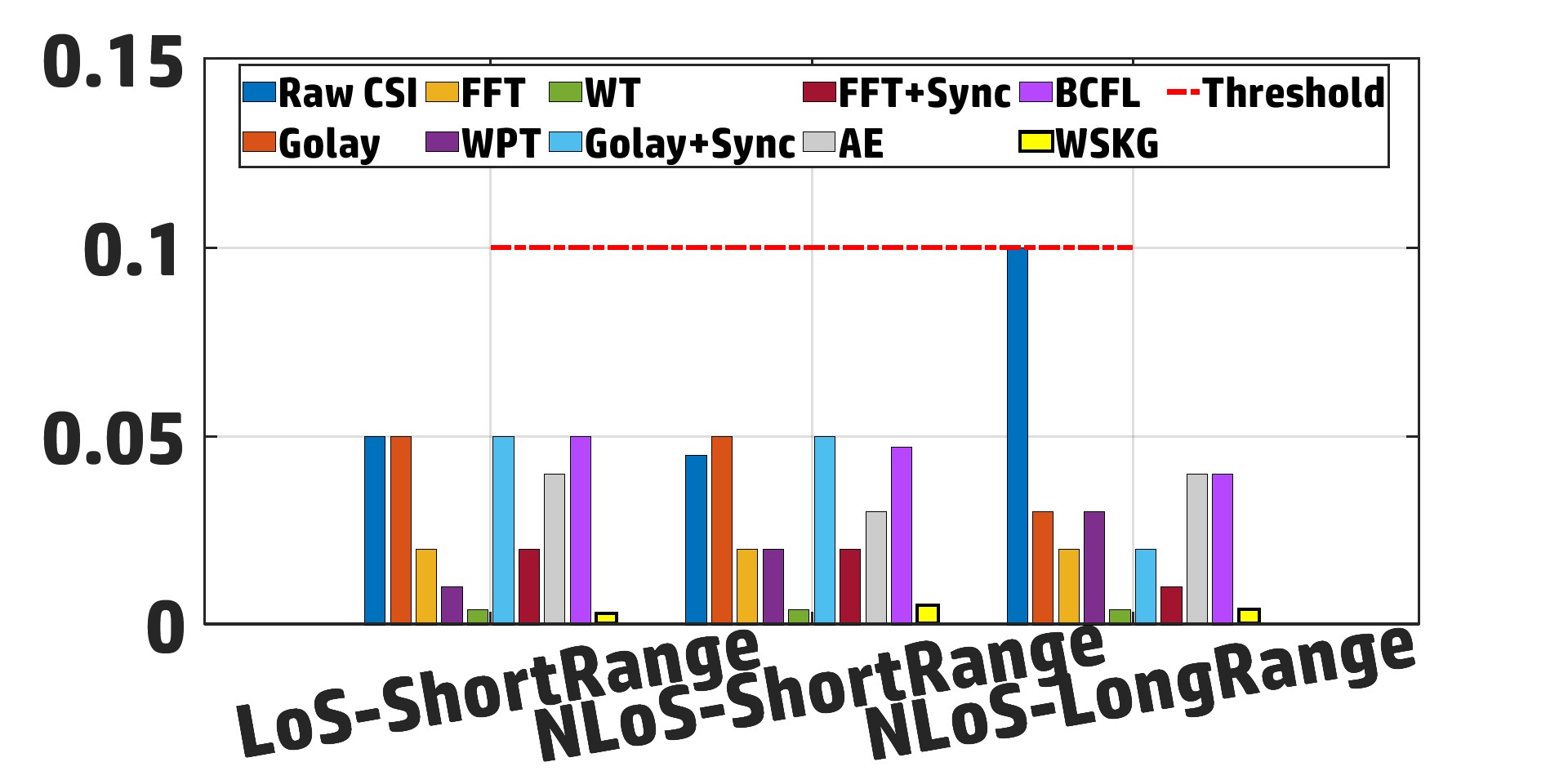}}
        }
   
   \caption{  Comparison of BER performances.  }
  \label{fig:BER} 
\end{figure*}

\begin{figure}
    \centering
    \includegraphics[width=1\linewidth]{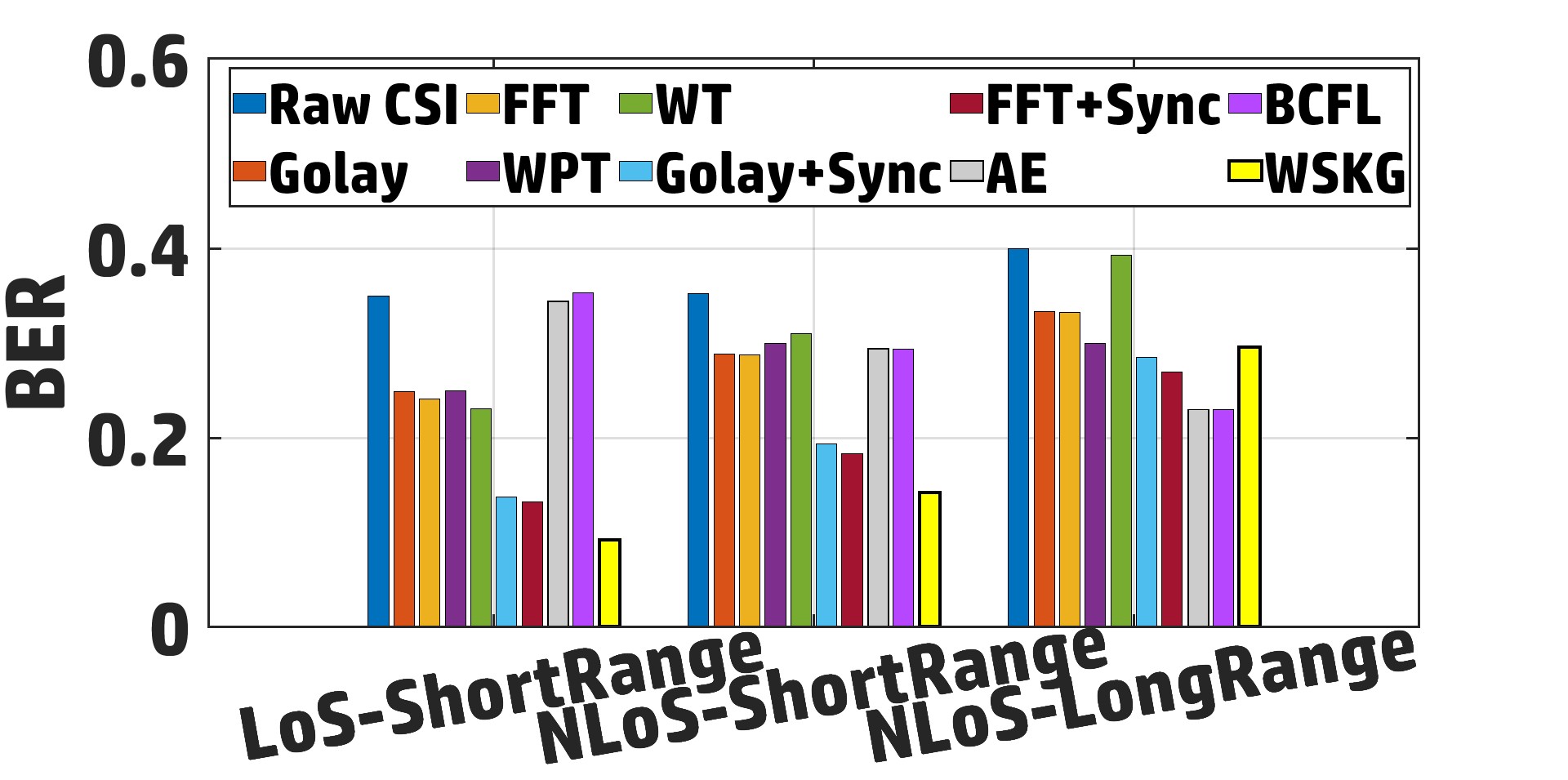}
    \caption{  Average overall BER for all generated sequences. }
    \label{fig:BER_overall}
\end{figure}

Fig. \ref{fig:KGR} depicts the performance of the proposed Wavelet-based Secret-Key Generation (WSKG) technique in terms of KGR at different bit-error threshold values under each of the three studied location scenarios. Fig. \ref{fig:BER} depicts the BER under each of the three studied location scenarios and for different bit-error threshold values, while Fig. \ref{fig:BER_overall} shows the average BER for all the generated keys regardless of the maximum allowed bit-error threshold values.The figures also compare WSKG to key generation using (baseline) raw CSI, FFT-based and Golay filtering-based CSI constructions with and without synchronization, as well as WPT, AE key generation, and BCFL-based key generation. 
%
%
\subsubsection{WSKG Performance Analysis}
\label{subsubsec:WSKG}
Overall, Figs.~\ref{fig:KGR},~\ref{fig:BER}, and \ref{fig:BER_overall} clearly show that the proposed \WSKG~scheme outperforms all other key generation schemes in both KGR and BER in almost all location scenarios and all error thresholds. 
For instance, under the \lab~scenario and compared to raw CSI, WSKG increases KGR from $4\times 10^{-4}$ to $3.5\times 10^{-3}$ bits per packet (Fig.~\ref{subfig:KGR_15}) and reduces BER from $0.04$ to $0.002$ (Fig.\ref{subfig:BER_15}).
The improved performance for the proposed \WSKG~scheme is also observed under \lc~and \lo.

Figs.~\ref{subfig:KGR_5} and \ref{subfig:BER_5} confirm \WSKG's superior performance under \lo~with a $5$-bit error threshold.
The figures consistently show \WSKG's superiority over synchronized FFT and Golay filtering CSI constructions across locations and error thresholds. \WSKG~utilizes WC's time and frequency information effectively, reconstructing CSI signals with coherent frequency components, while FFT and Golay filtering constructions remove high-frequency components, leading to limited reciprocity enhancement, repeated key agreement failures, and low KGR. 
The figures also show that the proposed WT- based CSI reconstruction without synchronization, has higher KGR and lower BER compared key generation using the Golay filtered CSI and the FFT reconstructed CSI under all location scenarios and bit-error thresholds.
Figs.~\ref{fig:KGR},~\ref{fig:BER}, and \ref{fig:BER_overall} also demonstrate that the proposed WSKG outperforms WPT key generation in KGR and BER in all location scenarios. The lower performance of WPT is attributed to its approach of nullifying small-valued coefficients of the discrete wavelet packet transformed CSI that fall below the median of the coefficients. Although this method helps mitigate some of the impact of noise on CSI, it fails to take advantage of the time and frequency information of WC, resulting in performance similar to Golay and FFT filtering.

%
Figs.~\ref{fig:KGR} and \ref{fig:BER} show FFT reconstruction's marginal improvement over Golay filtering across error thresholds and locations, consistent with their shared focus on high-frequency component elimination.
Figs.~\ref{fig:KGR} and~\ref{fig:BER} also show significantly low KGR and high BER for AE and BCFL key generation compared to Golay filtering, FFT reconstruction and the proposed WSKG scheme under \lo~and \lc. This observation is justified by the wireless channel temporal variation. During the training phase, the auto-encoder in the AE key generation scheme learns the correlated features between AP and STA using the available CSI training dataset. Similarly, the parameters of the bidirectional convolution network in the BCFL key generation scheme are optimized using the available CSI training dataset. As the channel varies over time, the trained models fail to extract the correlated features between the AP's and STA's recent CSIs estimated at the deployment time, resulting in degraded performance for the AE and BCFL schemes, making them equivalent to raw CSI key generation.
This observation suggests that a key generation scheme with varying thresholds or parameters that adapt with the channel variations (Golay filtering, FFT-reconstruction, and WSKG) is expected to have a better, non-degradable performance compared to deep learning schemes that do not account for the channel variations during the training phase.

\subsubsection{Impact of Synchronization}
Figs.~\ref{fig:KGR} and \ref{fig:BER} demonstrate synchronization's consistent enhancement of KGR and BER across locations and error thresholds. This improvement remains consistent across various preprocessing techniques, emphasizing the effectiveness of synchronization in aligning AP and STA CSIs through the time shift estimated through time-lagged cross-correlation, subsequently boosting CSI correlation and key generation performance. The improvement in KGR and BER due to synchronization is also observed for NLoS scenarios where high packet loss exists.
%
\subsubsection{Impact of AP \& STA Locations}
In Figs.\ref{fig:KGR}, and \ref{fig:BER}, key generation is influenced by AP and STA locations. \lab~exhibits the highest KGR and lowest BER, while \lo~shows the opposite due to harsher channel conditions and higher packet losses. This pattern holds across all key generation schemes.

Fig. \ref{fig:BER_overall} also illustrates the overall bit disagreement between the generated bit sequences under different location scenarios and techniques, and  shows a high overall BER under \lo~compared to \lc~and \lab~for all techniques due to noise and packet loss. Additionally, the figure demonstrates comparable overall BER for WSKG, AE, BCFL, synced Golay filtering and synced FFT-reconstruction of about $0.3$ due to harsher channel conditions.

\subsubsection{Impact of Error Thresholds}
%
%
%
%

Fig.~\ref{fig:KGR} shows that as the error threshold decreases, KGR decreases across all scenarios and the decrease is significant for the key generation schemes other than the proposed \WSKG. This observation is expected since a low error threshold value will cause many generated keys to be discarded if the number of bits in error between the keys of AP and STA exceeds the error threshold. The observation demonstrates the robustness of the proposed \WSKG~ to CSI noise. Fig. \ref{fig:BER} also demonstrates the impact of error thresholds on BER where a lower permissible bit-error threshold value produces keys with lower BER for all techniques under all location scenarios.

\subsection{Computational Complexity}
\begin{figure}
\centerline{
  \subfloat[Execution time in milliseconds \label{subfig:exe_time}]{%
       \includegraphics[width=0.6\columnwidth]{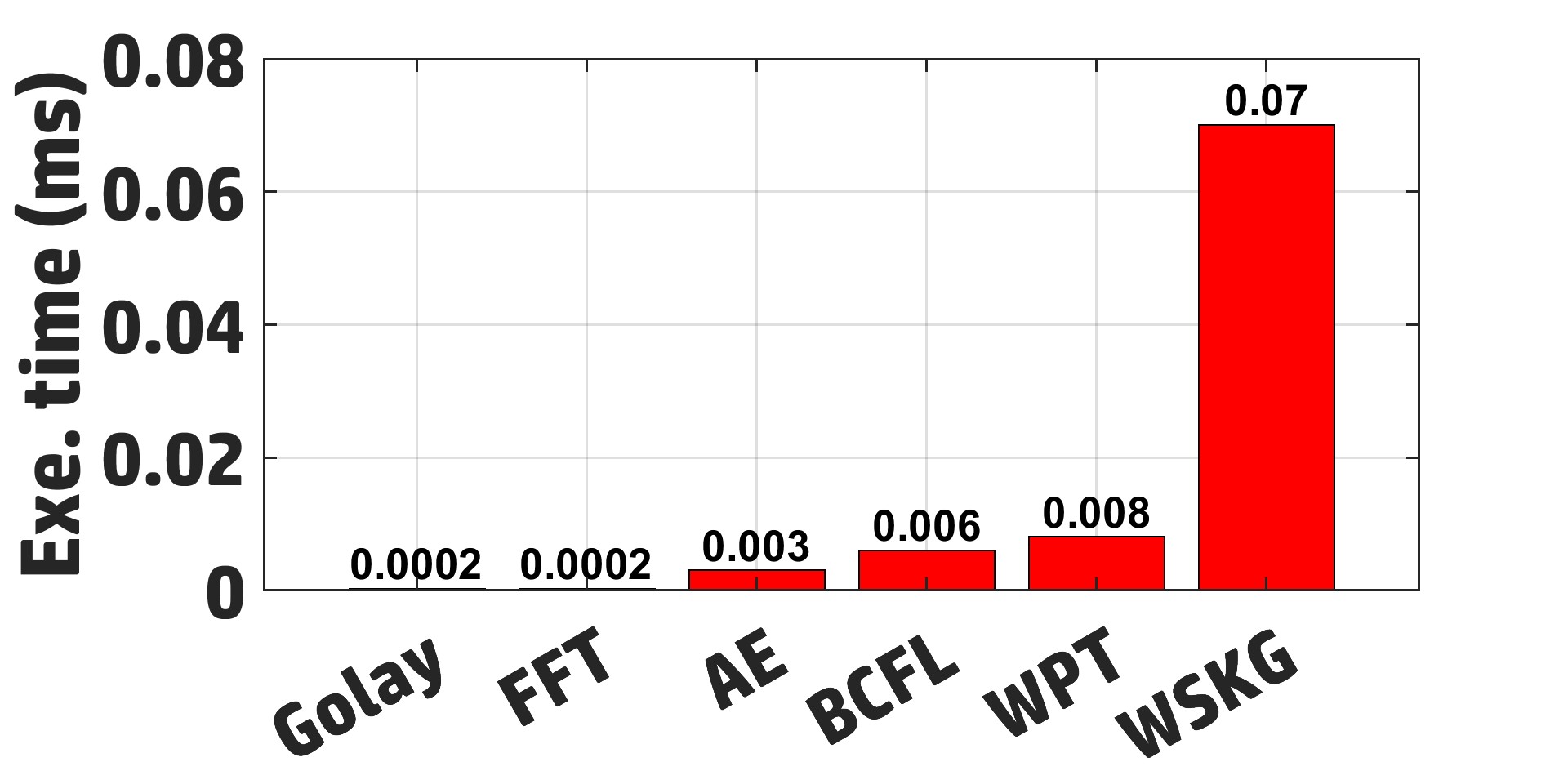}}
  \subfloat[Time composition\label{subfig:pi_chart}]{%
        \includegraphics[width=0.4\columnwidth]{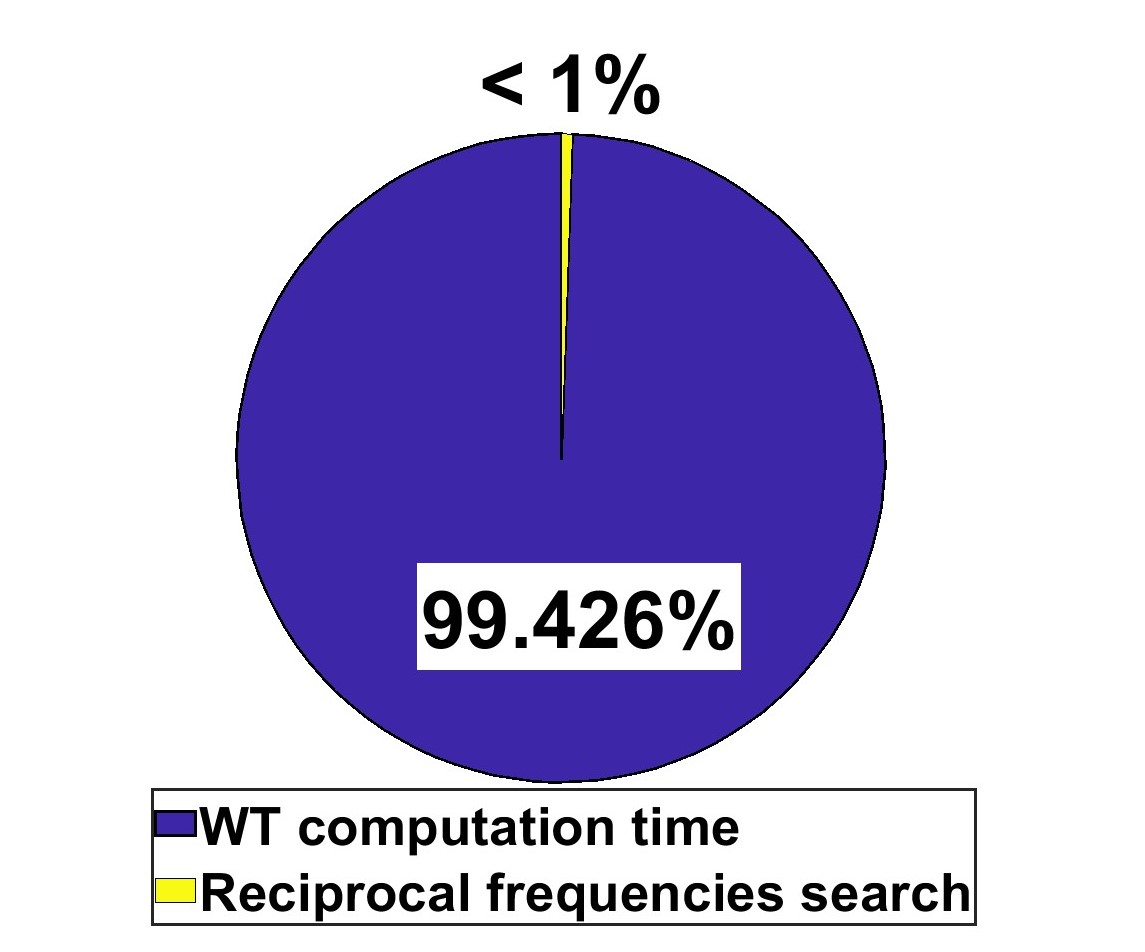}}}
\caption{  Computation complexity analysis of WSKG scheme. }
\label{fig:pi}
\end{figure}

In this section, we evaluate the computational complexity of the proposed WSKG approach and compare it to the previously described key generation benchmark schemes (Raw CSI, Golay filtering, FFT-reconstruction, AE, BCFL, and WPT). We measure the computational complexity of a key generation scheme by the average time required to preprocess the collected CSI samples before the key generation steps are executed. 
For each key generation technique, we computed the execution time of $41500$ CSI samples for $10$ rounds. The average execution time per sample in milliseconds is depicted in Fig.~\ref{subfig:exe_time}.
%
The figure shows that the computational complexity of the proposed WSKG approach is significantly higher than the benchmark techniques.
Fig.~\ref{subfig:pi_chart} depicts further analysis of the proposed WSKG scheme computational complexity and illustrates that the execution time of WSKG is primarily dominated by the computation of wavelet coefficients, while searching for the reciprocal frequency components between the AP and STA consumes only about $1\%$ of the execution time. Despite the high execution time of the proposed WSKG approach compared to the benchmarks, using our CSI data, we generated a 256-bit key using $128$ CSI samples in $10$ ms of execution time before information reconciliation. $10$ ms is the average execution time per key over $1125$ 256-bit sequences. This time is a reasonable delay for communication systems. 
Recent research works have been also proposed to provide fast wavelet transform computations for real time data processing \cite{arts_fast_2022}.

\subsection{Hardware Configurations and Environmental Parameters}
We now expand our experimental study and analyze the effectiveness of the proposed WSKG approach compared to the benchmark schemes while using multiple devices with different hardware configurations. We also study the impact of environmental parameters and deployment conditions on secret key generation.

\subsubsection{Expanded Experimental Setting}
\begin{figure}
\centerline{
\subfloat[LuatOS ESP32-C3\label{subfig:luatos}]{%
\includegraphics[keepaspectratio, height = 3 cm]{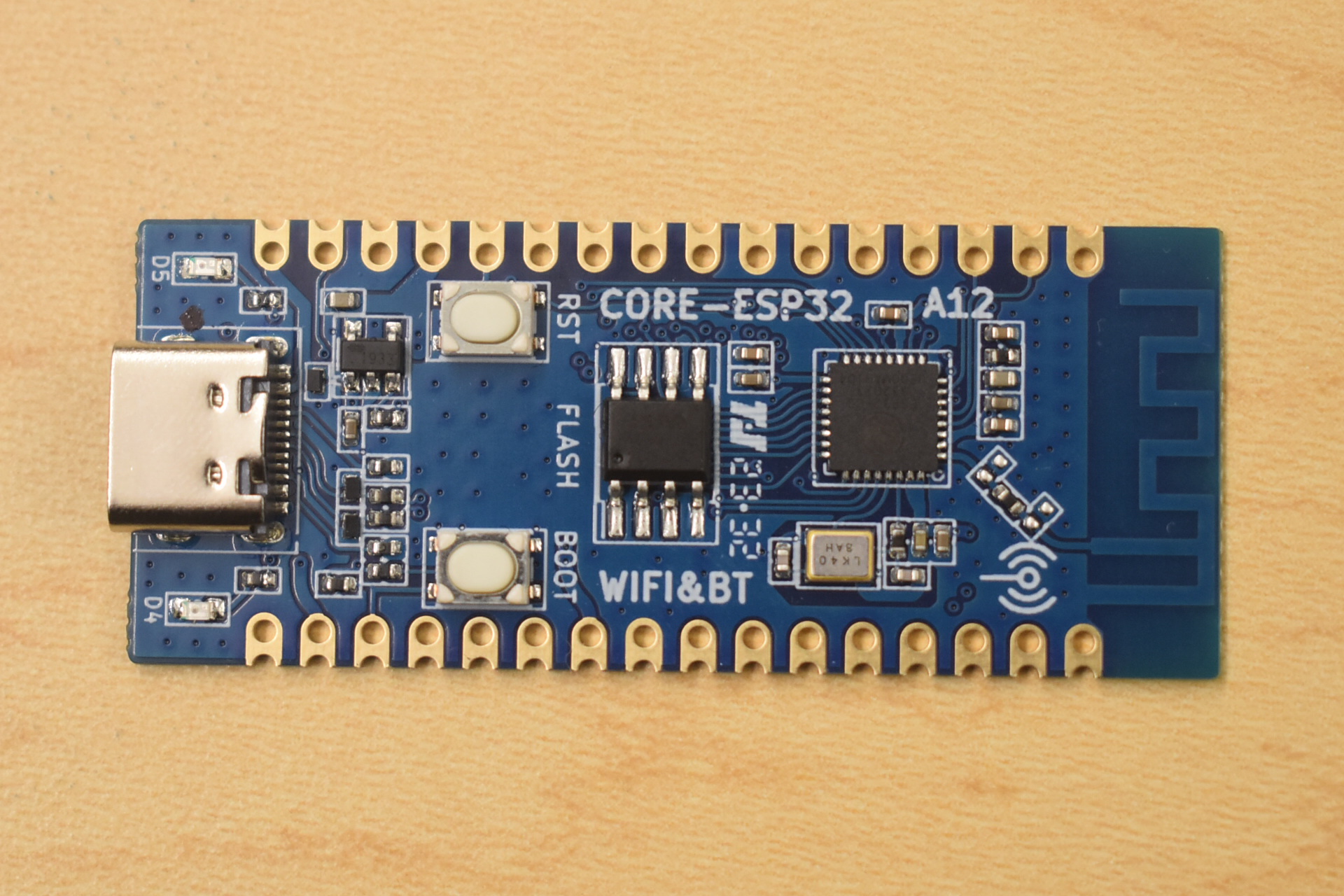} }
\subfloat[XIAO ESP32-C3\label{subfig:xiao}]{%
\includegraphics[keepaspectratio, height=3 cm]{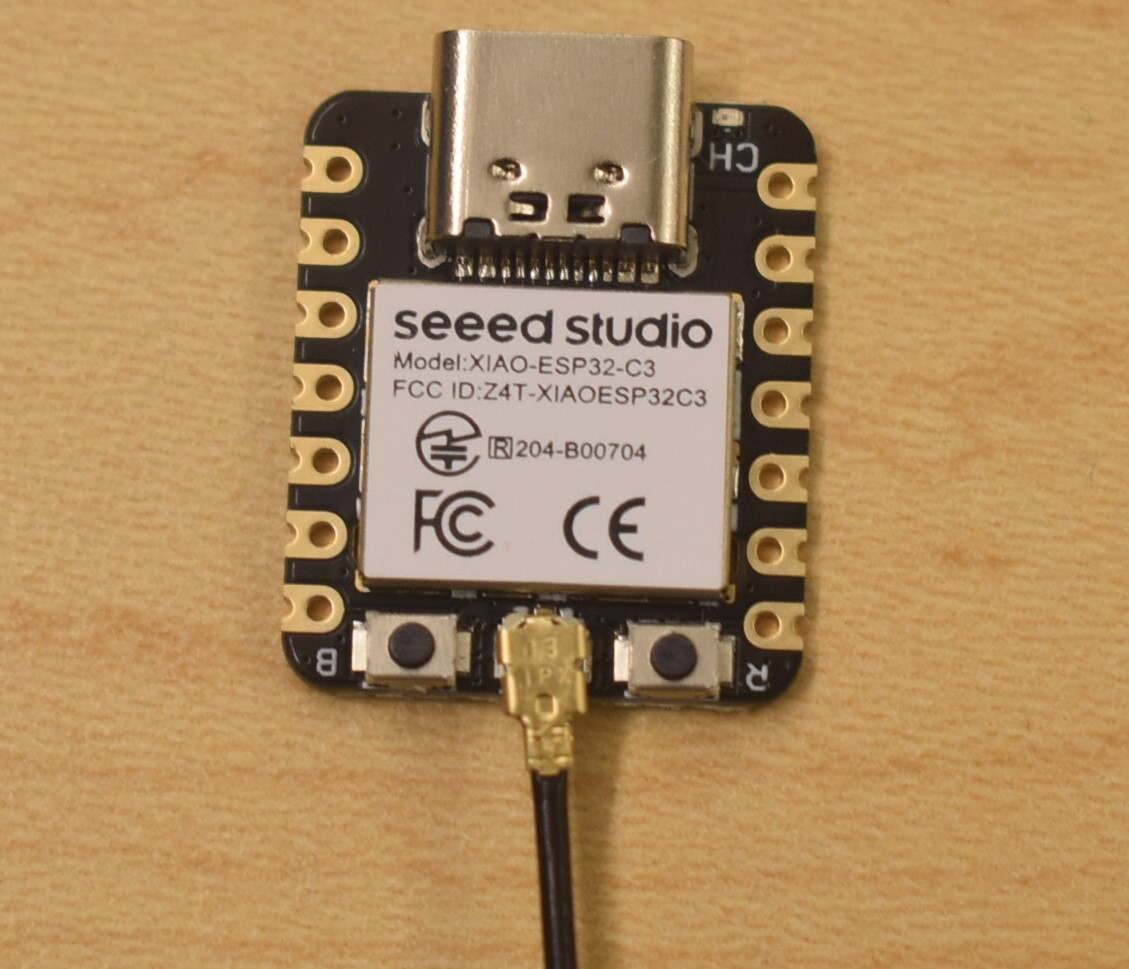}}
}
\caption{Expanded experiment setting devices.}
\label{fig:expanded}
\end{figure}

The setting consists of a WiFi network of $4$ devices with the following hardware configurations:
\begin{itemize}
    \item Two \textbf{Pycom} devices (Fig.~\ref{subfig:device}), one serving as AP and another as STA. The Pycom devices are development boards based on the Espressif ESP32 System on Chip (SoC).
    \item One \textbf{LuatOS} device (Fig.~\ref{subfig:luatos}) serving as STA. LuatOS development boards are based on the Espressif ESP32-C3 SoC.
    \item One Seeed Studio \textbf{XIAO} device (Fig.~\ref{subfig:xiao}) serving as STA. Seeed Studio XIAO devices are compact development board based on the Espressif ESP32-C3 SoC with an external antenna to increase the signal strength.
\end{itemize}
AP and STAs communicate using IEEE 802.11n WiFi protocol at 2.427 GHz. AP exchanges with each STA packets at a rate of $10$ packets per second. During CSI data collection using Pycom, LuatOS, Xiao boards, we observed high packet loss at packet exchange rates higher than $10$ packets per second. We also observed that the packet loss was even worse when more than $2$ STAs were connected to AP. Therefore, and to mitigate this hardware limitation, the packet exchange rate is fixed at $10$ packets per second during CSI data collection.

The AP and STAs collect CSI (Channel State Information) using the ESP23 CSI toolkit. The devices are connected to two Windows machines for data collection and processing via USB ports at a baud rate of $115200$. The devices estimate CSI using both the non-HT Legacy Long Training field (L-LTF) and the High Throughput Long Training Field (HT-LTF) of the WiFi physical layer frame. 
The network is used to evaluate the proposed WT-based secret key generation approach in two different RF-rich environments:
\begin{itemize}
    \item \textbf{Indoor, RF-rich environment:} The devices are located in a room of size 9 meters × 9.6 meters in the RF-rich environment of Kelley Engineering Center within the range of the wireless service of OSU and other RF interference sources such as the Bluetooth building management system, and other personal Bluetooth and WiFi networks.
    \item \textbf{Outdoor, RF-rich environment:} The devices are located outdoors on OSU campus in an RF-rich environment within the range of the wireless service of OSU and other RF interference sources. The outdoor environment is exposed to sunlight and at $30$ \textdegree C. 
\end{itemize}
For both environments, the STAs are located at different distances from AP and at different distances from each other.
For each environment, we collected $4$ datasets of raw CSI data from each device in the network, and the data collection duration for each dataset is $30$ minutes. 
We evaluate the performance of key generation using KGR and BER at 20-bits error threshold, and compare the performance of the proposed WSKG approach against raw CSI as well as Golay filtering-based key generation proposed in \cite{junejoLoRaLiSKLightweightShared2022a}, and the denoising autoencoder (AE) technique proposed in \cite{zhouPhysicalLayerSecret2022}.
Figs.~\ref{fig:Indoor_multiple_devices} and~\ref{fig:outdoor_multiple_devices} show the KGR and BER  for all generated sequences for the proposed WSKG scheme, raw CSI, Golay filtering-based key generation, and AE key generation when deployed by a network of $4$ devices (AP and $3$ STAs) of different hardware configurations in indoor and outdoor, RF-rich environments.
\subsubsection{Impact of Environmental Parameters} 
Figs.~\ref{fig:Indoor_multiple_devices} and~\ref{fig:outdoor_multiple_devices} clearly show an overall degradation in both KGR and BER in the outdoor environment compared to the indoor environment for all devices and all key generation schemes. These results are expected and attributed to the high noise level in the outdoor environment which impacts the reciprocity of the channel measurements between STAs and AP, and hence reduces KGR and increases BER. 
Figs.~\ref{subfig:indoor_KGR}, and \ref{subfig:outdoor_KGR} depict KGR for indoor and outdoor environments. The figures demonstrate that the proposed WSKG scheme outperforms all benchmark schemes in terms of KGR for all devices configurations in indoor and outdoor environments. For instance, WSKG doubles KGR compared to the denoising autoencoder (AE) and Golay filtering for all devices in both indoor and outdoor environments.
The figures also illustrate the severely deteriorated channel reciprocity in the outdoor, RF-rich environment where key generation fails for raw CSI for all devices (Fig.~\ref{subfig:outdoor_KGR}).
Figs.~\ref{subfig:indoor_BER}, and~\ref{subfig:outdoor_BER} show the improved performance of the proposed WSKG scheme over the benchmark schemes in terms of BER. For instance, WSKG improves BER for the indoor for Pycom to $0.23$ compared to $0.37$ for the raw CSI, $0.28$ for Golay filtering, and $0.26$ for AE. Lower BER values for the proposed WSKG compared to the benchmark schemes are also observed for the outdoor environment.
Figs.~\ref{fig:Indoor_multiple_devices} and~\ref{fig:outdoor_multiple_devices} also show low KGR and high BER for the de-noising AE compared to Golay filtering and the proposed WSKG scheme. This observation is commensurate with the results in Sec.~\ref{subsubsec:WSKG} and is justified by the wireless channel temporal variation. As the channel varies with time, the auto-encoder's learned latent space no longer represents the current channel, and the auto-encoder fails to extract the correlated features between the AP and STA CSIs.
\subsubsection{Impact of Hardware Configurations}
\begin{figure} 
        \centerline{
      
  \subfloat[BER\label{subfig:indoor_BER}]{%
        \includegraphics[keepaspectratio, width=0.25\textwidth]{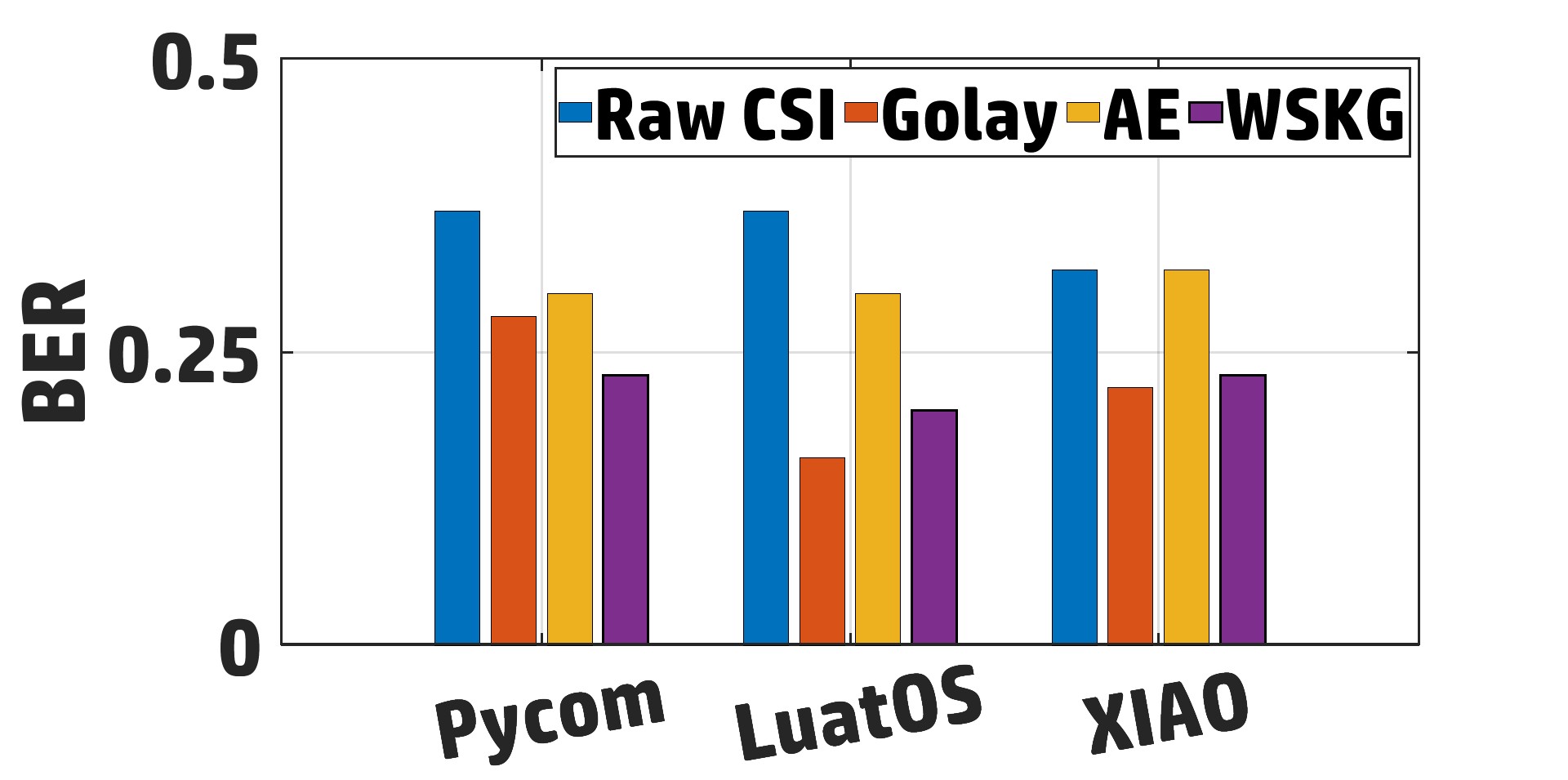}}
\hspace{-0.1in}
    
  \subfloat[KGR\label{subfig:indoor_KGR}]{%
        \includegraphics[keepaspectratio, width=0.25\textwidth]{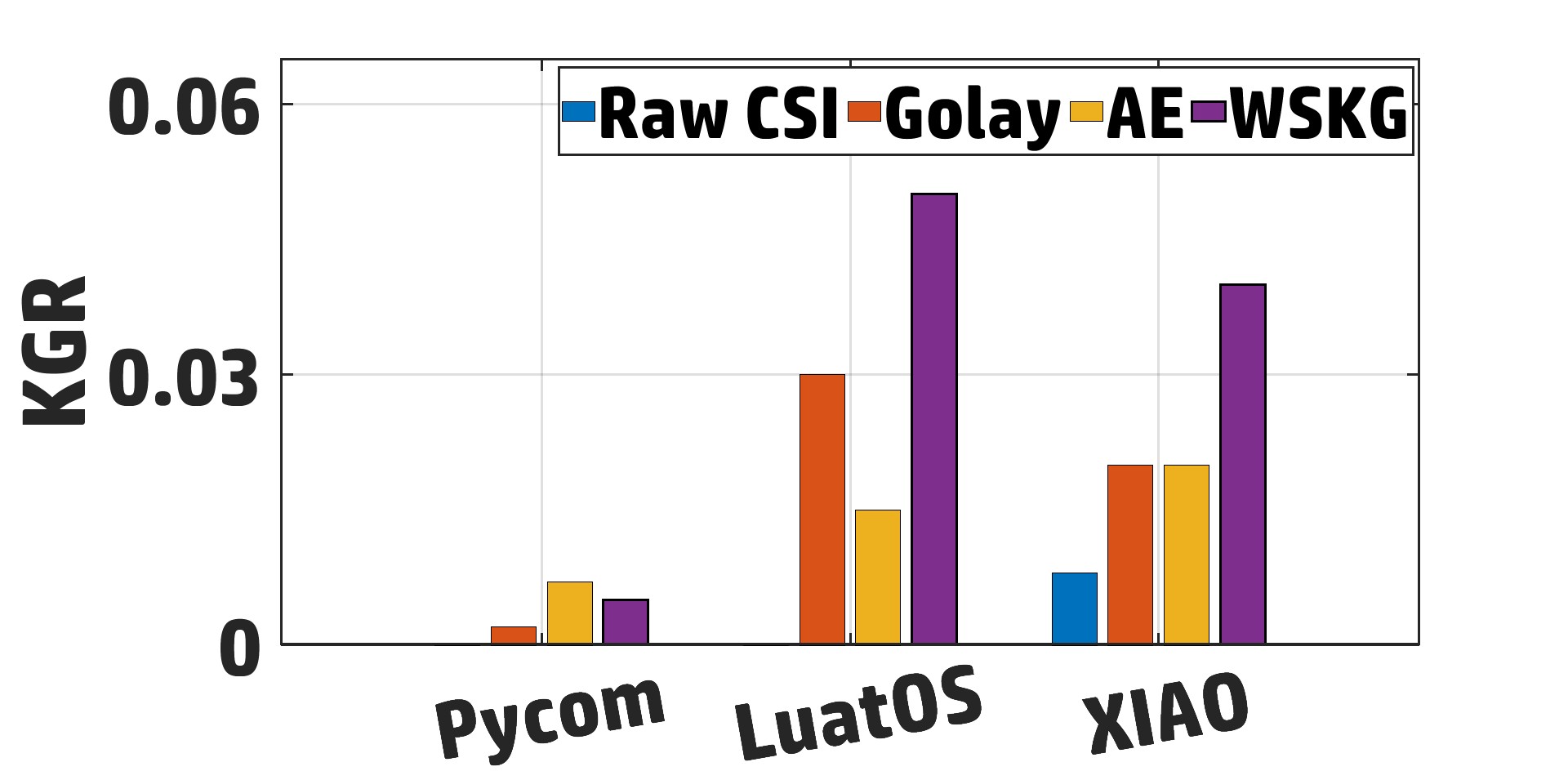}}}
        
   \caption{  Key generation performance for different hardware configurations in \textbf{Indoor, RF-rich environment}. }
  \label{fig:Indoor_multiple_devices} 
\end{figure}

\begin{figure} 
        \centerline{
      
  \subfloat[BER\label{subfig:outdoor_BER}]{%
        \includegraphics[keepaspectratio, width=0.25\textwidth]{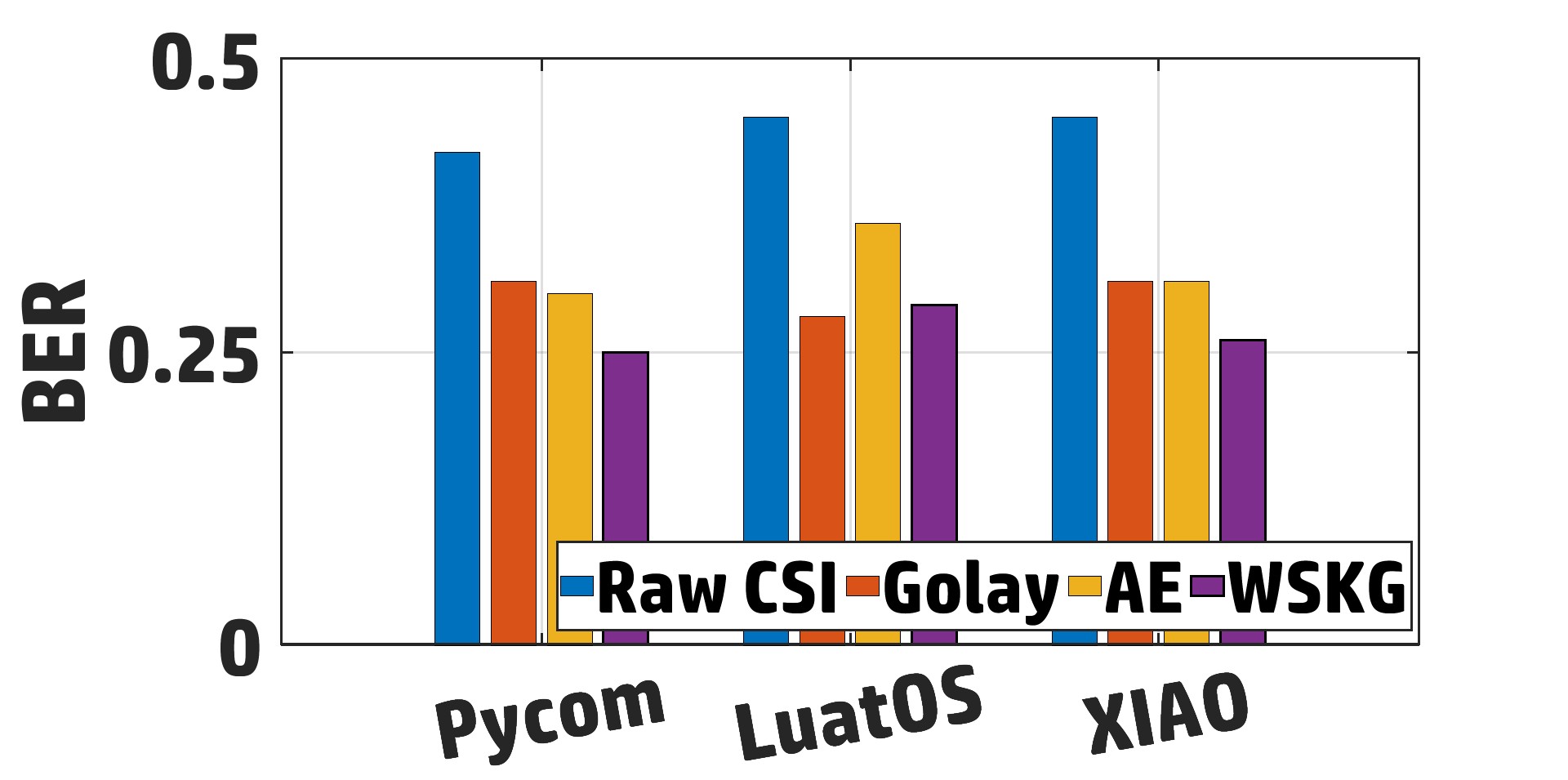}}
\hspace{-0.1in}
    
  \subfloat[KGR\label{subfig:outdoor_KGR}]{%
        \includegraphics[keepaspectratio, width=0.25\textwidth]{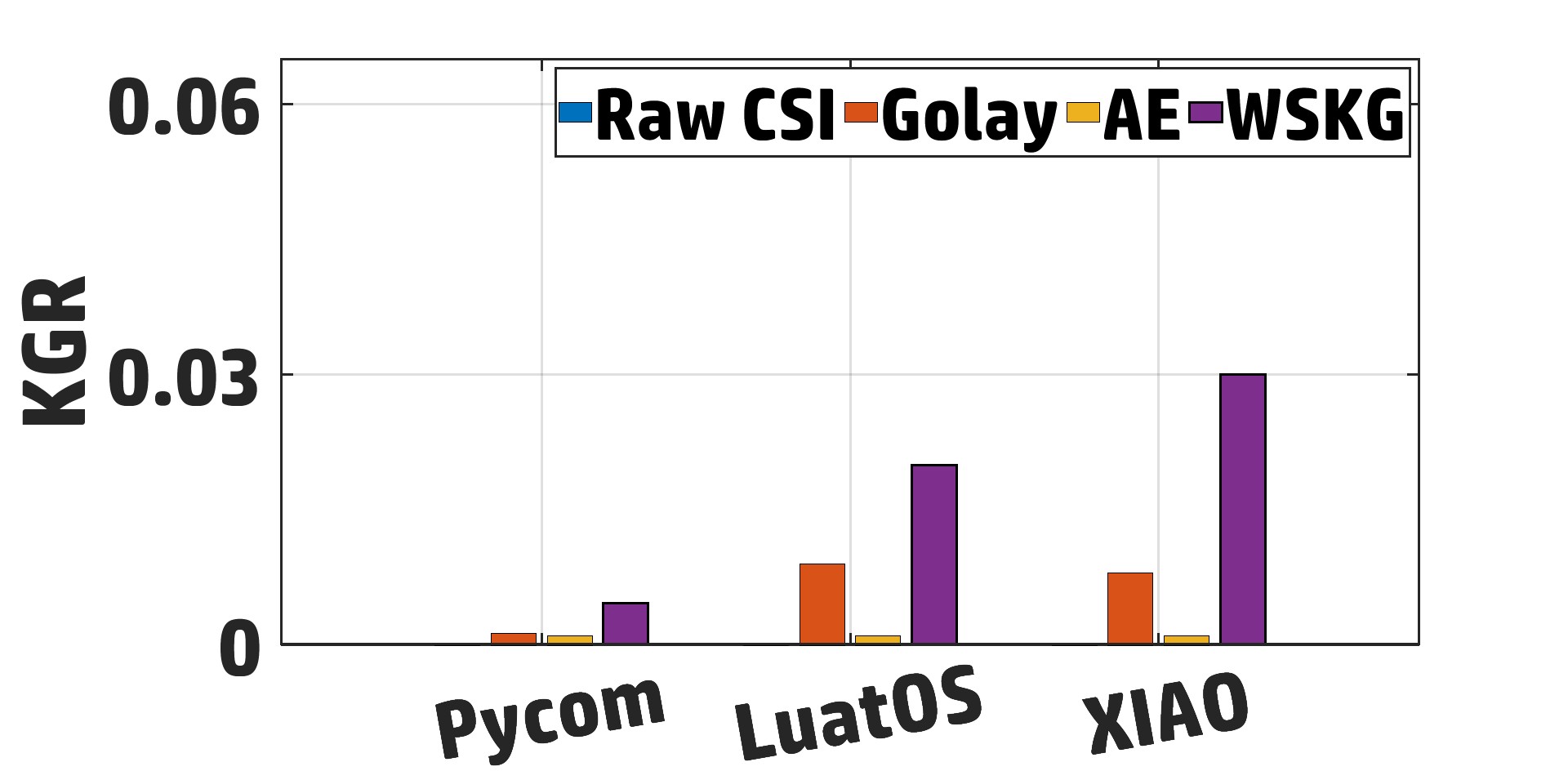}}}
        
   \caption{  Key generation performance for different hardware configurations an \textbf{Outdoor, RF-rich environment}. }
  \label{fig:outdoor_multiple_devices} 
\end{figure}

Figs.~\ref{fig:Indoor_multiple_devices} and~\ref{fig:outdoor_multiple_devices} also capture the impact of hardware configurations on KGR and BER. The figures show an overall comparable KGR and BER for LuatOS and XIAO devices compared to Pycom which has significantly lower KGR and higher BER for all key generation schemes as well as the raw CSI. This trend is observed in both indoor and outdoor environments. 
Raw CSI key generation is expected to marginalize the impact of CSI preprocessing and provide a clearer representation of the impact of hardware configurations on key generation. Fig.~\ref{subfig:indoor_KGR} which depicts KGR for all devices for the indoor environment shows that XIAO device has the highest KGR ($0.008$) compared to Pycom and LuatOS ($0.0002$) when the raw, unprocessed CSI is used for key generation. slightly lower BER is also observed for XIAO compared to LuatOS and Pycom for raw CSI in Fig.~\ref{subfig:indoor_BER}.
However, XIAO device's improved key generation for raw CSI is not observed in the outdoor environment due to the high noise level which limits KGR with raw CSI to $0$ for Pycom, LuatOS, and XIAO devices.
Figs.~\ref{fig:Indoor_multiple_devices} and~\ref{fig:outdoor_multiple_devices} also show that despite the improved performance of the proposed WSKG compared to raw CSI key generation, WSKG key generation performance is limited by the hardware configuration and the device key generation performance. For instance, in Fig.~\ref{subfig:outdoor_KGR}, WSKG improves KGR to $0.005$ compared to $0.0002$ for the raw CSI for Pycom. However, LuatOS and XIAO devices has about $10$ times higher KGR ($0.05$) for WSKG for the indoor environment. The same trend is observed for the outdoor environment as well as all the key generation benchmark schemes.
Our experimental results show that different hardware configurations have different key generation performances. The available documentations for the Espressif ESP32 chip in Pycom and the more recent ESP32-C3 chip in LuatOS and XIAO indicate that both chips have similar WiFi capabilities. However, according to the Espressif website, the ESP32-C3 chip (LuatOS and XIAO) maintains better RF performance at higher operating temperatures. This enhanced RF performance might contribute to the better key generation performance observed in LuatOS and XIAO boards. Additionally, the devices used in the experiments are development boards provided by different manufacturers and based on the ESP32 or the ESP32-C3 chips. The development boards provide an entire WiFi subsystem interfaced with ESP32 in Pycom and with ESP32-C3 in LuatOS and XIAO. This WiFi systems variations are expected to impact the channel estimation accuracy and thereby the key generation performance. Unfortunately, detailed information on channel estimation and the specifications of the WiFi radio of the development boards are not provided in the available documentations. This challenges the full understanding of how key generation performance is impacted by the channel estimation methods and the existing RF impairments.

%

%
\section{Secure Device Authentication} 

\label{sec:replay}
Packets/frames replay is a necessary step of sophisticated attacks on WiFi networks such as KRACK and multi-channel man-in-the-middle that require the establishment of a rogue AP \cite{vanhoefAdvancedWiFiAttacks2014}. Those attacks impact WiFi networks utilizing WPA2 and even WPA3 \cite{vanhoefKeyReinstallationAttacks2017,vanhoefOperatingChannelValidation2018}. In this section, we propose a CSI-based secure device authentication technique that enables the detection of those kinds of attacks based on the channel extracted features. 
\begin{figure} 
    \centerline{
          \includegraphics[keepaspectratio, height=5cm,width=0.5\textwidth]{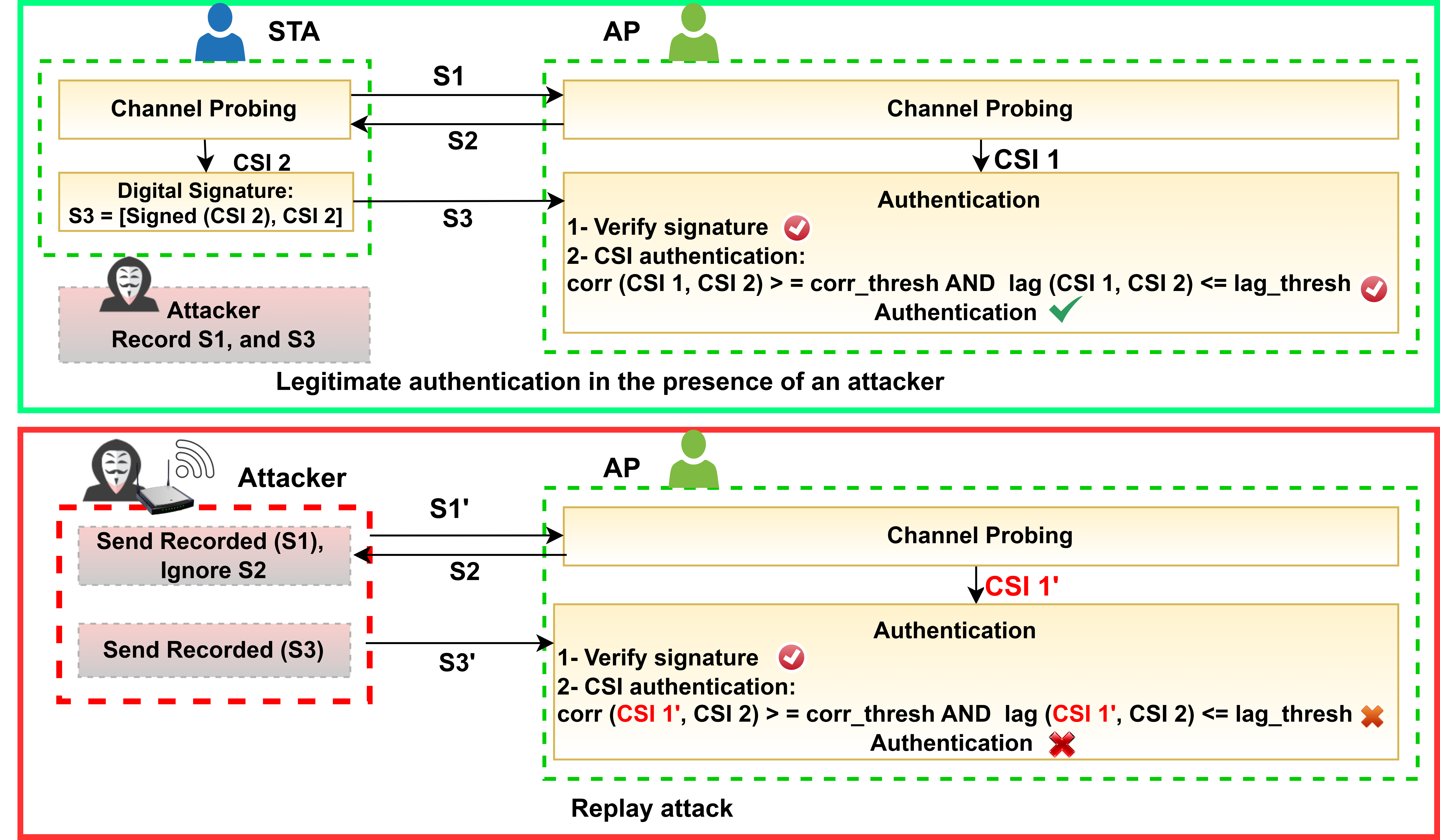}}
        
       \caption{CSI-assisted authentication.}
  \label{fig:replay attack} 
\end{figure}
CSI reciprocity and temporal variation can be exploited to enable device identity verification and replayed signal detection by using the proposed CSI-assisted authentication depicted in Fig.~\ref{fig:replay attack}. 
During (legitimate) authentication, both AP and STA estimate their CSI, $CSI_1$ and $CSI_2$, by exchanging probing signals, $S_1$ and $S_2$. Next, STA sends $S_3=[Signed(CSI_2),CSI_2]$ to AP. Then, after signature verification, the AP correlates $CSI_2$ with $CSI_1$. Because of channel reciprocity at the time of exchanging the probing signals, the correlation between AP's CSI and STA's CSI is expected to be high and the time shift is expected to be small. 
Now an attacker aiming to launch a replay attack records signals $S_1$ and $S_3$, exchanged between AP and STA earlier, and uses them to authenticate to AP at a later time; see Fig.~\ref{fig:replay attack}. 
More specifically, during the replay attack, the attacker sends $S_1^{'}=Recorded (S_1)$ and AP estimates new $CSI_1^{'}$ from the replayed signal. 
Next, the attacker sends $S_3^{'} = Recorded (S_3)$. Upon receiving $S_3^{'}$, AP then correlates $CSI_1^{'}$ and $CSI_2$ that is included in $S_3^{'}$. Due to the CSI temporal variation, the AP can then detect the replay attack, as the resulting correlation is going to be low and the time shift is going to be high.
%
Note that an attacker can still estimate new CSI after receiving $S_2$ from AP (which is expected to be highly correlated with $CSI_1^{'}$), but it cannot sign it as it lacks the proper (i.e., legitimate user's) signing key. 

\subsection{CSI Temporal Variation}
The proposed CSI handshake authentication approach relies on channel reciprocity and CSI temporal variation between AP and STA to prevent an illegitimate STA (an attacker) from replaying a previously recorded authentication message to authenticate with AP as a legitimate STA. In this section, we demonstrate through experimental measurements how CSI temporal variation can be used to detect such replay attacks. We used the testbed setting shown in Fig.~\ref{fig:exp1} to have each of AP and STA collect $7$ blocks of CSI samples, each of duration $10$ minutes, at different times over two days. In this setting, STA and AP are located $60$ cm apart (LoS). 
Comparing CSI blocks collected by AP and STA at approximately the same time mimics what would occur when the legitimate STA wants to authenticate with AP using the proposed CSI handshake authentication, whereas comparing new AP's CSI with an old STA's CSI mimics an illegitimate STA (attacker) launching a replay attack and trying to authenticate with AP, as illustrated in Fig.~\ref{fig:replay attack}.

\begin{figure}
\includegraphics[keepaspectratio, height = 4 cm]{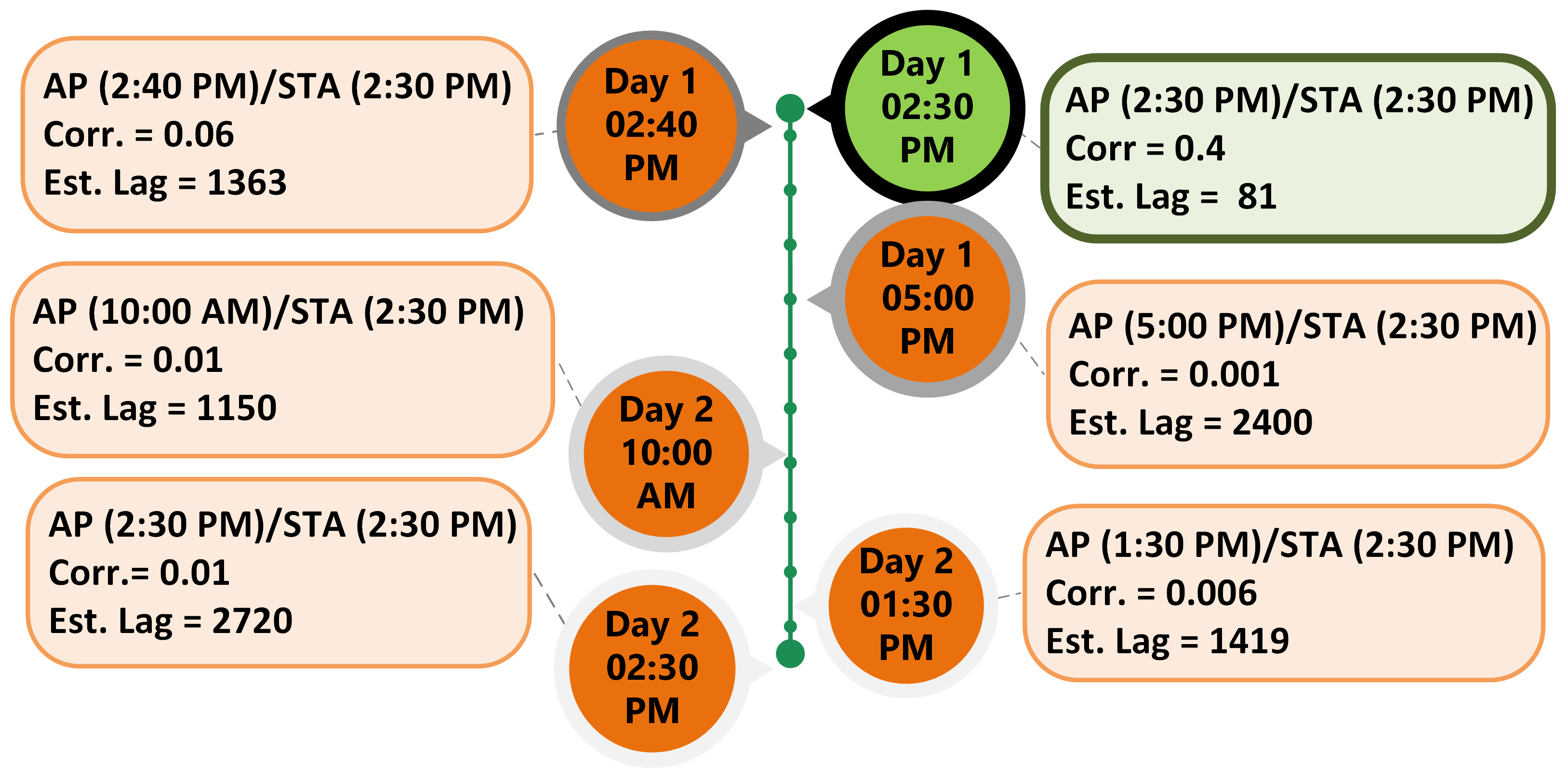}
\vspace*{-4mm}
\caption{Correlation and estimated time shift/lag at different time gaps between AP's and STA's CSI collections.}
\label{fig:timeline}
\end{figure}

Fig.~\ref{fig:timeline} shows the correlation and time shift of STA's and AP's CSIs when both are collected at 2:30 PM Day 1 (green ring) and when STA's CSI block is collected at 2:30 PM Day 1 but AP's CSI block is collected at a later time (orange rings). 
Observe that while same-time (concurrent) collections of CSI blocks (green ring) yield high correlation with relatively small time shift, when the AP's collection takes place after that of STA (orange rings) the correlation and the time shift deviate significantly. Note that even when the AP's collection takes place only 10 minutes after that of STA (top orange ring), the correlation drops to $0.06$ and time shift goes up to $1363$ from just $81$.
The decrease in the correlation value and the increase in the time shift persist regardless of the time when the AP's CSI collections occur, all due to the time-varying changes in the channel conditions. In the next section, we demonstrate how such a dependence of the correlation and time shift on the collection times will be exploited to detect replay attacks.

%
       
%
\subsection{Replay Attack Detection}
\begin{figure}
\centerline{ 
\subfloat[HackRF One and USRP \label{subfig:devices}]{%
\includegraphics[keepaspectratio, height = 3 cm]{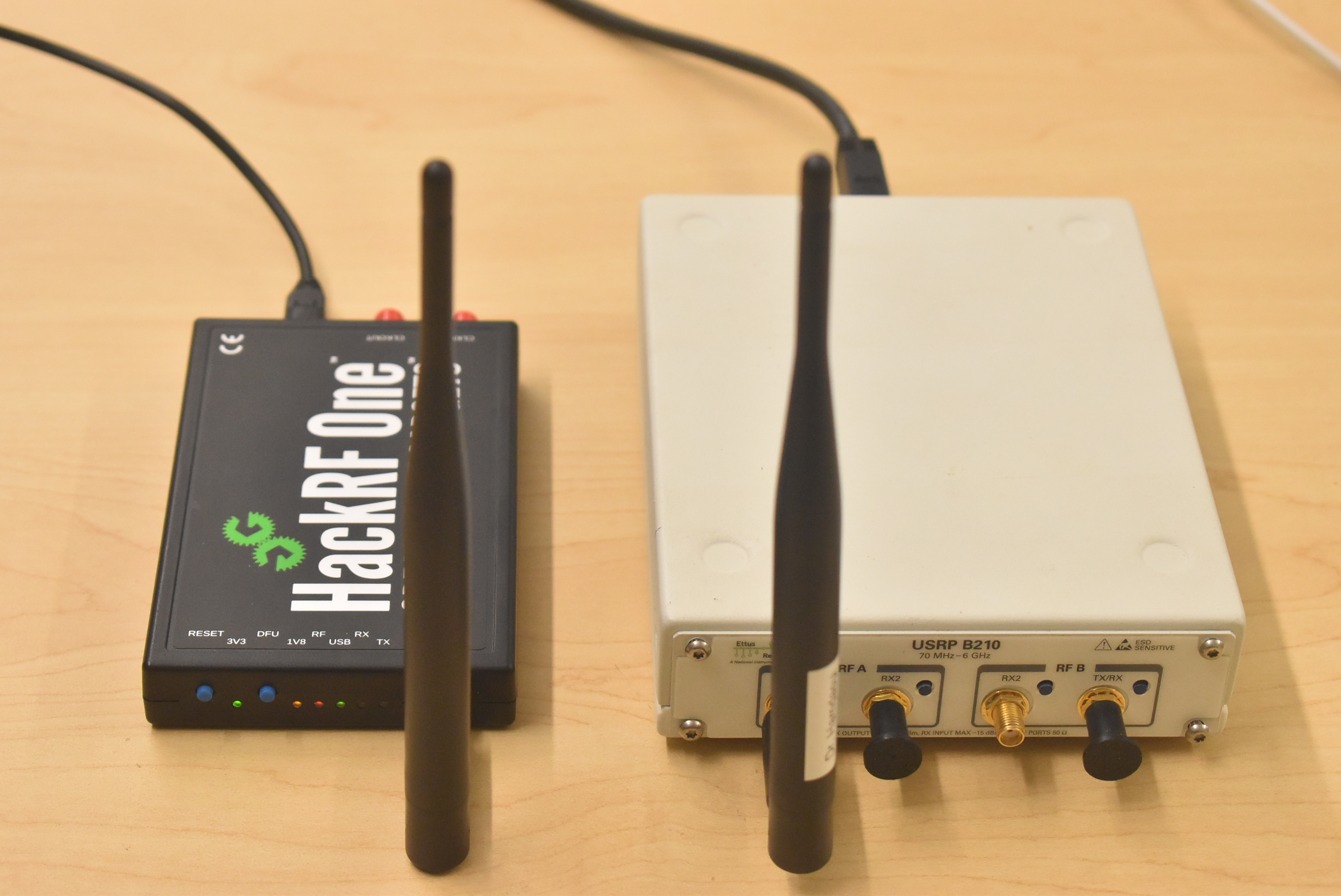} }
\subfloat[Replay attack setting\label{subfig:replay_setting}]{
\includegraphics[keepaspectratio, height= 3 cm ]{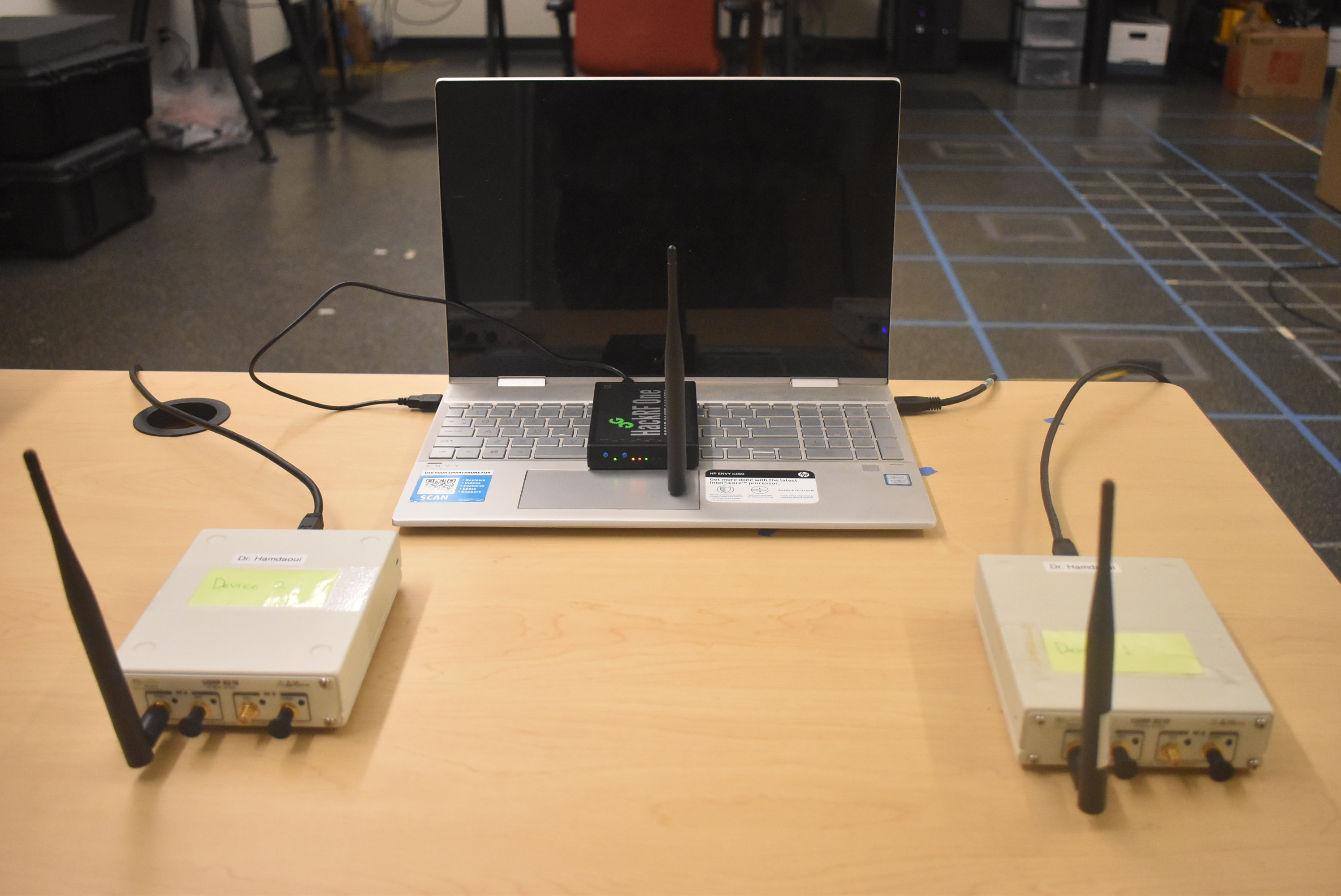}}}

\caption{CSI handshake authentication testbed setup.}
\label{fig:exp3}
\end{figure}
%
In this section, we implement the proposed CSI handshake authentication approach and demonstrate its resiliency against replay attacks.
For this, we use two USRP B210 and one HackRF One devices (shown in Fig.~\ref{subfig:devices}) to serve as an AP, an STA, and an attacker, respectively. Utilizing IEEE 802.11g, the USRP devices exchange packets at a rate of 1 per second at $2.427$ GHz and estimate CSI for 1 minute under the LoS setting, as depicted in Fig.~\ref{subfig:replay_setting}. The HackRF attacker records RF signals exchanged between AP and STA and later attempts to replay them to launch the replay attack, as described in Fig.~\ref{fig:replay attack}. 
Table \ref{Tb3} summarizes results from 12 legitimate authentication and 12 replay attack experiments conducted over three days. Legitimate authentication results in correlation of about $0.8$ and time shift of approximately $1$ sample, while the replay attack yields lower correlation of $0.04$ and higher time shift of $330$ samples. Establishing appropriate threshold values for legitimate CSI correlation and time shift can safeguard against replay and identity theft attacks.

\begin{table}
\centering 
\resizebox{1\columnwidth}{!}{\tiny \begin{tabular}{|l|c|c|} 
       \hline 
       Scenario  & Correlation & Time shift\\
       \hline
       \textbf{Legitimate Authentications}  & 0.8 & 1 \\
       \hline
       \textbf{Replay Attacks} & 0.04 &  330\\
       \hline
\end{tabular}}
\caption{\small Correlation and time shift values calculated under legitimate authentications and replay attacks.}
\label{Tb3}
\end{table}
%

\section{Conclusion}
\label{sec:conc}
This work emphasizes the critical role of channel reciprocity in enabling Physical Layer Security for resource-constrained devices. Our experimental investigations show that raw CSI's correlation drops dramatically, degrading the channel reciprocity substantially. We experimentally demonstrated that channel reciprocity is influenced by channel impairments and asynchronous measurements and is best quantified using wavelet coherence, Pearson's correlation, and time-lagged cross-correlation. We proposed a Wavelet-based secret-key generation scheme employing wavelet transform-based reconstruction and synchronization that doubles the key generation rate compared to Golay-filtered CSI. CSI's temporal variations impact Pearson's correlation and estimated time shift, a physical limitation that we leveraged to propose a CSI handshake-based protocol that increases the robustness of device authentication against replay attacks.

\bibliographystyle{IEEEtran}
\bibliography{IEEEabrv,References}

\end{document}